\pdfoutput=1
\newcommand*{\ATLASLATEXPATH}{}
\documentclass[PAPER, cernpreprint, texlive=2016, USenglish]{\ATLASLATEXPATH atlasdoc}
 
\usepackage{\ATLASLATEXPATH atlaspackage}
\usepackage{\ATLASLATEXPATH atlasbiblatex}
 
\usepackage{\ATLASLATEXPATH atlasphysics}
\AtlasJournalRef{Phys. Rev. D 101 (2020) 012002}
\AtlasDOI{10.1103/PhysRevD.101.012002}
 
\graphicspath{{logos/}{figures/}}
 
\usepackage{ANA-HIGG-2018-57-PAPER-defs}
\usepackage{multirow,bigdelim}
\usepackage{pdflscape}
\usepackage{cellspace}
\usepackage{amssymb}
\usepackage[shortlabels]{enumitem}
\usepackage{siunitx}
\usepackage{float}

\AtlasTitle{Combined measurements of Higgs boson production and decay using up to $80$ fb$^{-1}$ of proton--proton collision data at $\sqrt{s}=$ 13 TeV collected with the ATLAS experiment}

\AtlasJournal{Phys.\ Rev.\ D}

\AtlasAbstract{
Combined measurements of Higgs boson production cross sections and branching fractions are presented. The combination is based on the analyses of the Higgs boson decay modes $H \to \gamma\gamma$, $ZZ^\ast$, $WW^\ast$, $\tau\tau$, $b\bar{b}$, $\mu\mu$, searches for decays into invisible final states, and on measurements of off-shell Higgs boson production. Up to $79.8$ fb$^{-1}$ of proton--proton collision data collected at $\sqrt{s}=$ 13 TeV with the ATLAS detector are used. Results are presented for the gluon--gluon fusion and vector-boson fusion processes, and for associated production with vector bosons or top-quarks. The global signal strength is determined to be $\mu = 1.11^{+0.09}_{-0.08}$. The combined measurement yields an observed (expected) significance for the vector-boson fusion production process of $6.5\sigma$ ($5.3\sigma$). Measurements in kinematic regions defined within the simplified template cross section framework are also shown. The results are interpreted in terms of modifiers applied to the Standard Model couplings of the Higgs boson to other particles, and are used to set exclusion limits on parameters in two-Higgs-doublet models and in the simplified Minimal Supersymmetric Standard Model. No significant deviations from Standard Model predictions are observed.
}
 
\author{The ATLAS Collaboration}
 
\AtlasRefCode{HIGG-2018-57}
 
\PreprintIdNumber{CERN-EP-2019-097}

\AtlasCoverSupportingNote{Supporting documentation for the combined measurements of Higgs boson production at $\sqrt{s} = 13$ TeV using up to 79.8 fb$^{-1}$ of proton--proton collision data with the ATLAS detector}{https://cds.cern.ch/record/2650495}

\AtlasCoverAnalysisTeam{Haider~Abidi, Tim~Adye, Rahul~Balasubramanian, Fabian~Becherer, Nicolas~Berger, Claudia~Bertella, Carsten~Burgard, Fabio~Cerutti, Valerio~Dao, Jennet~Dickinson, Roberto~Di~Nardo, Michael~Duehrssen, Saskia~Falke, Stefan~Gadatsch, Guillermo~Hamity, Yanping~Huang, Zihao~Jiang, Shan~Jin, Jelena~Jovicevic, Karsten~Koeneke, Sau~Lan~Wu, Changqiao~Li, Haifeng~Li, Zhijun~Liang, Jeanette~Lorenz, Theo~Megy, Jianming~Qian, Kunlin~Ran, Nikolaos~Rompotis, Xiaohu~Sun, Wouter~Verkerke, Trevor~Vickey, Alex~Wang, Zirui~Wang, Tim~Wolf, Lailin~Xu, Haijun~Yang, Hongtao~Yang, Chen~Zhou}
 
\AtlasCoverEdBoardMember{Oleg Brandt (Heidelberg KIP) }
\AtlasCoverEdBoardMember{Glen Cowan (London RHBNC) }
\AtlasCoverEdBoardMember{Marumi Kado (Orsay LAL), chair}
\AtlasCoverEdBoardMember{Bruno Mansouli\'{e} (Saclay CEA) }

\AtlasCoverEgroupEditors{atlas-HIGG-2018-57-editors@cern.ch}
 
\AtlasCoverEgroupEdBoard{atlas-HIGG-2018-57-editorial-board@cern.ch}

\hypersetup{pdftitle={ATLAS document},pdfauthor={The ATLAS Collaboration}}
 
\begin{document}
 
\LEcontact{Pippa Wells pippa.wells@cern.ch}
 
\maketitle
\tableofcontents

\section{Introduction}
Following the discovery of the Higgs boson~$H$~\cite{Englert:1964et,Higgs:1964ia,Higgs:1964pj,Guralnik:1964eu,Higgs:1966ev,Kibble:1967sv} by the ATLAS~\cite{ATLAS:2012obs} and CMS~\cite{CMS:2012obs} experiments, its properties have been probed using proton--proton ($pp$) collision data produced by the Large Hadron Collider (LHC) at CERN. The coupling properties of the Higgs boson to other Standard Model (SM) particles, such as its production cross sections in $pp$ collisions and decay branching fractions, can be precisely computed within the SM, given the value of the Higgs boson mass. Measurements of these properties can therefore provide stringent tests of the validity of the SM.
 
Higgs boson production and decay rates have been determined using the Run~1 dataset collected in the years 2011 and 2012, through the combination of ATLAS and CMS measurements~\cite{HIGG-2015-07}. More recently, these measurements have been extended using the Run~2 dataset recorded by the ATLAS detector in 2015, 2016 and 2017, using up to $79.8\, \ifb$ of $pp$ collision data produced by the LHC. The analyses target several production and decay modes, including: the \hgg\ and \hfourl\footnote{Throughout the paper $\ell$ denotes the light leptons $e$ and $\mu$.} decay channels following the same methodologies as those presented in Ref.~\cite{HIGG-2016-21} and Ref.~\cite{HIGG-2016-22} respectively, with improved selections for Higgs boson production in association with a top--antitop pair, described in Ref.~\cite{HIGG-2018-13}; the \hww~\cite{HIGG-2016-07} and \htt~\cite{HIGG-2017-07} decay channels; \Hbb\ in associated production with a weak vector boson $V=W$ or~$Z$~(\VH)~\cite{HIGG-2018-04,HIGG-2018-50} and in the weak vector-boson fusion (\VBF) production process~\cite{HIGG-2016-30}; associated production with a top--antitop pair (\ttH)~\cite{HIGG-2017-02,HIGG-2017-03,HIGG-2018-13}; the \hmm\ decay channel following the same methodology as presented in Ref.~\cite{HIGG-2016-10}, applied to the larger 2015--2017 input dataset; Higgs decays into invisible final states~\cite{EXOT-2016-37,HIGG-2016-28,EXOT-2016-23,HIGG-2018-54}; and off-shell production of Higgs bosons~\cite{HIGG-2017-06}. This paper presents measurements of Higgs boson properties at $\sqrt{s}=13$\,\TeV\ obtained from the combination of these results, using techniques similar to those in Ref.~\cite{HIGG-2015-07}. A Higgs boson mass value of $m_H=125.09\,\GeV$, corresponding to the central value of the combination of ATLAS and CMS measurements in Run 1~\cite{HIGG-2014-14}, is used for SM predictions. The uncertainty in the measured Higgs boson mass is considered in the \hgg\ and \hfourl\ analyses. Similar measurements~\cite{CMS-HIG-16-040,CMS-HIG-16-042,CMS-HIG-16-043,CMS-HIG-16-044,CMS-HIG-17-035,CMS-HIG-18-016,CMS-HIG-18-002}, as well as their combination~\cite{CMS-HIG-17-031}, have been reported by the CMS Collaboration.
 
All the input analyses except those for the \hmm\ and the \VBF, \Hbb\ processes use a parameterization of the Higgs boson signal yields based on the Stage 1 simplified template cross-section (STXS) framework~\cite{YR4,LesHouches} described in Section~\ref{sec:stxs_framework}. These cross sections are defined in the fiducial region $|y_H|<2.5$, where $y_H$ is the Higgs boson rapidity, partitioned within each Higgs boson production process into multiple kinematic regions based on the transverse momentum of the Higgs boson, the number of associated jets, and the transverse momentum of associated $W$~or~$Z$~bosons. The \hmm\ and \VBF, \Hbb\ analyses use a coarser description based on the Higgs boson production mode only.
 
The paper is structured as follows: Section~\ref{sec:sample} describes the data and simulation samples and Section~\ref{sec:channels} presents the analyses in individual decay channels which are used as inputs to the combination. Section~\ref{sec:model} provides a short description of the statistical procedures. The measurement of the signal strength $\mu$, defined as the ratio of the total Higgs boson signal yield to its SM prediction, is presented in Section~\ref{sec:mu}. Measurements of the cross sections of the main production processes within $|y_H|<2.5$, assuming SM predictions for the branching fractions, are then shown in Section~\ref{sec:fivexs}. The production modes considered are gluon--gluon fusion (\ggF), \VBF, \VH, \ttH\ and associated production with a single top quark (\tH). Measurements of cross sections times branching fractions for Higgs boson production and decay processes are shown in Section~\ref{sec:fivetimesfour}. Section~\ref{sec:ratios} presents a parameterization where the measured quantities are the cross section times branching fraction of the process $gg \to \hzz$, together with ratios of production cross sections and ratios of branching fractions. Common systematic uncertainties and modeling assumptions partially cancel out in these ratios, reducing the model dependence of the result. Section~\ref{sec:stxs} presents results in the STXS framework. Potential deviations from SM predictions are then probed in a framework of multiplicative modifiers $\kappa$ applied to the SM values of Higgs boson couplings~\cite{Heinemeyer:2013tqa}, presented in Section~\ref{sec:kappa}. Finally, Section~\ref{sec:bsm} presents an interpretation of the data within two benchmark models of beyond-the-SM (BSM) phenomena. Indirect limits on model parameters are set following a methodology similar to that of Ref.~\cite{HIGG-2015-03}. Section~\ref{sec:conclusion} summarizes the results.

\section{Data and simulated event samples}
\label{sec:sample}
 
The results of this paper are based on $pp$ collision data collected by the ATLAS experiment\footnote{\AtlasCoordFootnote}~\cite{Aad:2008zzm,ATLAS-TDR-2010-19,PIX-2018-001} in the years 2015, 2016 and 2017, with the LHC operating at a center-of-mass energy of 13\,\TeV. The integrated luminosities of the datasets used in each analysis are shown in Table~\ref{tab:lumi}. The uncertainty in the combined 2015--2016 integrated luminosity is $2.1\%$ and $2.0\%$ in the combined 2015--2017 integrated luminosity~\cite{ATLAS-CONF-2019-021}, obtained using the LUCID-2~detector~\cite{LUCID2} for the primary luminosity measurements.

\begin{table}[!htbp]
\caption{Dataset and integrated luminosity ($\mathcal{L}$) used for each input analysis to the combination. The last column provides the references for published analyses. The references in parentheses indicate analyses similar to the ones used in the combination but using a smaller dataset, in the cases where the analyses were not published separately.}
\begin{center}
\begin{tabular}{lccc}
\hline \hline
Analysis                           & Dataset & $\mathcal{L}$ [fb$^{-1}$] & Ref. \\
\hline
\Hyy\ (including \ttH, \Hyy)       & \multirow{4}{*}{2015--2017} & $79.8$ & (\cite{HIGG-2016-21}), \cite{HIGG-2018-13} \\
\hfourl\ (including \ttH, \hfourl) &                             & $79.8$ & (\cite{HIGG-2016-22}), \cite{HIGG-2018-13} \\
\VH, \hbb                          &                             & $79.8$ & \cite{HIGG-2018-04,HIGG-2018-50} \\
\hmm                               &                             & $79.8$ & (\cite{HIGG-2016-10}) \\
\hline
\hwwenmun                          & \multirow{6}{*}{2015--2016} & $36.1$ & \cite{HIGG-2016-07} \\
\htt                               &                             & $36.1$ & \cite{HIGG-2017-07} \\
\VBF, \hbb                         &                             & $24.5$ -- $30.6$ & \cite{HIGG-2016-30} \\
\ttH, \hbb\ and \ttH\ multilepton  &                             & $36.1$ & \cite{HIGG-2017-03,HIGG-2017-02,HIGG-2018-13}\\
\Hinv                              &                             & $36.1$ & \cite{EXOT-2016-37,HIGG-2016-28,EXOT-2016-23,HIGG-2018-54}\\
Off-shell \Hllll\ and \Hllvv\      &                             & $36.1$ & \cite{HIGG-2017-06} \\
\hline \hline
\end{tabular}
\end{center}
\label{tab:lumi}
\end{table}

Most analyses use a consistent set of simulated Higgs boson samples to describe the signal processes, which is detailed in the following paragraphs.
Exceptions are the \VBF, \Hbb\ and off-shell production analyses, described in Sections~\ref{sec:channels:hbb} and~\ref{sec:channels:offshell} respectively, and the measurements targeting decays of the Higgs boson into invisible final states described in Section~\ref{sec:channels:hinv}. The samples used for these analyses are described separately at the end of this section.
For each Higgs boson decay mode, the branching fraction used corresponds to higher-order state-of-the-art theoretical calculations~\cite{YR4}.
The simulated background samples vary channel by channel and are described in the individual references for the input analyses.
 
Higgs boson production via gluon--gluon fusion was simulated using the \powhegbox~\cite{Nason:2004rx,Frixione:2007vw,Alioli:2010xd,Alioli:2008tz} NNLOPS implementation~\cite{Hamilton:2013fea,Hamilton:2015nsa}. The event generator uses the HNNLO formalism~\cite{Catani:2007vq} to reweight the inclusive Higgs boson rapidity distribution produced by the next-to-leading order (NLO) generation of $pp\to H+\textrm{parton}$, with the scale of each parton emission determined using the \minlo\ procedure~\cite{Hamilton:2012np,Campbell:2012am,Hamilton:2012rf}. The PDF4LHC15~\cite{Butterworth:2015oua} parton distribution functions (PDFs) were used for the central prediction and uncertainty. The sample is normalized such that it reproduces the total cross section predicted by a next-to-next-to-next-to-leading-order (N\textsuperscript{3}LO) QCD calculation with NLO electroweak corrections applied~\cite{YR4,Anastasiou:2015ema,Anastasiou:2016cez,Dulat:2018rbf,Harlander:2009mq,Harlander:2009bw,Harlander:2009my,Actis:2008ug,Actis:2008ts,Anastasiou:2008tj,Aglietti:2004nj}. The NNLOPS generator reproduces the Higgs boson $p_{\textrm T}$ distribution predicted by the next-to-next-to-leading-order (NNLO) plus next-to-next-to-leading-logarithm (NNLL) calculation of \hres~\cite{Bozzi:2005wk,deFlorian:2011xf,Grazzini:2013mca}, which includes the effects of top- and bottom-quark masses and uses dynamical renormalization and factorization scales.
 
The \VBF\ production process was simulated to NLO accuracy in QCD using the \powhegbox~\cite{Nason:2009ai} generator with the PDF4LHC15 set of PDFs. The sample is normalized to an approximate-NNLO QCD cross section with NLO electroweak corrections applied~\cite{YR4,Ciccolini:2007jr,Ciccolini:2007ec,Bolzoni:2010xr}.
 
The $qq \to \VH$ production processes were simulated to NLO accuracy in QCD using the \powhegbox, \gosam~\cite{Cullen:2011ac} and \minlo~\cite{Hamilton:2012np,Luisoni:2013kna} generators with the PDF4LHC15 set of PDFs. The samples are normalized to cross sections calculated at NNLO in QCD with NLO electroweak corrections~\cite{Brein:2012ne,Harlander:2018yio,Brein:2003wg,Brein:2011vx,Harlander:2014wda,Ferrera:2017zex,Caola:2017xuq,Ciccolini:2003jy,Denner:2011id,Denner:2014cla}. The $gg\to ZH$ process was generated only at leading order (LO), using \powhegbox\ and NLO PDFs and normalized to an NLO computation with next-to-leading-logarithm (NLL) corrections~\cite{YR4,Altenkamp:2012sx}.
 
Higgs boson production in association with a top--antitop pair was simulated at NLO accuracy in QCD using the \powhegbox~\cite{Hartanto:2015uka} generator with the PDF4LHC15 set of PDFs for the \hgg\ and \hfourl\ decay processes. For other Higgs boson decays, the \amc~\cite{Alwall:2014hca,Artoisenet:2012st} generator was used with the \nnpdf~\cite{Ball:2014uwa} set of PDFs. In both cases the sample is normalized to a calculation with NLO QCD and electroweak corrections~\cite{YR4,Beenakker:2002nc,Dawson:2003zu,Yu:2014cka,Frixione:2014qaa}.
 
In addition to the primary Higgs boson processes, separate samples are used to model lower-rate processes. Higgs boson production in association with a \bbbar\ pair (\bbH) was simulated using \amc~\cite{Wiesemann:2014ioa} with \nnpdflo\ PDFs~\cite{Ball:2012cx} and is normalized to a cross section calculated to NNLO in QCD~\cite{YR4,Dawson:2003kb,Dittmaier:2003ej,Harlander:2003ai}. The sample includes the effect of interference with the \ggF\ production mechanism. Higgs boson production in association with a single top quark and a $W$ boson (\WtH) was produced at LO accuracy using \amc\ with the CTEQ6L1 PDF set~\cite{Pumplin:2002vw}. Finally, Higgs boson production in association with a single top quark in the t-channel (\tHq) was generated at LO accuracy using \amc\ with \ctten~\cite{Lai:2010vv} PDFs. The \tH\ samples are normalized to NLO QCD calculations~\cite{YR4,Demartin:2015uha,Demartin:2016axk}.
 
The parton-level events were input to \pythia~\cite{Sjostrand:2007gs} or \herwigpp~\cite{Gieseke:2003hm} to model the Higgs boson decay, parton showering, hadronization, and multiple parton interaction (MPI) effects. The generators were interfaced to \pythia\ for all samples except $\WtH$. For \pythia\ the AZNLO~\cite{STDM-2012-23} and A14~\cite{ATL-PHYS-PUB-2014-021} parameter sets were used, and for \herwigpp\ its UEEE5 parameter set was used.
 
Higgs boson decay branching fractions were computed using HDECAY~\cite{Djouadi:1997yw,Spira:1997dg,Djouadi:2006bz} and PROPHECY4F~\cite{Bredenstein:2006ha,Bredenstein:2006rh,Bredenstein:2006nk}.
 
In the all-hadronic channel of the \VBF, \Hbb\ analysis, the \powhegbox\ generator with the CT10~\cite{Lai:2010vv} set of PDFs was used to simulate the \ggF~\cite{Bagnaschi:2011tu} and \VBF\ production processes, and interfaced with \pythia\ for parton shower. In the photon channel of the \VBF, \Hbb\ analysis, \VBF\ and \ggF\ production in association with a photon was simulated using the \amc\ generator with the PDF4LHC15 set of PDFs, and also using \pythia\ for parton shower. For both channels, contributions from \VH\ and \ttH\ production were generated using the \pythia\ generator with the \nnpdf\ set of PDFs, and using the \amc\ generator interfaced with \herwigpp\ and the NLO CT10 set of PDFs, respectively.
 
In the analyses targeting Higgs boson decays into invisible final states, the \ggF, \VBF\ and \ZH\ signals were simulated in a similar way to the general procedure described above, but for the \VBF\ production process the \nnpdf\ PDF set was used instead of PDF4LHC15, while for the \ZH\ process the CT10 PDF set was used.
 
In the off-shell production analysis, the $gg \to H^\ast \to ZZ$ process was generated together with the corresponding irreducible continuum production, using the \textsc{Sherpa}~2.2.2~+~\textsc{OpenLoops}~\cite{Cascioli:2013gfa,Gleisberg:2008ta,Cascioli:2011va,Denner:2014gla} generator and the \nnpdf\ PDF set. The generation was performed at leading order with up to one additional jet in the final state, and interfaced with the \textsc{Sherpa} parton shower~\cite{Schumann:2007mg}. The cross-section calculations take into account $K$-factors following the methodology described in Ref.~\cite{HIGG-2017-06}.

The particle-level Higgs boson events were passed through a \geant~\cite{geant-1} simulation of the ATLAS detector~\cite{SOFT-2010-01} and reconstructed using the same analysis software as used for the data. Event pileup is included in the simulation by overlaying inelastic $pp$ collisions, such that the average number of interactions per bunch crossing reproduces that observed in the data. The inelastic $pp$ collisions were simulated with \pythia\ using the \mstwlo~\cite{Martin:2009iq} set of PDFs with the A2~\cite{ATL-PHYS-PUB-2012-003} set of tuned parameters or using the \nnpdflo\ set of PDFs with the A3~\cite{ATL-PHYS-PUB-2016-017} set of tuned parameters.


\begin{landscape}
\begin{table}[htbp]
\caption{Summary of the signal regions entering the combined measurements.
"Leptonic" and "hadronic" refers to \ttH\ and \VH\ processes where the associated \ttbar\ pair or vector boson decays to final states with respectively at least one lepton or no leptons. "Resolved" and "boosted" refers to configurations in which hadronic Higgs boson decay products are reconstructed respectively as two or more jets, or a single jet.
In the \VBF, \hgg\ mode, \pTggjj\ is the transverse momentum of the system of the \VBF\ jets and the photon candidates. In the \VBF, \hgg\ analysis, \ptlmet\ is the transverse momentum of the system composed of the leading lepton and the missing transverse momentum. Other notations are defined in Section~\ref{sec:channels}.
Each 0-jet and 1-jet \hww\ entry corresponds to two categories for a leading lepton flavor of either $e$ or $\mu$. For \htt, each entry corresponds to three categories for \ttll, \ttlh\ and \tthh, unless otherwise specified. \textquote{Multilepton} refers to decays of the Higgs boson with one or more leptons, and encompasses \hww, \htt, and \hzz\ excluding \hfourl. The selections targeting \hmm,\ \Hinv\ and off-shell Higgs boson production are not included this table.}
\vspace*{-0.4cm}
\begin{center}
\resizebox{\columnwidth}{!}{
\footnotesize
\begin{tabular}{l|l|l|l|l|l}
\hline
\hline
& \hgg & \hzz & \hww & \htt & \hbb \\
\hline
\multirow{7}{*}{\ttH}
& \ttH\ leptonic (3 categories) & \multicolumn{3}{l|}{\ttH\ multilepton 1 $\ell$ + 2 \thad}                                   & \ttH\ 1 $\ell$, boosted                  \\
& \ttH\ hadronic (4 categories) & \multicolumn{3}{l|}{\ttH\ multilepton 2 opposite-sign $\ell$ + 1 \thad}                     & \ttH\ 1 $\ell$, resolved (11 categories) \\
&                               & \multicolumn{3}{l|}{\ttH\ multilepton 2 same-sign $\ell$ (categories for $0$ or $1$ \thad)} & \ttH\ 2 $\ell$ (7 categories)            \\
&                               & \multicolumn{3}{l|}{\ttH\ multilepton 3 $\ell$ (categories for $0$ or $1$ \thad)}           & \\
&                               & \multicolumn{3}{l|}{\ttH\ multilepton 4 $\ell$ (except \hfourl)}                            & \\
&                               & \multicolumn{3}{l|}{\ttH\ leptonic, \hfourl}                                                & \\
&                               & \multicolumn{3}{l|}{\ttH\ hadronic, \hfourl}                                                & \\
\hline
\multirow{8}{*}{\VH}
& \VH\ 2 $\ell$                             & \VH\ leptonic                     & & & 2 $\ell$, $75 \leq \pTV < 150~\GeV$, $\Njets=2$      \\
& \VH\ 1 $\ell$, \ptlmet $\geq 150$~\GeV    &                                   & & & 2 $\ell$, $75 \leq \pTV < 150~\GeV$, $\Njets \geq 3$ \\
& \VH\ 1 $\ell$, \ptlmet$<$150~\GeV         &                                   & & & 2 $\ell$, $\pTV \geq 150~\GeV$, $\Njets=2$           \\
& \VH\ \met, \met $\geq 150$~\GeV           & 0-jet, $\ptfourl \geq 100~\GeV$   & & & 2 $\ell$, $\pTV \geq 150~\GeV$, $\Njets \geq 3$      \\
& \VH\ \met, \met$<$150~\GeV                &                                   & & & 1 $\ell$  $\pTV \geq 150~\GeV$, $\Njets=2$           \\
& \VH+\VBF\ \ptleadingj $\geq 200$~\GeV     &                                   & & & 1 $\ell$  $\pTV \geq 150~\GeV$, $\Njets=3$           \\
& \VH\ hadronic (2 categories)              & 2-jet, $\mjj < 120~\GeV$          & & & 0 $\ell$, $\pTV \geq 150~\GeV$, $\Njets=2$           \\
&                                           &                                   & & & 0 $\ell$, $\pTV \geq 150~\GeV$, $\Njets=3$           \\
\hline
\multirow{4}{*}{\VBF}
& \VBF, \pTggjj $\geq 25$~\GeV\ (2 categories) & 2-jet \VBF, $\ptleadingj\geq 200~\GeV$ & 2-jet \VBF & \VBF\ $\pttautau>140~\GeV$ & \VBF, two central jets  \\
& \VBF, \pTggjj$<$25~\GeV\ (2 categories)      & 2-jet \VBF, \ptleadingj$<$200~\GeV     &            & \quad  (\tthh\ only)       & \VBF, four central jets \\
&                                              &                                        &            & \VBF\ high-$\mjj$          & \VBF $+\gamma$          \\
&                                              &                                        &            & \VBF\ low-$\mjj$           &                         \\
\hline
\multirow{9}{*}{\ggF}
& 2-jet, \pTgg $\geq 200$~\GeV          & 1-jet, $\ptfourl \geq 120~\GeV$ & 1-jet, $\mll< 30~\GeV$, $\ptsubleadl<20~\GeV$      & Boosted, $\pttautau > 140~\GeV$  &  \\
& 2-jet, 120~\GeV $\leq$ \pTgg<200~\GeV & 1-jet, 60~\GeV$\leq\ptfourl$<120~\GeV  & 1-jet, $\mll< 30~\GeV$, $\ptsubleadl\geq 20~\GeV$  & Boosted, $\pttautau \leq 140~\GeV$  & \\
& 2-jet, 60~\GeV $\leq$ \pTgg<120~\GeV  & 1-jet, $\ptfourl < 60~\GeV$     & 1-jet, $\mll\geq 30~\GeV$, $\ptsubleadl<20~\GeV$      &                               &  \\
& 2-jet, \pTgg $< 60$~\GeV              & 0-jet, $\ptfourl < 100~\GeV$    & 1-jet, $\mll\geq 30~\GeV$, $\ptsubleadl\geq 20~\GeV$  &                               &  \\
& 1-jet, \pTgg $\geq$ 200~\GeV          &                                 & 0-jet, $\mll<    30~\GeV$, $\ptsubleadl<20~\GeV$      &                               &  \\
& 1-jet, 120~\GeV $\leq$ \pTgg<200~\GeV &                                 & 0-jet, $\mll<    30~\GeV$, $\ptsubleadl\geq 20~\GeV$  &                               &  \\
& 1-jet, 60~\GeV $\leq$ \pTgg<120~\GeV  &                                 & 0-jet, $\mll\geq 30~\GeV$, $\ptsubleadl<20~\GeV$      &                               &  \\
& 1-jet, \pTgg $<$ 60~\GeV              &                                 & 0-jet, $\mll\geq 30~\GeV$, $\ptsubleadl\geq 20~\GeV$  &                               &  \\
& 0-jet (2 categories)                  &                                 &                                                       &                               &  \\
\hline
\hline
\end{tabular}
}
\end{center}
\label{tab:regions}
\end{table}
\end{landscape}

 
\section{Individual channel measurements}
\label{sec:channels}
Brief descriptions of the input analyses to the combination are given below. More details can be found in the individual analysis references listed in each section. The categorization is summarized in Table~\ref{tab:regions}. The overlap between the event selections of the analyses included in the combination is found to be negligible.
 
\subsection{\hgg}
\label{sec:channels:hgg}
The \hgg\ analysis~\cite{HIGG-2016-21,HIGG-2018-13} requires the presence of two isolated photons~\cite{PERF-2017-02} within the pseudorapidity range $|\eta| < 2.37$, excluding the region $1.37 < |\eta| < 1.52$ corresponding to the transition between the barrel and endcap sections of the electromagnetic calorimeter. The transverse momenta of the leading and subleading photons are required to be greater than $0.35\mgg$ and $0.25\mgg$ respectively, where \mgg\ is the invariant mass of the diphoton system.
The event reconstruction and selection procedures are largely unchanged from the ones described in Ref.~\cite{HIGG-2016-21}. The only significant change concerns the reconstruction of the calorimeter energy clusters associated with the photons; a dynamical, topological cell clustering-based algorithm~\cite{PERF-2014-07,EGAM-2018-01} is now used instead of a sliding-window technique~\cite{PERF-2017-02,ATL-LARG-PUB-2008-002}.
 
Selected events are separated into 29~mutually exclusive categories based on the kinematics of the diphoton system and associated particles, chosen to approximately match those of the Stage~1 STXS regions described in Section~\ref{sec:stxs_framework}.
Seven categories are defined to select \ttH\ production, including both semileptonic and hadronic top-quark decay processes through various selections on the multiplicities and kinematics of leptons~\cite{PERF-2017-01,PERF-2017-03,PERF-2015-10}, jets~\cite{PERF-2016-04}, and jets tagged as containing $b$-hadrons~\cite{PERF-2016-05}.
These categories are described in Ref.~\cite{HIGG-2018-13}.
The remaining events are classified into categories targeting the \VH, \VBF\ and \ggF\ production modes, described in Ref.~\cite{HIGG-2016-21}. Five categories are defined to select $WH$ and $ZH$ production with leptonic decays of the $W$ or $Z$, based on the presence of leptons and missing transverse momentum $\ETmiss$~\cite{PERF-2016-07}. Seven categories cover the \VBF\ and \VH\ processes: one category requires the presence of two jets, with the leading jet transverse momentum $\ptleadingj > 200\,\GeV$; two categories select hadronic vector-boson decays by requiring two jets with an invariant mass compatible with the $W$ or $Z$ boson mass; and four categories enrich \VBF\ production by requiring forward jets in a \VBF-like topology. The requirement of a second jet for the $\ptleadingj > 200\,\GeV$ category is a change compared to Ref.~\cite{HIGG-2016-21} where only one jet was required, and helps to reduce contamination from \ggF\ production.
The remaining events are split into 10 categories, separating events with $0$, $1$, and ~$\geq 2$-jets and classifying them further according to the pseudorapidity of the two photons (for $0$-jet events) or the transverse momentum of the diphoton system \ptgg\ (for $1$ and~$\geq 2$-jet events).
The distribution of \mgg\ is used to separate the Higgs boson signal from continuum background processes in each category.
 
\subsection{\hZZllll}
\label{sec:channels:hzz}
The \hfourl\ analysis requires the presence of at least two same-flavor and opposite-charge light-lepton pairs, with a four-lepton invariant mass $m_{4\ell}$ in the range $115\,\GeV < m_{4\ell} < 130\,\GeV$. The analysis follows the strategy described in the previous publication~\cite{HIGG-2016-22}, but employs improved event reconstruction and electron reconstruction~\cite{EGAM-2018-01} techniques, and defines additional event categories to enhance sensitivity to the production of the SM Higgs boson associated with a vector boson ($\VH$, $V \to \ell\nu/\nu\nu$) and with a top-quark pair~\cite{HIGG-2018-13}.
 
To distinguish the \ttH, \VH, \VBF, and \ggF\ production modes and to enhance the purity of each kinematic selection, 11~mutually exclusive reconstructed event categories based on the presence of jets and additional leptons in the final state are defined. Candidate events with at least one $b$-tagged jet and three or more additional jets, or one additional lepton and at least two additional jets are classified into categories enriched in \ttH\ production with fully hadronic or semileptonic top-quark decays respectively~\cite{HIGG-2018-13}. Events failing these requirements but containing at least one additional lepton are assigned to a \VH-enriched category with leptonic vector boson decays. The remaining events are classified according to their jet multiplicity ($0$-jet, $1$-jet, and $\geq 2$-jet). Events with at least two jets are divided into a \VBF-enriched region, for which the dijet invariant mass $\mjj$ is required to be above $120\,\gev$, and a region enriched in \VH\ events with a hadronically decaying vector boson for $\mjj < 120\,\gev$. The \VBF-enriched region is further split into two categories, in which the transverse momentum of the leading jet \ptleadingj\ is required to be either above or below $200\,\gev$.
The selected $0$-jet and $1$-jet events are further separated according to the transverse momentum \ptfourl\ of the four-lepton system: the $0$-jet events are split into two categories with a boundary at $\ptfourl = 100\,\gev$, with the lower $\ptfourl$ selection being enriched in Higgs boson events produced via \ggF\ and the higher $\ptfourl$ selection being enriched in Higgs boson events produced in association with a weak vector boson. The $1$-jet events are split into three categories, each containing predominantly Higgs boson events produced via \ggF, with boundaries at $\ptfourl = 60$ and $120\,\gev$ to match the STXS selections described in Section~\ref{sec:stxs_framework}.
Boosted decision trees (BDTs) are employed to separate the signal from the background processes and to enhance the sensitivity to the various Higgs boson production modes.
 
\subsection{\hwwenmun}
The \hwwenmun\ analysis~\cite{HIGG-2016-07} included in the combination targets the \ggF\ and \VBF\ production modes. Signal candidates are selected by requiring the presence of an isolated $e^{\pm}\mu^{\mp}$ pair, with transverse momentum thresholds at $22$~and~$15$~GeV for the leading and subleading lepton. Events with jets tagged as containing $b$-hadrons are rejected to suppress background contributions originating from top-quark production. Contributions from $\wtotv$ decays in which the $\tau$-leptons subsequently decay into electrons or muons are also included.

Selected events are classified according to the number of associated jets ($\Njets$). Exclusive $\Njets = 0$ and $\Njets = 1$ selections are enriched in signal events produced via \ggF. To isolate regions with higher sensitivity, they are each further split into eight categories apiece, based on the flavor of the leading lepton ($e$ or $\mu$), two bins of the invariant mass of the dilepton system $m_{\ell\ell}$ and two bins of the transverse momentum of the subleading lepton $\ptsubleadl$.
The distribution of the transverse mass of the dilepton plus \ETmiss\ system is used to separate the Higgs boson signal from background in each category.
The $\Njets \geq 2$ category is naturally sensitive to the \VBF\ process. A central-jet veto is applied to suppress the multijet background and the contribution from \ggF\ production. The output of a BDT exploiting the kinematic properties of the two leading jets and the two leptons is used to separate \VBF\ Higgs boson production from background processes, including Higgs boson production via \ggF.
 
\subsection{\htt}
The \htt\ analysis~\cite{HIGG-2017-07} measures the Higgs boson production cross section in the \VBF\ production process or in \ggF\ production with large Higgs boson transverse momentum \ptH. Final states with both leptonic ($\tlep$) and hadronic ($\thad$) decays of the $\tau$-lepton are considered. Selected lepton candidates are required to be of opposite charge, meet identification and isolation criteria and satisfy the $\pt$ thresholds of the triggers used. Three mutually exclusive analysis channels, $\ttll$, $\ttlh$, and $\tthh$, are defined according to the number of selected electron, muon and $\thad$ candidates.
All channels require the presence of at least one jet with high transverse momentum.
 
To exploit signal-sensitive event topologies, candidate events are divided into three categories targeting the \VBF\ process and two categories for high-\ptH\ Higgs production. The \VBF\ categories collect events with two jets with a large pseudorapidity separation and a high invariant mass ($\mjj$). The Higgs boson decay products are required to be in the central rapidity region. One \VBF\ category is defined by requiring the transverse momentum of the $\tau\tau$ system $\pttautau$ to be above $140\,\GeV$, for $\tthh$ events only. The two remaining \VBF\ categories are defined for lower and higher values of $\mjj$, with definitions that differ between the $\ttll$, $\ttlh$, and $\tthh$ channels. The high-\ptH\ categories select events with large values of \pttautau, with contributions mainly from the \ggF\ process. Events failing the \VBF\ selection and with $\pttautau > 100\, \GeV$ are selected. In order to improve the sensitivity of the analysis, two categories are defined for $\pttautau > 140\, \GeV$ and $\pttautau \leq 140\, \GeV$, with additional selections on the angular separation between the $\tau$-leptons.
The distribution of the invariant mass of the $\tau\tau$~system is used to separate the Higgs boson signal from background in each category.
 
 
\subsection{\hbb}
\label{sec:channels:hbb}
The \Hbb\ decay channel is used to measure the production cross section in the \VH, \VBF\ and \ttH\ production modes, the latter described in Section~\ref{sec:channels:ttH}.

The search for \Hbb\ in the \VH\ production mode~\cite{HIGG-2018-04,HIGG-2018-50} considers final states containing at least two jets, of which exactly two must be tagged as containing $b$-hadrons. Either zero, one or two charged leptons are also required, exploring the associated production of a Higgs boson with a $W$ or $Z$ boson decaying leptonically as $\ztovv$, $\wtolv$, or $\ztoll$. Contributions from $\wtotv$ and $\ztott$ decays in which the $\tau$-leptons subsequently decay into electrons or muons are also included.
 
To enhance the signal sensitivity, selected candidate events are classified according to the charged-lepton multiplicity, the vector-boson transverse momentum $\pTV$, and the jet multiplicity. For final states with zero or one lepton, $\pTV > 150\,\GeV$ is required.
In two-lepton final states, two regions are considered, $75\, \GeV < \pTV < 150\, \GeV$ and $\pTV > 150\, \GeV$. The \pTV\ thresholds are chosen to select regions with strong experimental sensitivity, and match the STXS definitions described in Section~\ref{sec:stxs_framework}.
Each of these regions is finally separated into a category with exactly two reconstructed jets and another with three or more. In the zero- and one-lepton channel, events with four or more jets are rejected.
Topological and kinematic selection criteria are applied within each of the resulting categories.
BDTs incorporating the event kinematics and topology, in addition to the dijet invariant mass, are employed in each lepton channel and analysis region to separate the signal process from the sum of the expected background processes.

The \Hbb\ mode is also used to measure the \VBF\ production process~\cite{HIGG-2016-30}. Three orthogonal selections are employed, targeting two all-hadronic channels and a photon-associated channel. Each selection requires the presence of at least two jets tagged as containing $b$-hadrons in the central pseudorapidity region $|\eta|<2.5$ as well as at least two additional jets used to identify the \VBF\ topology.
 
The first of the two all-hadronic selections requires the $b$-tagged jets to have transverse momenta larger than $95\,\GeV$ and $70\,\GeV$, while one of the additional jets is required to be in the forward region $3.2 < |\eta| < 4.4$ and have a transverse momentum larger than $60\,\GeV$ and another must satisfy $p_\mathrm{T} > 20\,\GeV$ and $|\eta|<4.4$. The transverse momentum $p_\mathrm{T}^{bb}$ of the system composed of the two $b$-tagged jets must be larger than $160\,\GeV$.
 
The second all-hadronic selection with four central jets is defined by the presence of two jets with $|\eta|<2.8$ in addition to the $b$-tagged jets with $|\eta|<2.5$. All selected jets must pass a common threshold requirement of $55\,\GeV$ on their transverse momenta. The $p_\mathrm{T}$ of the $\bbbar$-system is required to be larger than $150\,\GeV$. Events containing at least one forward jet satisfying the selection criteria of the first all-hadronic channel are removed.
 
A \VBF$+\gamma$ selection is defined by the presence of a photon with transverse momentum $\pt > 30\,\GeV$ and $|\eta| < 2.37$, excluding the region $1.37 < |\eta| < 1.52$, which suppresses the dominant background from non-resonant $\bbbar jj$ production. Events must have at least four jets, all satisfying $\pt > 40\,\GeV$ and $|\eta|<4.4$, with at least two jets in $|\eta|<2.5$ passing the $b$-tag requirements. The invariant mass of the \VBF\ jets is required to be higher than $800\,\GeV$, and $p_\mathrm{T}^{bb} > 80\,\GeV$.
 
In all three selections a BDT built from variables describing jet and photon kinematics is used to enhance the sensitivity, and the distribution of the invariant mass $m_{bb}$ of the two $b$-tagged jets is used to separate the Higgs boson signal from background.
 
The \VBF, \Hbb\ channels are included in all the measurements except for those presented in Section~\ref{sec:stxs}.
 
 
\subsection{\hmm}
The \hmm\ search uses a similar technique to the \hgg\ analysis, requiring a pair of opposite-charge muons. The analysis closely follows the \hmm\ search described in Ref.~\cite{HIGG-2016-10}, which used a smaller dataset collected in the years 2015 and 2016 only.
 
Events are classified into eight categories. The output of a BDT exploiting the kinematic properties of the two leading jets and the two muons is used to define two categories targeting the \VBF\ process. In order to enhance the sensitivity of the analysis, the remaining events are classified into three ranges of the transverse momentum $\pT^{\mu\mu}$ of the dimuon system ($\pT^{\mu\mu} < 15\,\GeV$, $15\,\GeV \leq \pT^{\mu\mu} < 50\,\GeV$ and $\pT^{\mu\mu} \geq 50\,\GeV$) and two ranges of the muon pseudorapidities $\eta^{\mu}$ (both muons within $|\eta^{\mu}|\leq 1$, or at least one muon outside this range), for a total of six categories.
The distribution of the invariant mass $m_{\mu\mu}$ of the two muons is used to separate signal from background in each category.
 
The analysis is not sensitive at the level of the Higgs boson signal expected in the SM, and is only included in the results presented in Section~\ref{sec:kappa:moneyplot}.
 
\subsection{\ttH, \hbb\ and \ttH\ multilepton analyses}
\label{sec:channels:ttH}
Searches for the associated production of the Higgs boson with a \ttbar~pair have been performed using Higgs boson decays into \bb~\cite{HIGG-2017-03} and in multilepton final states, targeting Higgs boson decays into \wwstar,~\zzstar\ and~$\tau\tau$~\cite{HIGG-2017-02,HIGG-2018-13}. These analyses complement the selections sensitive to \ttH\ production defined in the analyses of the \hgg\ and \hfourl\ decay channels, described in Sections~\ref{sec:channels:hgg} and~\ref{sec:channels:hzz}.
 
The search for \ttH\ production with \Hbb\ employs two selections, optimized for single-lepton and dilepton final states of \ttbar\ decays. In the single-lepton channel, events are required to have one isolated electron or muon and at least five jets, of which at least two 
must be identified as containing $b$-hadrons.
In the dilepton channel, events are required to have two opposite-charge leptons and at least three jets, of which at least two must be identified as containing $b$-hadrons.
Candidate events are classified into 11~(7)~orthogonal categories in the single-lepton (dilepton) channel,  according to the jet multiplicity and the values of the $b$-tagging discriminant for the jets. In the single-lepton channel, an additional category, referred to as \textit{boosted}, is designed to select events with large transverse momenta for the Higgs candidate ($\ptH > 200~\GeV$) and one of the top-quark candidates ($\pt^t > 250~\GeV$).
In each region, a BDT exploiting kinematic information of the events is employed to separate \ttH\ production from background processes.

The \ttH\ search with Higgs boson decays into \wwstar, \zzstar\ and $\tau\tau$ exploits several multilepton signatures resulting from leptonic decays of vector bosons and/or the presence of $\thad$ candidates. Seven final states, categorized by the number and flavor of reconstructed charged-lepton candidates, are examined. They are: one lepton with two $\thad$ candidates, two same-charge leptons with zero or one $\thad$ candidates, two opposite-charge leptons with one $\thad$ candidate, three leptons with zero or one $\thad$ candidates, and four leptons, excluding events from \hfourl\ decays. Events in all channels are required to have at least two jets, at least one of which must be $b$-tagged. Additional requirements are employed for each final state.
Multivariate analysis techniques exploiting the kinematic properties and topologies of the selected events are applied in most channels to improve the discrimination between the signal and the background.
 
\subsection{Searches for invisible Higgs boson decays}
\label{sec:channels:hinv}
Searches for decays of the Higgs boson into invisible final states select events with large missing transverse momentum; backgrounds are suppressed by requiring in addition either jets with a \VBF\ topology~\cite{EXOT-2016-37}, an associated $Z$ boson decaying into charged leptons~\cite{HIGG-2016-28} or an associated $W$ or $Z$ boson decaying into hadronic final states~\cite{EXOT-2016-23}.

Production in the \VBF\ topology is identified by requiring two jets with a pseudorapidity difference $|\Delta\eta_{jj}| > 4.8$ and invariant mass $m_{jj} > 1\,\TeV$. The missing transverse momentum is required to be larger than $180\,\GeV$. Events with isolated lepton candidates or additional jets are rejected. Three signal regions are defined for $1 < m_{jj} < 1.5\,\TeV$, $1.5 < m_{jj} < 2\,\TeV$ and $m_{jj} > 2\,\TeV$.

Production in association with a leptonically decaying $Z$ boson is identified by requiring the presence of a pair of isolated electrons or muons with an invariant mass close to $m_Z$. The missing transverse momentum is required to be larger than $90\,\GeV$. It must also be larger than $60\%$ of the scalar sum of the transverse momenta of the identified leptons and jets, and must be oriented back-to-back with the dilepton system in the transverse plane.

Two event topologies are considered in order to identify production in association with a hadronically decaying $W$ and $Z$ boson. The \textit{resolved} topology is defined by the presence of two jets compatible with originating from the hadronic decay of a $W$ or $Z$ boson, reconstructed using the anti-$k_t$ algorithm~\cite{AntiKt} with a radius parameter of $0.4$. The \textit{merged} topology identifies $W$ or $Z$ bosons with large transverse momentum through the presence of a single jet, reconstructed using the anti-$k_t$ algorithm with a radius parameter of $1$. The missing transverse momentum is required to be larger than $150\,\GeV$ and $250\,\gev$ for the resolved and boosted topologies respectively. In both cases, events are categorized according to the multiplicity of jets tagged as containing $b$-quarks. A separate category is also defined for events in which the mass of the jet system, defined as the dijet mass in the resolved topology and the mass of the large-radius jet in the merged topology, is compatible with a hadronic $W$ or $Z$ decay.
 
The statistical combination of these analyses~\cite{HIGG-2018-54} yields an observed (expected) upper limit on the branching fraction for Higgs boson decays into invisible final states of $\Binv < 0.38$~($0.21$) at $95\%$~confidence level. In this paper, these analyses are only included in the coupling measurements presented in Sections~\ref{sec:kappa:kg_ky} and~\ref{sec:kappa:all}.
 
\subsection{Off-shell Higgs boson production}
\label{sec:channels:offshell}
Measurements of the $H^\ast \to ZZ$ final state in the mass range above the $2m_Z$ threshold (off-shell region) provide an opportunity to measure the off-shell coupling strength of the observed Higgs boson, as discussed in Refs.~\cite{Kauer:2012hd,Caola:2013yja,Campbell:2013una,Campbell:2013wga}. The $ZZ \to 4\ell$ and $ZZ\to2\ell2\nu$ decay channels, detailed in Ref.~\cite{HIGG-2017-06}, are used in these measurements.
 
Assuming that the coupling modifiers are identical for on-shell and off-shell production, the total width of the Higgs boson can be constrained from a combination with the on-shell measurements. It is also assumed that the coupling modifiers are independent of the momentum transfer of the Higgs boson production mechanism considered in the analysis, and that
any new physics which modifies the off-shell signal strength and the off-shell couplings does not modify the relative phase of the interfering signal and background processes. Further, it is assumed that there are neither sizable kinematic modifications to the off-shell signal nor new sizable signals in the search region of this analysis unrelated to an enhanced off-shell signal strength~\cite{Englert:2014ffa,Logan:2014ppa}.
 
The analysis in the $ZZ\to 4\ell$ final state closely follows the Higgs boson measurements in the same final state, described in Section~\ref{sec:channels:hzz}, with the same event reconstruction, trigger and event selections and background estimation methods. The off-peak region is defined to cover the range $220\,\GeV < m_{4\ell} < 2000\,\GeV$.
The distribution of a matrix-element-based discriminant constructed to enhance the $gg \to H^\ast \to ZZ$ is used to separate the Higgs boson signal from background processes.
 
The analysis in the $ZZ\to 2\ell 2\nu$ channel is similar to the one designed to search for heavy $\zz$ resonances~\cite{HIGG-2016-19} with the same object definitions. The analysis is performed inclusively in the number of final-state jets and kinematic selections are optimized accordingly.
Sensitivity to the off-shell Higgs boson signal is obtained through the distribution of the transverse mass $m_{\mathrm{T}}^{ZZ}$ reconstructed from the momentum of the dilepton system and the missing transverse momentum~\cite{HIGG-2017-06}, within the range $250\,\GeV < m_{\mathrm{T}}^{ZZ} <2000\,\GeV$.
 
These off-shell analyses are only included in the coupling measurements presented in Section~\ref{sec:kappa:all}.
 

\section{Statistical model}
\label{sec:model}
The statistical methods used in this paper follow those of Ref.~\cite{HIGG-2015-07}.
The results of the combination are obtained from a likelihood function defined as the product of the likelihoods of each input analysis. These are themselves products of likelihoods computed in mutually exclusive regions selected in the analysis, referred to as analysis categories.
 
The number of signal events in each analysis category $k$ is expressed as
\begin{equation}
n^{\text{signal}}_k = \mathcal{L}_k \sum_i \sum_f (\sigma \times \BR)_{if} (A \times \epsilon)_{if,k}
\label{eq:yields}
\end{equation}
where the sum runs over production modes $i$~($i = \ggF, \VBF, \WH, \ZH, \ttH, \ldots$) and decay final states $f$~($f = \gamma\gamma, ZZ^\ast, WW^\ast, \tau\tau, b\bbar, \mu\mu$), $\mathcal{L}_k$ is the integrated luminosity of the dataset used in category $k$, and $(A \times \epsilon)_{if,k}$ is the acceptance times efficiency factor in category $k$ for production mode $i$ and final state $f$.
The  cross section times branching fraction $(\sigma \times \BR)_{if}$ for each relevant pair ($i$, $f$) are the parameters of interest of the model. The measurements presented in this paper are obtained from fits in which these products are free parameters (Section~\ref{sec:fivetimesfour}), or in which they are re-expressed in terms of smaller sets of parameters: of a single signal-strength parameter~$\mu$~(Section~\ref{sec:mu}), of the cross sections $\sigma_i$ in each of the main production modes (Section~\ref{sec:fivexs}), of ratios of cross sections and branching fractions (Sections~\ref{sec:ratios} and Section~\ref{sec:stxs_results}) or of coupling modifiers (Section~\ref{sec:kappa}).
Additional parameters, referred to as nuisance parameters, are used to describe systematic uncertainties and background quantities that are constrained by sidebands or control regions in data.
 
Systematic uncertainties that affect multiple analyses are modeled with common nuisance parameters to propagate the effects of these uncertainties coherently to all measurements. The assessment of the associated uncertainties varies between data samples, reconstruction algorithms and software releases, leading to differences particularly between analyses performed using the 2017 dataset and those using 2015 and 2016 data only. Between these two sets of analyses, components of systematic uncertainties in the luminosity, the jet energy scale, the electron/photon resolution and energy scale, and in the electron reconstruction and identification efficiencies are also treated as correlated.
Uncertainties due to the limited number of simulated events used to estimate expected signal and background yields are included using the simplified version of the Beeston--Barlow technique~\cite{BeestonBarlow} implemented in the \textsc{HistFactory} tool~\cite{HistFactory}. They are counted among the systematic uncertainties.
 
Theory uncertainties in the signal, such as missing higher-order QCD corrections and PDF-induced uncertainties, affect the expected signal yields of each production and decay process, as well as the signal acceptance in each category. These uncertainties are modeled by a common set of nuisance parameters in most channels. For the signal-strength (Section~\ref{sec:mu}) and coupling modifier (Section~\ref{sec:kappa}) results and constraints on new phenomena (Section~\ref{sec:bsm}), which rely on  the comparison of measured and SM-expected yields, both the acceptance and signal yield uncertainties are included. For the cross-section and branching fraction results of Sections~\ref{sec:fivexs} and~\ref{sec:stxs}, only acceptance uncertainties are considered.
The effects of correlations between Higgs boson branching fractions are modeled using the correlation model specified in Ref.~\cite{YR4}. Uncertainties due to dependencies on SM parameter values and missing higher-order effects are applied to the partial decay widths and propagated to the branching fractions. The uncertainties due to modeling of background processes are typically treated as uncorrelated between analyses.
 
The measurement of the parameters of interest is carried out using a statistical test based on the profile likelihood ratio~\cite{Cowan:2010st},
\begin{equation*}
\Lambda(\vec{\alpha}) =
\frac{L(\vec{\alpha}, \hat{\hat{\vec{\theta}}}(\vec{\alpha})) }
{L(\hat{\vec{\alpha}}, \hat{\vec{\theta}})} \;,
\end{equation*}
where $\vec{\alpha}$ and $\vec{\theta}$ are respectively the parameters of interest and the nuisance parameters.
In the numerator, the nuisance parameters are set to their profiled values $\hat{\hat{\vec{\theta}}}(\vec{\alpha})$, which maximize the likelihood function for fixed values of the parameters of interest $\vec{\alpha}$. In the denominator, both the parameters of interest and the nuisance parameters are set to the values $\hat{\vec{\alpha}}$ and $\hat{\vec{\theta}}$ respectively which jointly maximize the likelihood.
 
In the asymptotic regime, in which the likelihood is approximately Gaussian, the value of $-2\ln \Lambda(\vec{\alpha})$ follows a $\chi^2$ distribution with a number of degrees of freedom $n$ equal to the dimensionality of the vector $\vec{\alpha}$~\cite{Cowan:2010st}. This property is assumed to hold for all the results presented in the following sections.
Confidence intervals for a confidence level (CL) $1 - p$ are then defined as the regions with values of $-2\ln \Lambda(\vec{\alpha})$ below a threshold $F^{-1}_{\chi^2_n}(1-p)$, where $F^{-1}_{\chi^2_n}$ is the quantile function 
of the $\chi^2$ distribution with $n$~degrees of freedom.
 
The CL$_\text{s}$ prescription~\cite{Read:2002hq} is applied when setting an upper limits on a single parameter directly related to measured event rates, for instance a production cross section. When setting limits in more than one dimension, the CL$_\text{s}$ procedure is not applied.
 
For relevant parameters of interest, a physical bound on the parameter values is included in the statistical interpretation. For example, branching fraction parameters cannot conceptually be smaller than zero. The 95\% confidence interval quoted for such parameters is then based on the profile likelihood ratio restricted to the allowed region of parameter space, using the $\tilde{t}_{\mu}$ test statistic of Ref.~\cite{Cowan:2010st}. The confidence interval is defined by the standard $\chi^2$ cutoff, which leads to some over-coverage near the boundaries.
 
Uncertainties in the measurement parameters are in some cases broken down into separate components for theory uncertainties affecting the background processes, theory uncertainties affecting the Higgs boson signal production, experimental uncertainties including Monte Carlo (MC) statistical uncertainties, and statistical uncertainties. Each component is derived by fixing the associated nuisance parameters to their best-fit values $\hat{\theta}$ in both the numerator and denominator of $\Lambda$, and computing again the uncertainty in the measurement parameters. This is done for each component in turn, following the order in which they are listed above.
The uncertainty obtained at each step is then subtracted in quadrature from the uncertainty obtained in the previous step (in the first step, from the total uncertainty) to obtain the corresponding uncertainty component.
The statistical uncertainty component is obtained in the last step, with all nuisance parameters fixed except for the ones that are only constrained by data, such as parameters used to describe data-driven background estimates.
 
For the systematic uncertainties reported in the detailed breakdowns shown for instance in Table~\ref{tab:sys:mu}, a simpler procedure is used: in each case the corresponding nuisance parameters are fixed to their best-fit values, while other nuisance parameters are left free, and the resulting uncertainty is subtracted in quadrature from the total uncertainty.
 
The probability of compatibility with the Standard Model is quantified using the test statistic $\lambda_{\text{SM}} = -2 \ln \Lambda(\vec\alpha=\vec\alpha_\textrm{SM})$, where $\vec\alpha_\textrm{SM}$ are the Standard Model values of the parameters of interest. A $p$-value\footnote{The $p$-value is defined as the probability to obtain a value of the test statistic that is at least as high as the observed value under the hypothesis that is being tested.} $p_{\text{SM}}$ for the probability of compatibility is computed in the asymptotic approximation as $p_{\text{SM}} = 1 - F_{\chi^2_n}(\lambda_{\text{SM}})$, with $n$ equal to the number of free parameters of interest. For the cross-section and branching fraction measurements reported
in this paper, this definition does not account for the uncertainties in the SM values used as reference and may therefore lead to an underestimate of the probability of compatibility with the SM.
 
Results for expected significances and limits are obtained using the Asimov dataset technique~\cite{Cowan:2010st}.
 
The correlation coefficients presented in this paper are constructed to be symmetric around the observed best-fit values of the parameters of interest using the second derivatives of the negative log-likelihood ratio. Hence, the correlation matrices shown are not fully representative of the observed asymmetric uncertainties in the measurements. While the reported information is sufficient to reinterpret the measurements in terms of other parameterizations of the parameters of interest, this provides only an approximation to the information contained in the full likelihood function. For this reason, results for a number of commonly used parameterizations are also provided in Sections~\ref{sec:mu_XS_BR}~to~\ref{sec:kappa}.


\section{Combined measurements of signal strength, production cross sections and branching ratios}
\label{sec:mu_XS_BR}
\subsection{Global signal strength}
\label{sec:mu}
The global signal strength $\mu$ is determined following the procedures used for the measurements performed at $\sqrt{s}=7$ and 8~\TeV~\cite{HIGG-2015-07}.
For a specific production mode $i$ and decay final state $f$, the signal yield is expressed in terms of a single modifier $\mu_{if}$, as the production cross section $\sigma_i$ and the branching fraction $B_f$ cannot be separately measured without further assumptions.
The modifiers are defined as the ratios of the measured Higgs boson yields and their SM expectations, denoted by the superscript \textquote{SM},
\begin{equation}
\mu_{if} = \frac{\sigma_i}{\sigma_i^\text{SM}} \times \frac{\BR_f}{\BR_f^\text{SM}} .
\label{eq:mu}
\end{equation}
The SM expectation by definition corresponds to $\mu_{if} = 1$. The uncertainties in the SM predictions are included as nuisance parameters in the measurement of the signal strength modifiers, following the methodology
introduced in Section~\ref{sec:model}, where the procedures to decompose the uncertainties are also described.
 
In the model used in this section, all the $\mu_{if}$ are set to a global signal strength $\mu$, describing a common scaling of the expected Higgs boson yield in all categories. Its combined measurement is
\begin{equation*}
\mu = 1.11^{+0.09}_{-0.08}=1.11 \pm 0.05 \, (\text{stat.})\,^{+0.05}_{-0.04} \, (\text{exp.})\,^{+0.05}_{-0.04}\, (\text{sig.\ th.})\,\pm 0.03 \, (\text{bkg.\ th.})
\end{equation*}
where the total uncertainty is decomposed into components for statistical uncertainties, experimental systematic uncertainties, and theory uncertainties in signal and background modeling.
The signal theory component includes uncertainties due to missing higher-order perturbative QCD and electroweak corrections in the MC simulation, uncertainties in PDF and $\alpha_\mathrm{s}$ values, the treatment of the underlying event, the matching between the hard-scattering process and the parton shower, choice of hadronization models, and branching fraction uncertainties.
The measurement is consistent with the SM prediction with a $p$-value of $p_\text{SM}=18\%$, computed using the procedure defined in Section~\ref{sec:model} with one degree of freedom.
The value of $-2\ln\Lambda(\mu)$ as a function of $\mu$ is shown in Figure~\ref{fig:Lambdaexample}, for the full likelihood and the versions with sets of nuisance parameters fixed to their best-fit values to obtain the components of the uncertainty.
 
Table~\ref{tab:sys:mu} shows a summary of the leading uncertainties in the combined measurement of the global signal strength.
The dominant uncertainties arise from the theory modeling of the signal and background processes in simulation. Further important uncertainties relate to the luminosity measurement; the selection efficiencies, energy scale and energy resolution of electrons and photons; the estimate of lepton yields from heavy-flavor decays, photon conversions or misidentified hadronic jets (classified as \textit{background modeling} in the table); the jet energy scale and resolution, and the identification of heavy-flavor jets.
 
\begin{figure}[htbp]
\centering
\includegraphics[width=.6\textwidth]{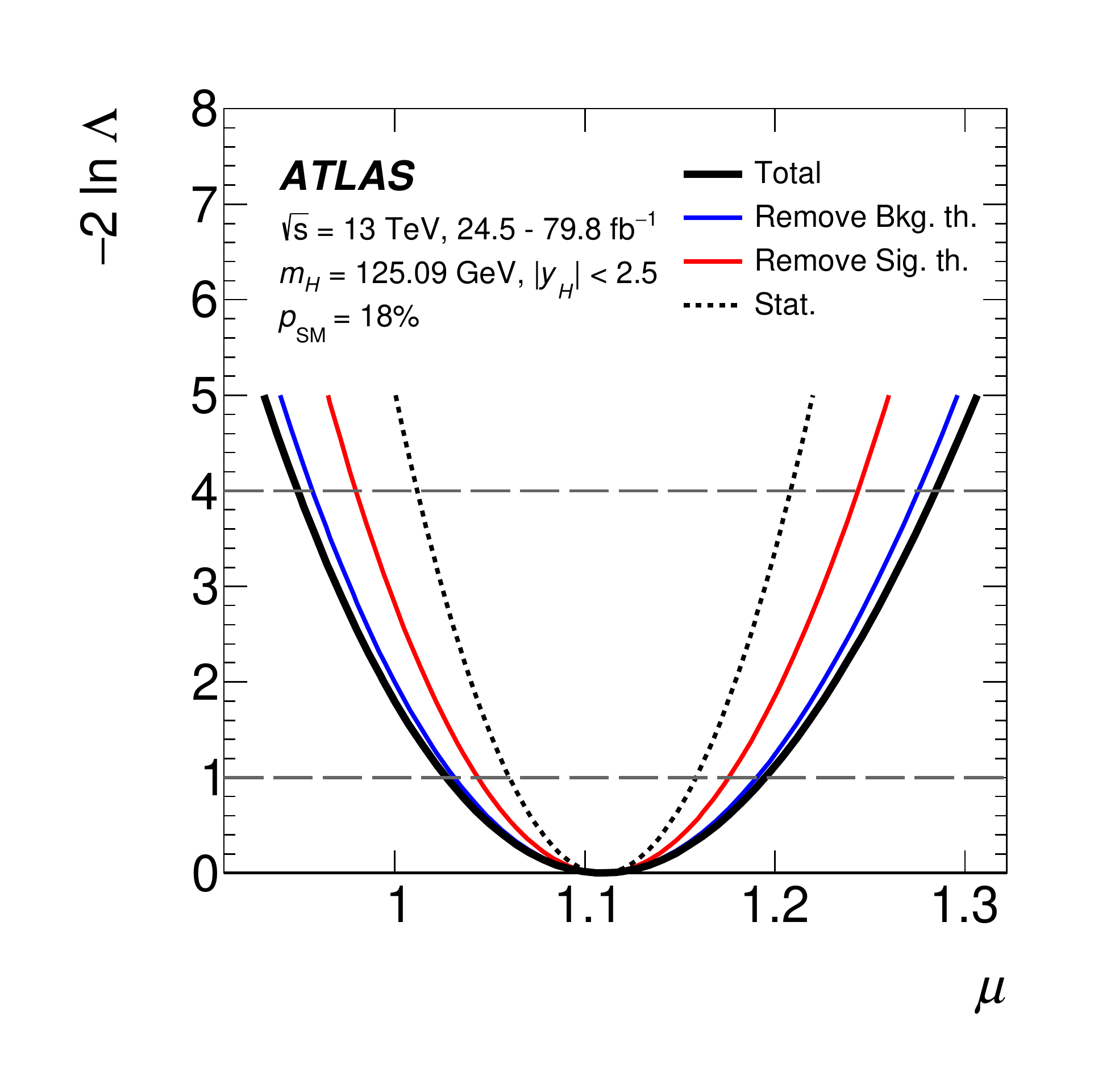}
\caption{Variations of $-2\ln\Lambda(\mu)$ as a function of $\mu$ with all systematic uncertainties included (solid black line), with parameters describing theory uncertainties in background processes fixed to their best-fit values (solid blue line), with the same procedure also applied to theory uncertainties in the signal process (solid red line) and to all systematic uncertainties, so that only statistical uncertainties remain (dotted black line). The dashed horizontal lines show the levels $-2\ln\Lambda(\mu)=1$ and $-2\ln\Lambda(\mu)=4$ which are used to define, respectively, the $1\sigma$ and $2\sigma$ confidence intervals for $\mu$.}
\label{fig:Lambdaexample}
\end{figure}
 
\begin{table}[htbp]
\centering
\caption{\label{tab:sys:mu} Summary of the relative uncertainties $\Delta\mu/\mu$ affecting the measurement of the combined global signal strength $\mu$.
\textquote{Other} refers to the combined effect of the sources of experimental systematic uncertainty not explicitly listed in the table.
The sum in quadrature of systematic uncertainties from individual sources differs from the uncertainty evaluated for the corresponding group in general, due to the presence of small correlations between nuisance parameters describing the different sources and other effects which are not taken into account in the procedure described in Section~\ref{sec:model}.
}
\renewcommand{\arraystretch}{1.3}
\begin{tabular}{lr}
 
\hline\hline
Uncertainty source                               & $\Delta\mu/\mu$ [\%] \\
\hline
Statistical uncertainty                          & 4.4 \\
\hline
Systematic uncertainties                         & 6.2 \\
~~~~~Theory uncertainties                        & 4.8 \\
~~~~~~~~~~Signal                                 & 4.2 \\
~~~~~~~~~~Background                             & 2.6 \\
~~~~~Experimental uncertainties (excl. MC stat.) & 4.1 \\
~~~~~~~~~~Luminosity                             & 2.0 \\
~~~~~~~~~~Background modeling                    & 1.6 \\
~~~~~~~~~~Jets, \MET                             & 1.4 \\
~~~~~~~~~~Flavor tagging                         & 1.1 \\
~~~~~~~~~~Electrons, photons                     & 2.2 \\
~~~~~~~~~~Muons                                  & 0.2 \\
~~~~~~~~~~$\tau$-lepton                          & 0.4 \\
~~~~~~~~~~Other                                  & 1.6 \\
~~~~~MC statistical uncertainty                  & 1.7 \\
\hline
Total uncertainty                                & 7.6 \\
\hline\hline
\end{tabular}
\end{table}
 
\FloatBarrier
 
\subsection{Production cross sections}
\label{sec:fivexs}
Higgs boson production is studied in each of its main production modes. The production mechanisms considered are \ggF, \VBF, \WH, \ZH\ (including \ggZH), and the combination of \ttH\ and \tH\ ($\ttH$+$\tH$). In cases where several processes are combined, the combination assumes the relative fractions of each component to be as in the SM, with theory uncertainties assigned. The small contribution from \bbH\ is grouped with \ggF. Cross sections are reported in the region $|y_H| < 2.5$ of the Higgs boson rapidity $y_H$. Results are obtained in a simultaneous fit to the data, with the cross sections of each production mechanism as parameters of interest. Higgs boson decay branching fractions are set to their SM values, within the uncertainties specified in Ref.~\cite{YR4}.
 
The results are shown in Figure~\ref{fig:fivexs} and Table~\ref{tab:fivexs}. The leading sources of uncertainty in the production cross-section measurements are summarized in Table~\ref{tab:sys:5xs}, with uncertainties computed as described in Section~\ref{sec:model}. The measured $\ttH$+$\tH$ production cross section differs from the \ttH\ cross section reported in Ref.~\cite{HIGG-2018-13}, even after accounting for the difference between the $|y_H|<2.5$ region used in this paper and the inclusive phase space considered in Ref.~\cite{HIGG-2018-13}. This is due in part to the inclusion of \tH, which in Ref.~\cite{HIGG-2018-13} is fixed to the SM expectation and not included in the reported \ttH\ cross section, as well as to better control of systematic effects, especially those related to photon energy scale and resolution, due to the \hgg\ categories targeting other processes which are included in this combination, as described in Section~\ref{sec:channels:hgg}.
The correlations between the measured cross sections, shown in Figure~\ref{fig:fivexs_corr}, are significantly reduced relative to previous analyses~\cite{HIGG-2015-07,HIGG-2014-06}.
\begin{figure}[bp]
\centering
\includegraphics[width=.7\textwidth]{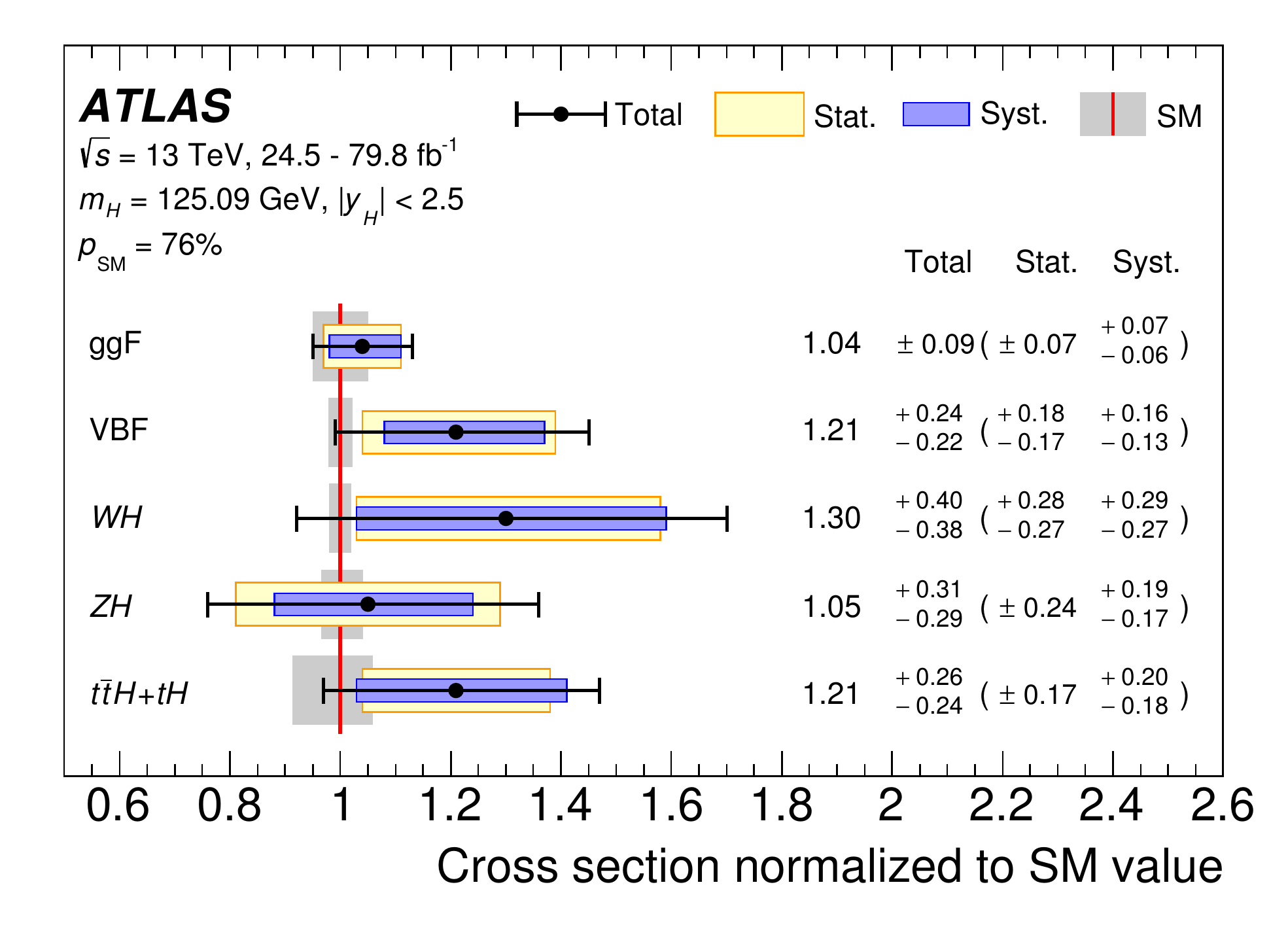}
\caption{Cross sections for \ggF, \VBF, \WH, \ZH\ and $\ttH$+$\tH$ normalized to their SM predictions,
measured with the assumption of SM branching fractions. The black error bars, blue boxes and yellow
boxes show the total, systematic, and statistical uncertainties in the measurements, respectively.
The gray bands indicate the theory uncertainties in the cross-section predictions.
}
\label{fig:fivexs}
\end{figure}
\begin{table}[p]
\caption{Best-fit values and uncertainties for the production cross sections of the Higgs boson, assuming SM values for its decay branching fractions. The total uncertainties are decomposed into components for data statistics (Stat.), experimental systematic uncertainties (Exp.), and theory uncertainties in the modeling of the signal (Sig.\ th.) and background (Bkg.\ th.) processes.
SM predictions are shown for the cross section of each production process. They are obtained from the inclusive cross-sections and associated uncertainties reported in Ref.~\cite{YR4}, multiplied by an acceptance factor for the region $|y_H| < 2.5$ computed using the Higgs boson simulation samples described in Section~\ref{sec:sample}. The observed (obs.) and expected (exp.) significances of the observed signals relative to the no-signal hypothesis are also shown for all processes except \ggF, which was observed in Run 1. For the \WH\ and \ZH\ modes, a combined \VH\ significance is reported assuming the SM value of the ratio of \WH\ to \ZH\ production.
}
\begin{center}
\renewcommand{\arraystretch}{1.3}
\resizebox{\columnwidth}{!}{
\begin{tabular}{Cr|CrClClClClCl|Cl|CrCr}
\hline \hline
Process       & \multicolumn{1}{c}{Value} & \multicolumn{5}{c|}{Uncertainty [pb] } & \multicolumn{1}{c|}{SM pred.} & \multicolumn{2}{c}{Significance} \\
($|y_H|<2.5$) & \multicolumn{1}{c}{[pb]}  & Total  & \small{Stat.}      & \small{Exp.}       & \small{Sig.\ th.}   & \small{Bkg.\ th.}   & \multicolumn{1}{c|}{[pb]} & \multicolumn{2}{c}{obs. (exp.)} \\
\hline
$\ggF$        &$46.5$ & ${}\pm{} 4.0$          & ${}\pm{} 3.1$ & ${}\pm{} 2.2$ & ${}\pm{} 0.9$ & ${}\pm{} 1.3$                                     & $\044.7 \pm 2.2$ & \multicolumn{1}{c}{-} & \\
$\VBF$        &$4.25$ & $\,\,^{+\;\,0.84}_{-\;\,0.77}$ & $\,\,^{+\;\,0.63}_{-\;\,0.60}$ & $\,\,^{+\;\,0.35}_{-\;\,0.32}$ & $\,\,^{+\;\,0.42}_{-\;\,0.32}$ & $\,\,^{+\;\,0.14}_{-\;\,0.11}$ & $3.515\pm 0.075$ & $6.5$ ($5.3$) & \\
$\WH$         &$1.57$ & $\,\,^{+\;\,0.48}_{-\;\,0.46}$ & $\,\,^{+\;\,0.34}_{-\;\,0.33}$ & $\,\,^{+\;\,0.25}_{-\;\,0.24}$ & $\,\,^{+\;\,0.11}_{-\;\,0.07}$ & ${}\pm{} 0.20$  & $1.204 \pm  0.024$   & $3.5$ ($2.7$) & \rdelim\}{2}{17.5mm}[\parbox{25mm}{\ $5.3$ ($4.7$)}] \\
$\ZH$         &$0.84$ & $\,\,^{+\;\,0.25}_{-\;\,0.23}$ & ${}\pm{} 0.19$ & ${}\pm{} 0.09$ & $\,\,^{+\;\,0.07}_{-\;\,0.04}$ & ${}\pm{} 0.10$               & $0.797\,\,^{+\;\,0.033}_{-\;\,0.026}$ & $3.6$ ($3.6$) & \\
$\ttH$+$\tH$  &$0.71$ & $\,\,^{+\;\,0.15}_{-\;\,0.14}$ & ${}\pm{} 0.10$ & $\,\,^{+\;\,0.07}_{-\;\,0.06}$ & $\,\,^{+\;\,0.05}_{-\;\,0.04}$ & $\,\,^{+\;\,0.08}_{-\;\,0.07}$   & $0.586\,\,^{+\;\,0.034}_{-\;\,0.049}$ & $\ 5.8$ ($5.4$) & \\
\hline \hline
\end{tabular}
}
\end{center}
\label{tab:fivexs}
\end{table}
\begin{figure}[tpb!]
\centering
\includegraphics[width=.7\textwidth]{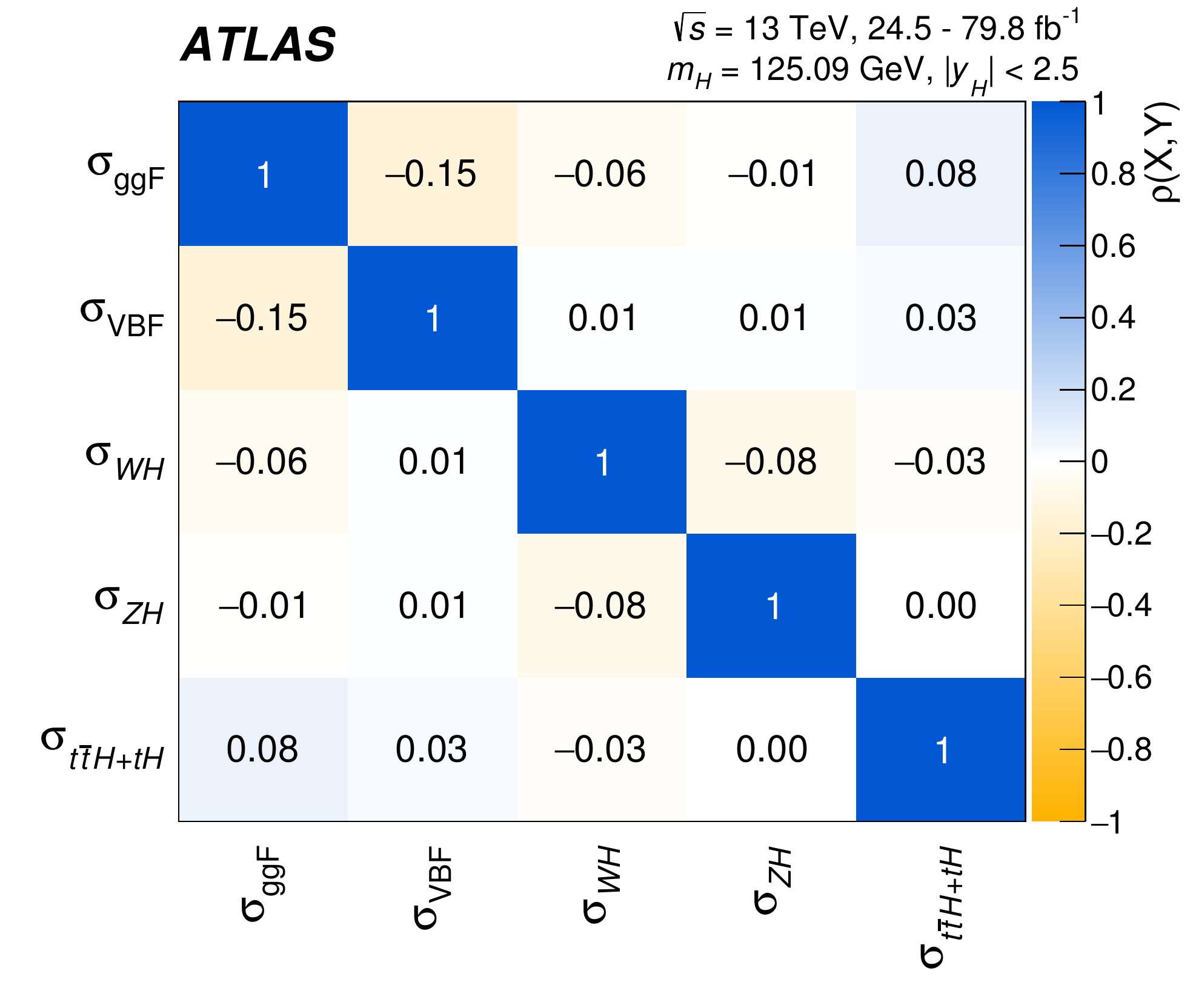}
\caption{Correlation matrix for the measurement of production cross sections of the Higgs boson, assuming SM values for its decay branching fractions.
}
\label{fig:fivexs_corr}
\end{figure}

\begin{table}
\centering
\caption{\label{tab:sys:5xs} Summary of the uncertainties affecting the production cross-section measurements.
\textquote{Other} refers to the combined effect of the sources of experimental systematic uncertainty not explicitly listed in the table.
The sum in quadrature of systematic uncertainties from individual sources differs from the uncertainty evaluated for the corresponding group in general, due to the presence of small correlations between nuisance parameters describing the different sources and other effects which are not taken into account in the procedure described in Section~\ref{sec:model}.
}
\resizebox{\columnwidth}{!}{
\begin{tabular}{Cl Cr Cr Cr Cr Cr}
 
\hline\hline
Uncertainty source   & $\frac{\Delta\sigma_{\ggF}}{\sigma_{\ggF}}$ [\%] & $\frac{\Delta\sigma_{\VBF}}{\sigma_{\VBF}}$ [\%] & $\frac{\Delta\sigma_{\WH}}{\sigma_{\WH}}$ [\%] & $\frac{\Delta\sigma_{\ZH}}{\sigma_{\ZH}}$ [\%] & $\frac{\Delta\sigma_{\ttH+\tH}}{\sigma_{\ttH+\tH}}$ [\%] \\
\hline
Statistical uncertainties                        & 6.4 & 15\dpt\0  & 21\dpt\0  & 23\dpt\0  & 14\dpt\0  \\
\hline
Systematic uncertainties                         & 6.2 & 12\dpt\0  & 22\dpt\0  & 17\dpt\0  & 15\dpt\0  \\
~~~~~Theory uncertainties                        & 3.4 & 9.2 & 14\dpt\0  & 14\dpt\0  & 12\dpt\0  \\
~~~~~~~~~~Signal                                 & 2.0 & 8.7 & 5.8 & 6.7 & 6.3 \\
~~~~~~~~~~Background                             & 2.7 & 3.0 & 13\dpt\0  & 12\dpt\0  & 10\dpt\0  \\
~~~~~Experimental uncertainties (excl. MC stat.) & 5.0 & 6.5 & 9.9 & 9.6 & 9.2 \\
~~~~~~~~~~Luminosity                             & 2.1 & 1.8 & 1.8 & 1.8 & 3.1 \\
~~~~~~~~~~Background modeling                    & 2.5 & 2.2 & 4.7 & 2.9 & 5.7 \\
~~~~~~~~~~Jets, \MET                             & 0.9 & 5.4 & 3.0 & 3.3 & 4.0 \\
~~~~~~~~~~Flavor tagging                         & 0.9 & 1.3 & 7.9 & 8.0 & 1.8 \\
~~~~~~~~~~Electrons, photons                     & 2.5 & 1.7 & 1.8 & 1.5 & 3.8 \\
~~~~~~~~~~Muons                                  & 0.4 & 0.3 & 0.1 & 0.2 & 0.5 \\
~~~~~~~~~~$\tau$-lepton                          & 0.2 & 1.3 & 0.3 & 0.1 & 2.4 \\
~~~~~~~~~~Other                                  & 2.5 & 1.2 & 0.3 & 1.1 & 0.8 \\
~~~~~MC statistical uncertainties                & 1.6 & 4.8 & 8.8 & 7.9 & 4.4 \\
\hline
Total uncertainties                              & 8.9 & 19\dpt\0  & 30\dpt\0  & 29\dpt\0  & 21\dpt\0  \\
\hline\hline
\end{tabular}
}
\end{table}
 
A modest correlation of $-15\%$ between the \ggF\ and \VBF\ processes remains, however, because of contributions from \ggF\ production in the \VBF-enriched selections.
The probability of compatibility between the measurement and the SM prediction corresponds to a $p$-value of $p_\text{SM}=76\%$, computed using the procedure outlined in Section~\ref{sec:model} with five degrees of freedom.
 
Figure~\ref{fig:twod} shows the observed likelihood contours in the plane of $\sigma_{\ggF}$ versus $\sigma_{\VBF}$ from individual channels and the combined fit, together with the SM prediction. The cross sections for the other production modes are profiled.
\begin{figure}[h!]
\centering
\includegraphics[width=.8\textwidth]{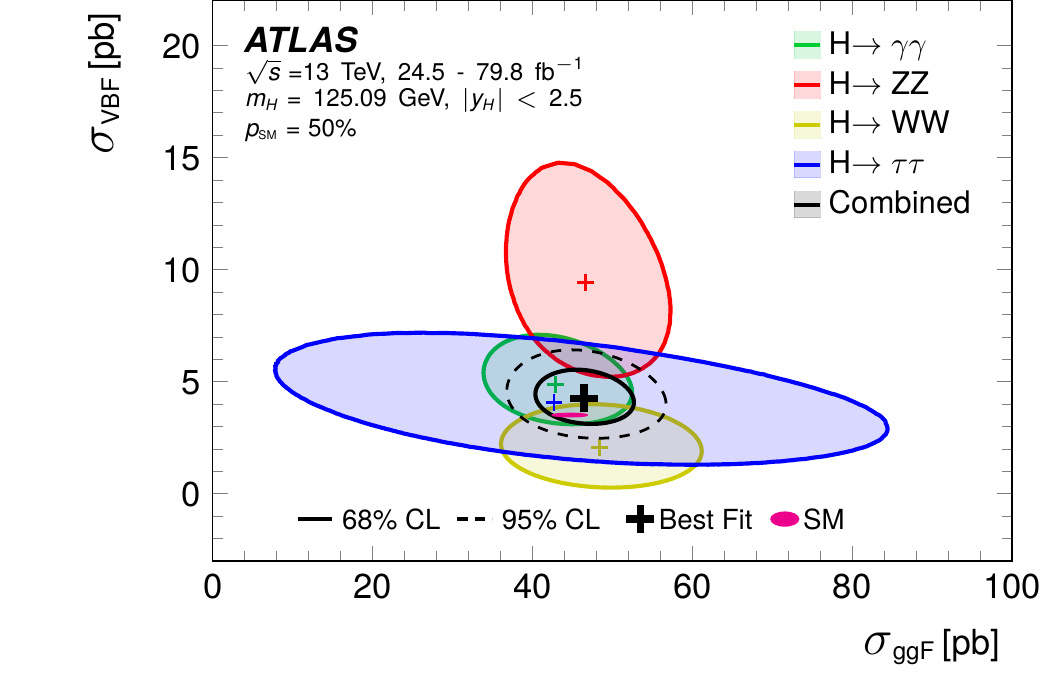}
\caption{Observed likelihood contours in the plane of $\sigma_{\VBF}$ versus $\sigma_{\ggF}$ from individual channels and the combined fit. Contours for 68\% CL, defined in the asymptotic approximation by $-2 \ln \Lambda = 2.28$, are shown as solid lines. The 95\% CL contour for the combined fit, corresponding to $-2 \ln \Lambda = 5.99$, is also shown as a dashed line. The crosses indicate the best-fit values, and the solid ellipse the SM prediction. Higgs boson branching fractions are fixed to their SM values within theory uncertainties. The probability of compatibility between the combined measurement and the SM prediction, estimated using the procedure outlined in the text with two degrees of freedom, corresponds to a p-value of $p_\text{SM} = 50\%$.}
\label{fig:twod}
\end{figure}

Significances above $5\sigma$ are observed for the combined measurements of the \ggF, \VBF, \VH\ and \ttH+\tH\ production processes. For the \VBF\ process, the observed (expected) significance is $6.5\sigma$ ($5.3\sigma$). For the \WH\ and \ZH\ modes, these are respectively $3.5\sigma$ ($2.7\sigma$) and $3.6\sigma$ ($3.6\sigma$). Combining \WH\ and \ZH\ production into a single \VH\ process, with the ratio of \WH\ to \ZH\ production set to its SM value leads to an observed (expected) significance for this process of $5.3\sigma$ ($4.7\sigma$). For the combination of \ttH\ and \tH\ production, the observed (expected) significance is $5.8\sigma$ ($5.4\sigma$).
 
\FloatBarrier
 
\subsection{Products of production cross sections and branching fractions}
\label{sec:fivetimesfour}
 
A description of both the production and decay mechanisms of the Higgs boson is obtained by considering the products $(\sigma \times \BR)_{if}$ of the cross section in production process $i$ and branching fraction to final state $f$. The production processes are defined as in Section~\ref{sec:fivexs} except for the fact that the \WH\ and \ZH\ processes, which cannot be reliably determined in all decay channels except \hbb, are considered together as a single \VH\ process, with the ratio of \WH\ to \ZH\ cross sections fixed to its SM value within uncertainties. The decay modes considered are \hgg, \hzz, \hww, \htt\ and \hbb. There are in total 20 such independent products, but the analyses included in the combination provide little sensitivity to \ggF\ production in the \hbb\ decay mode, and to \VH\ production in the \hww\ and \htt\ decay modes. The corresponding products are therefore fixed to their SM values within uncertainties. For the same reason, in \ttH\ production the \hzz\ decay mode is considered together with \hww\ as a single \hvv\ process, with the ratio of \hzz\ to \hww\ fixed to its SM value.
The results are obtained from a simultaneous fit of all input analyses, with the 16 independent $(\sigma \times \BR)$ products defined above as parameters of interest. They are shown in Figure~\ref{fig:fivetimesfour} and Table~\ref{tab:fivetimesfour}. The correlation matrix of the measurements is shown in Figure~\ref{fig:fivetimesfour_corr}. The largest terms in absolute value are between the \ttH, \hvv\ and \ttH, \htt\ processes, and between the \ggF, \htt\ and \VBF, \htt\ processes. In both cases, this is due to cross-contamination between these processes in the analyses providing the most sensitive measurements.
The probability of compatibility between the measurement and the SM prediction corresponds to a $p$-value of $p_\text{SM}=71\%$, computed using the procedure outlined in Section~\ref{sec:model} with 16 degrees of freedom.
 
\begin{figure}[h!]
\centering
\includegraphics[width=0.9\textwidth]{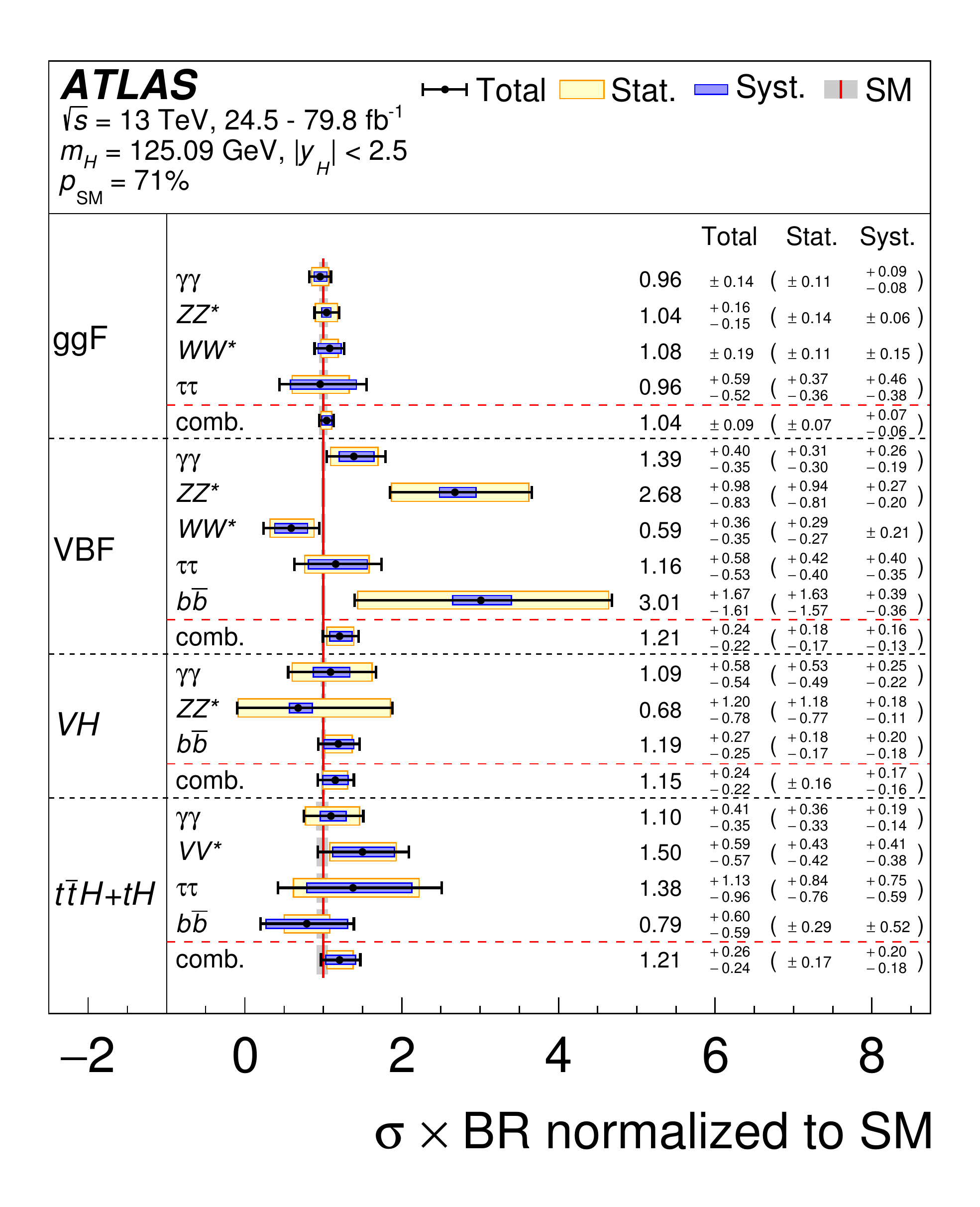}
\caption{Cross sections times branching fraction for \ggF, \VBF, \VH\ and $\ttH$+$\tH$ production in each relevant decay mode, normalized to their SM predictions. The values are obtained from a simultaneous fit to all channels. The cross sections of the \ggF, \Hbb, \VH, \hww\ and \VH, \Htt\ processes are fixed to their SM predictions. Combined results for each production mode are also shown, assuming SM values for the branching fractions into each decay mode.  The black error bars, blue boxes and yellow boxes show the total, systematic, and statistical uncertainties in the measurements, respectively. The gray bands show the theory uncertainties in the predictions.}
\label{fig:fivetimesfour}
\end{figure}
 
\begin{table}
\caption{Best-fit values and uncertainties for the production cross sections times branching fractions of the Higgs boson, for the combinations in which sufficient sensitivity is provided by the input analyses. Combinations not shown in the table are fixed to their SM values within uncertainties. For $\ttH$+$\tH$ production, \hvv\ refers to the combination of \hww\ and \hzz, with a relative weight fixed by their respective SM branching fractions.
The total uncertainties are decomposed into components for data statistics (Stat.), experimental systematic uncertainties (Exp.), and theory uncertainties in the modeling of the signal (Sig.\ th.) and background (Bkg.\ th.) processes. SM predictions~\cite{YR4} are shown for each process.
}
\begin{center}
\renewcommand{\arraystretch}{1.3}
\begin{tabular}{Cl|CrCrClClClCl|Cl}
\hline \hline
Process             & \multicolumn{1}{c}{Value}   & \multicolumn{5}{c|}{Uncertainty [fb] }                                                                 & \multicolumn{1}{c}{SM pred.}  \\
($|y_H|<2.5$)       & \multicolumn{1}{c}{[fb]}    & Total              & \small{Stat.}      &\small{Exp.}        & \small{Sig.\ th.}  &\small{Bkg.\ th.}   & \multicolumn{1}{c}{[fb]}      \\ \hline
$\ggF$, \hgg        &    $97$ & ${}\pm{} 14$             & ${}\pm{}   11$         & ${}\pm{}    8$         & ${}\pm{}  2$             & $\,\,^{+\;\,2}_{-\;\,1}$     & $     101.5 \pm 5.3$ \\
$\ggF$, \hzz        &  $1230$ & $\,\,^{+\;\,190}_{-\;\,180}$   & ${}\pm{}  170$         & ${}\pm{}   60$         & ${}\pm{} 20$             & ${}\pm{} 20$             & $\dpt 1181  \pm 61$ \\
$\ggF$, \hww        & $10400$ & ${}\pm{} 1800$           & ${}\pm{} 1100$         & ${}\pm{} 1100$         & ${}\pm{} 400$            & $\,\,^{+\;\,1000}_{-\;\,900}$ & $\dpt 9600  \pm 500$ \\
$\ggF$, \htt        & $ 2700$ & $\,\,^{+\;\,1700}_{-\;\,1500}$ & ${}\pm{} 1000$         & ${}\pm{}  900$         & $\,\,^{+\;800}_{-\;\,300}$ & ${}\pm{} 400$       & $\dpt 2800  \pm 140$ \\
\hline
$\VBF$, \hgg        & $ 11.1$ & $\,\,^{+\;\,3.2} _{-\;\,2.8}$  & $\,\,^{+\;\,2.5}_{-\;\,2.4}$   & $\,\,^{+\;\,1.4}_{-\;\,1.0}$ & $\,\,^{+\;\,1.5}_{-\;\,1.1}$ & $\,\,^{+\;\,0.3}_{-\;\,0.2}$ & $\0 7.98 \pm 0.21$ \\
$\VBF$, \hzz        &  $249$  & $\,\,^{+\;\,91}  _{-\;\,77}$   & $\,\,^{+\;\,87} _{-\;\,75}$    & $\,\,^{+\;\,16} _{-\;\,11}$  & $\,\,^{+\;\,17} _{-\;\,12}$  & $\,\,^{+\;\,9}  _{-\;\,7}$ & $\0     92.8 \pm 2.3$ \\
$\VBF$, \hww        &  $450$  & $\,\,^{+\;\,270} _{-\;\,260}$  & $\,\,^{+\;\,220}_{-\;\,200}$   & $\,\,^{+\;\,120}_{-\;\,130}$ & $\,\,^{+\;\,80} _{-\;\,70}$  & $\,\,^{+\;\,70} _{-\;\,80}$ & $\0\dpt  756 \pm 19$ \\
$\VBF$, \htt        &  $260$  & $\,\,^{+\;\,130} _{-\;\,120}$  & ${}\pm{} 90$               & $\,\,^{+\;\,80} _{-\;\,70}$  & $\,\,^{+\;\,30} _{-\;\,10}$  & $\,\,^{+\;\,30} _{-\;\,20}$ & $\0\dpt  220 \pm 6$ \\
$\VBF$, \hbb        & $6100$  & $\,\,^{+\;\,3400}_{-\;\,3300}$ & $\,\,^{+\;\,3300}_{-\;\,3200}$ & $\,\,^{+\;\,700}_{-\;\,600}$ & ${}\pm{} 300$          & ${}\pm{} 300$        & $\dpt   2040 \pm 50$ \\
\hline
$\VH$, \hgg        &  $5.0$  & $\,\,^{+\;\,2.6}_{-\;\,2.5}$   & $\,\,^{+\;\,2.4}_{-\;\,2.2}$  & $\,\,^{+\;\,1.0}_{-\;\,0.9}$ & ${}\pm{} 0.5$           & ${}\pm{} 0.1$        & $\0   4.54\,\,^{+\;\,0.13}_{-\;\,0.12}$ \\
$\VH$, \hzz        &   $36$  & $\,\,^{+\;\,63} _{-\;\,41}$    & $\,\,^{+\;\,62}_{-\;\,41}$   & $\,\,^{+\;\,5}  _{-\;\,4}$   & $\,\,^{+\;\,6}  _{-\;\,4}$  & $\,\,^{+\;\,4}_{-\;\,2}$ & $\0   52.8 \pm 1.4$ \\
$\VH$, \hbb        & $1380$  & $\,\,^{+\;\,310}_{-\;\,290}$   & $\,\,^{+\;\,210}_{-\;\,200}$  & ${}\pm{} 150$            & $\,\,^{+\;\,120}_{-\;\,80}$ & ${}\pm{} 140$        & $\dpt 1162 \,\,^{+\;\,31}_{-\;\,29}$ \\
\hline
$\ttH$+$\tH$, \hgg & $1.46$   & $\,\,^{+\;\,0.55}_{-\;\,0.47}$ & $\,\,^{+\;\,0.48}_{-\;\,0.44}$ & $\,\,^{+\;\,0.19}_{-\;\,0.15}$ & $\,\,^{+\;\,0.17}_{-\;\,0.11}$ & ${}\pm{} 0.03$       & $\0     1.33\,\,^{+\;\,0.08}_{-\;\,0.11}$ \\
$\ttH$+$\tH$, \hvv &  $212$   & $\,\,^{+\;\,84}  _{-\;\,81}$   & $\,\,^{+\;\,61}  _{-\;\,59}$   & $\,\,^{+\;\,47}  _{-\;\,44}$   & $\,\,^{+\;\,17}  _{-\;\,10}$   & $\,\,^{+\;\,31} _{-\;\,30}$ & $\0\dpt  142\,\,^{+\;\,8}   _{-\;\,12}$ \\
$\ttH$+$\tH$, \htt &   $51$   & $\,\,^{+\;\,41}  _{-\;\,35}$   & $\,\,^{+\;\,31}  _{-\;\,28}$   & $\,\,^{+\;\,26}  _{-\;\,21}$   & $\,\,^{+\;\,6}   _{-\;\,4}$    & $\,\,^{+\;\,8}  _{-\;\,6}$ & $\0     36.7\,\,^{+\;\,2.2} _{-\;\,3.1}$ \\
$\ttH$+$\tH$, \hbb &  $270$   & ${}\pm{} 200$              & ${}\pm{} 100$          & ${}\pm{} 80$           & $\,\,^{+\;\,40}  _{-\;\,10}$   & $\,\,^{+\;\,150}_{-\;\,160}$ & $\0\dpt  341\,\,^{+\;\,20}  _{-\;\,29}$ \\
\hline \hline
\end{tabular}
\end{center}
\label{tab:fivetimesfour}
\end{table}
 
\begin{figure}[tpb!]
\centering
\includegraphics[width=.9\textwidth]{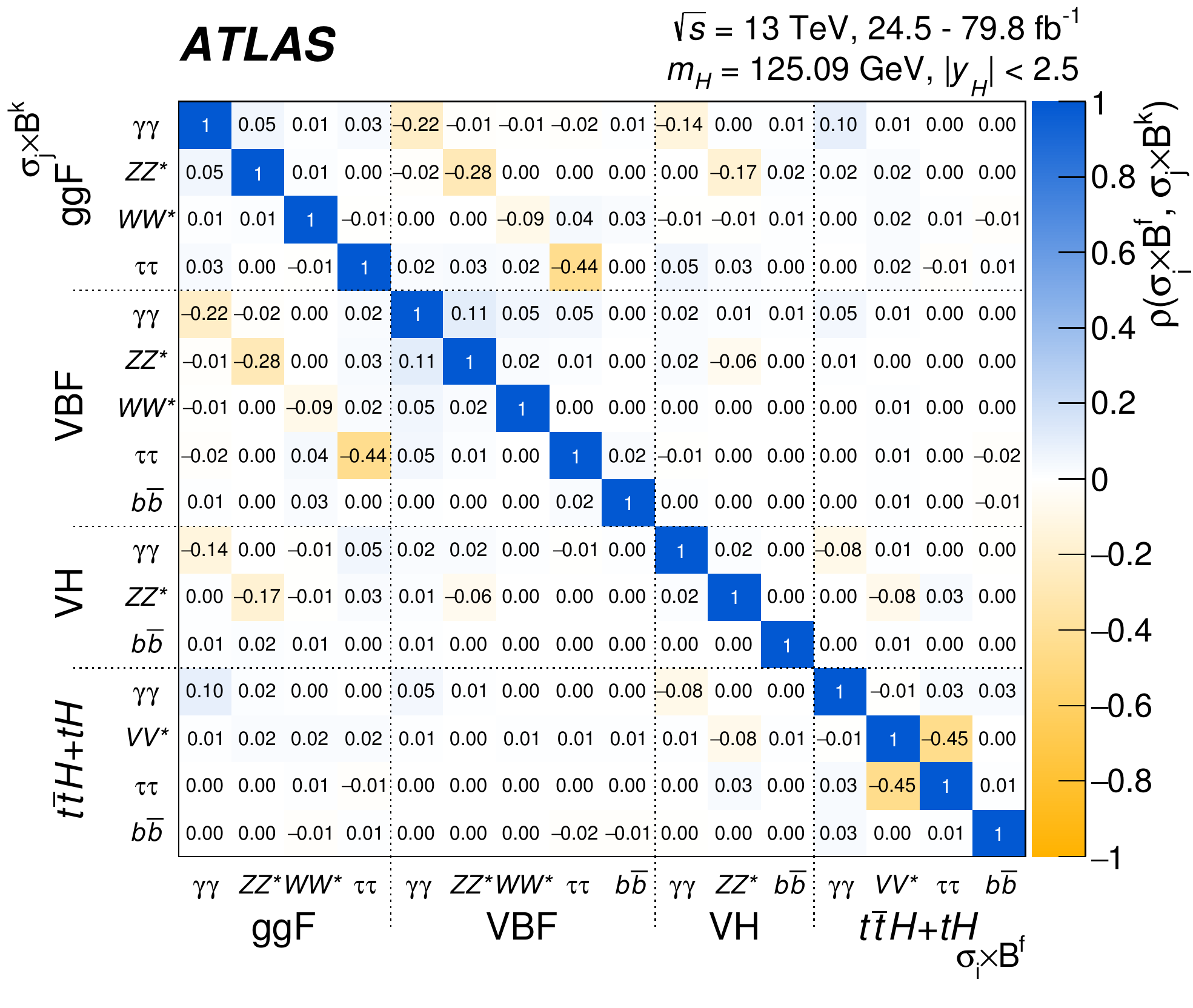}
\caption{Correlation matrix for the measured values of the production cross sections times branching fractions of the Higgs boson, for the combinations in which sufficient sensitivity is provided by the input analyses.}
\label{fig:fivetimesfour_corr}
\end{figure}

\FloatBarrier

\subsection{Ratios of cross sections and branching fractions}
\label{sec:ratios}
 
The products $(\sigma \times \BR)_{if}$ described in Section~\ref{sec:fivetimesfour} can be expressed as
\begin{equation*}
(\sigma \times \BR)_{if} = \sigma_{\ggF}^{\zz} \cdot
\left(\frac{\sigma_{i}}{\sigma_{\ggF}}\right)\cdot\left(\frac{\BR_{f}}{\BR_{\zz}}\right),
\end{equation*}
in terms of the cross section times branching fraction $\sigma_{\ggF}^{\zz}$ for the reference process $gg \to H \to ZZ^*$, which is precisely measured and exhibits small systematic uncertainties, ratios of production cross sections to that of \ggF, $\sigma_i/\sigma_{\ggF}$, and ratios of branching fractions to that of \hzz, $\BR_f/\BR_{\zz}$.
 
Results are shown in Figure~\ref{fig:ratio} and Table~\ref{tab:ratio}. The probability of compatibility between the measurements and the SM predictions corresponds to a $p$-value of $p_\text{SM}= 93\%$, computed using the procedure outlined in Section~\ref{sec:model} with nine degrees of freedom.
 
\begin{figure}[h!]
\centering
\includegraphics[width=.8\textwidth]{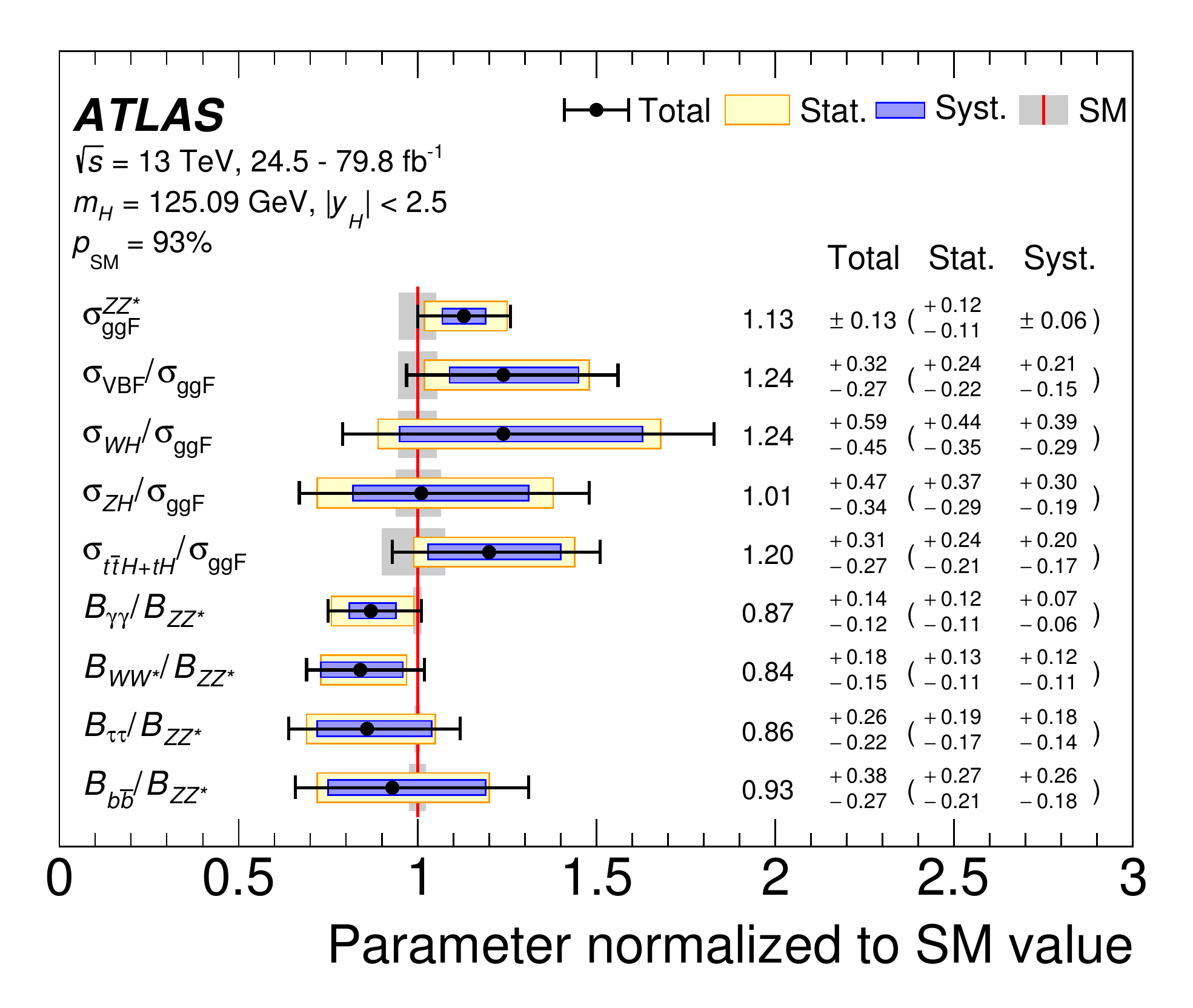}
\caption{Results of a simultaneous fit for $\sigma_{\ggF}^{\zz}$, $\sigma_{\VBF}/\sigma_{\ggF}$, $\sigma_{\WH}/\sigma_{\ggF}$, $\sigma_{\ZH}/\sigma_{\ggF}$, $\sigma_{\ttH+\tH}/\sigma_{\ggF}$, $\BR_{\gamma \gamma}/\BR_{\zz}$, $\BR_{\ww}/\BR_{\zz}$, $\BR_{\tau\tau}/\BR_{\zz}$,
and $\BR_{bb}/\BR_{\zz}$.  The fit results are normalized to the SM predictions.  The black error bars, blue boxes and yellow boxes show the total, systematic, and statistical uncertainties in the measurements, respectively. The gray bands show the theory uncertainties in the predictions.}
\label{fig:ratio}
\end{figure}
 
\begin{table}[!tpb]
\caption{Best-fit values and uncertainties for $\sigma_{\ggF}^{\zz}$, together with ratios of production cross sections normalized to $\sigma_{\ggF}$, and ratios of branching fractions normalized to $\BR_{\zz}$. The total uncertainties are decomposed into components for data statistics (Stat.), experimental systematic uncertainties (Exp.), and theory uncertainties in the modeling of the signal (Sig.\ th.) and background (Bkg.\ th.) processes. The SM predictions~\cite{YR4} are also shown with their total uncertainties.  }
\begin{center}
\renewcommand{\arraystretch}{1.35}
\resizebox{\columnwidth}{!}{
\begin{tabular}{Cl Cl | Cr Cl Cl Cl Cl Cl | Cl}
\hline \hline
\multirow{2}{*}{Quantity} &  & \multirow{2}{*}{Value}
& \multicolumn{5}{c|}{Uncertainty} & \multirow{2}{*}{SM prediction} \\
& & &
Total & Stat. & Exp. & Sig.\ th. & Bkg.\ th. & 
\\
\hline
$\sigma_{\ggF}^{\zz}$             & [pb] & $1.33$   & ${}\pm{} 0.15$   & $\,\,^{+\;\,0.14}  _{-\;\,0.13}$    & ${}\pm{} 0.06$             & $\,\,^{+\;\,0.02}  _{-\;\,0.01}$             & $\,\,^{+\;\,0.04}  _{-\;\,0.02}$             & $\01.181 \pm 0.061$    \\
$\sigma_{\VBF}/\sigma_{\ggF}$     &      & $0.097$  & $\,\,^{+\;\,0.025} _{-\;\,0.021}$  & $\,\,^{+\;\,0.019} _{-\;\,0.017}$   & $\,\,^{+\;\,0.010} _{-\;\,0.008}$  & $\,\,^{+\;\,0.011} _{-\;\,0.008}$  & $\,\,^{+\;\,0.006} _{-\;\,0.005}$   & $0.0786 \pm 0.0043$  \\
$\sigma_{\WH}/\sigma_{\ggF}$      &      & $0.033$  & $\,\,^{+\;\,0.016} _{-\;\,0.012}$  & $\,\,^{+\;\,0.012} _{-\;\,0.009}$   & $\,\,^{+\;\,0.007} _{-\;\,0.006}$  & $\,\,^{+\;\,0.003} _{-\;\,0.002}$  & $\,\,^{+\;\,0.007} _{-\;\,0.005}$   & $0.0269\,\,^{+\;\,0.0014}_{-\;\,0.0015}$ \\
$\sigma_{\ZH}/\sigma_{\ggF}$      &      & $0.0180$ & $\,\,^{+\;\,0.0084}_{-\;\,0.0061}$ & $\,\,^{+\;\,0.0066}_{-\;\,0.0052}$  & $\,\,^{+\;\,0.0034}_{-\;\,0.0021}$ & $\,\,^{+\;\,0.0016}_{-\;\,0.0009}$ & $\,\,^{+\;\,0.0037}_{-\;\,0.0025}$ & $0.0178\,\,^{+\;\,0.0011}_{-\;\,0.0010}$ \\
$\sigma_{\ttH+\tH}/\sigma_{\ggF}$ &      & $0.0157$ & $\,\,^{+\;\,0.0041}_{-\;\,0.0035}$ & $\,\,^{+\;\,0.0031}_{-\;\,0.0028}$  & $\,\,^{+\;\,0.0020}_{-\;\,0.0017}$ & $\,\,^{+\;\,0.0012}_{-\;\,0.0008}$ & $\,\,^{+\;\,0.0013}_{-\;\,0.0012}$ & $0.0131\,\,^{+\;\,0.0010}_{-\;\,0.0013}$ \\
$\BR_{\gamma\gamma}/\BR_{\zz}$    &      & $0.075$  & $\,\,^{+\;\,0.012} _{-\;\,0.010}$  & $\,\,^{+\;\,0.010} _{-\;\,0.009}$   & $\,\,^{+\;\,0.006} _{-\;\,0.005}$  & ${}\pm{} 0.001$  & ${}\pm{} 0.002$            & $0.0860\pm 0.0010$           \\
$\BR_{\ww}/\BR_{\zz}$             &      & $ 6.8$    & $\,\,^{+\;\,1.5}   _{-\;\,1.2}$    & $\,\,^{+\;\,1.1}   _{-\;\,0.9}$     & $\,\,^{+\;\,0.8}   _{-\;\,0.7}$    & ${}\pm{} 0.2$              & $\,\,^{+\;\,0.6} _{-\;\,0.5}$              & $\0\0 8.15\pm{} < 0.01$             \\
$\BR_{\tau\tau}/\BR_{\zz}$        &      & $2.04$   & $\,\,^{+\;\,0.62}  _{-\;\,0.52}$   & $\,\,^{+\;\,0.45}  _{-\;\,0.40}$    & $\,\,^{+\;\,0.36}  _{-\;\,0.31}$   & $\,\,^{+\;\,0.17}  _{-\;\,0.09}$   & $\,\,^{+\;\,0.12}_{-\;\,0.09}$     & $\02.369\pm 0.017$               \\
$\BR_{bb}/\BR_{\zz}$              &      & $20.5$      & $\,\,^{+\;\,8.4}   _{-\;\,5.9}$    & $\,\,^{+\;\,5.9}   _{-\;\,4.6}$     & $\,\,^{+\;\,3.7}   _{-\;\,2.4}$    & $\,\,^{+\;\,1.3}   _{-\;\,0.9}$    & $\,\,^{+\;\,4.2} _{-\;\,2.9}$       & $\022.00\pm 0.51$                \\
\hline \hline
\end{tabular}
}
\end{center}
\label{tab:ratio}
\end{table}
 
\FloatBarrier

\section{Combined measurements of simplified template cross sections}
\label{sec:stxs}
\subsection{Simplified template cross-section framework}
\label{sec:stxs_framework}
Simplified template cross sections~\cite{YR4,LesHouches} are defined through a partition of the phase space of the SM Higgs production process into a set of non-overlapping regions. These regions are defined in terms of the kinematics of the Higgs boson and, when they are present, of associated jets and~$W$ and $Z$~bosons, independently of the Higgs boson decay process. They are chosen according to three criteria: sensitivity to deviations from the SM expectation, avoidance of large theory uncertainties in the corresponding SM predictions, and to approximately match experimental selections so as to minimize model-dependent extrapolations. Analysis selections do not, however, necessarily correspond exactly to the STXS regions.
 
All regions are defined for a Higgs boson rapidity $y_H$ satisfying $|y_H| < 2.5$, corresponding approximately to the region of experimental sensitivity.
Jets are reconstructed from all stable particles with a lifetime greater than $10\,\ps$, excluding the decay products of the Higgs boson and leptons from $W$ and $Z$ boson decays, using the anti-$k_t$ algorithm with a jet radius parameter $R = 0.4$, and must have a transverse momentum $\pTj\ > 30\,\GeV$.
 
The measurements presented in this paper are based on the Stage~1 splitting of the STXS framework~\cite{YR4}. Higgs boson production is first classified according to the nature of the initial state and of associated particles, the latter including the decay products of $W$ and $Z$ bosons if they are present. These categories are, by order of decreasing selection priority: \ttH\ and \tH\ processes; \qqtoHqq\ processes, with contributions from both \VBF\ production and quark-initiated \VH\ production with a hadronic decay of the gauge boson; \ggtoZH\ with $Z\to q\bar{q}$; \VH\ production with a leptonic decay of the vector boson (\VHlep), including \ggtoZH\ production; and finally the gluon--gluon fusion process. The last is considered together with \ggtoZH, $Z\to q\bar{q}$ production, as a single \ggtoH\ process. The \bbH\ production mode is modeled as a $1\%$~\cite{YR4} increase of the \ggtoH\ yield in each STXS bin, since the acceptances for both processes are similar for all input analyses~\cite{YR4}. The \ttH\ and \tH\ processes are also combined in a single \ttH+\tH\ category, assuming the relative fraction of each component to be as in the SM, within uncertainties.
 
The analyses included in this paper provide only limited sensitivity to the cross section in some bins of the Stage~1 scheme, mainly due to limited data statistics in some regions. In other cases, they only provide sensitivity to a combination of bins, leading to strongly correlated measurements. To mitigate these effects, the results are presented in terms of a reduced splitting, with the measurement bins defined as merged groups of Stage~1 bins (and in the case of \VHlep\ with an additional splitting not present in the original Stage 1 scheme, as described below). These measurement bins are defined as follows for each process:
 
\begin{itemize}
\item \textbf{\ggtoH} is separated into regions defined by the jet multiplicity and the Higgs boson transverse momentum \ptH. A region is defined for events with one or more jets and $\ptH \ge 200\,\GeV$, providing sensitivity to deviations from the SM at high momentum transfer. The remaining events are separated into classes with 0, 1 and $\geq 2$ jets in the final state. The one-jet category is further split in bins of \ptH, probing perturbative QCD predictions and providing sensitivity to deviations from the SM. Three bins are defined with $\ptH < 60\,\GeV$, $60\,\GeV \leq \ptH <120\,\GeV$ and $120\,\GeV \leq \ptH <200\,\GeV$.
 
\item \textbf{\qqtoHqq} is separated into three regions. The first selects events in which the transverse momentum of the leading jet $p_\mathrm{T}^j$ is $\geq 200\,\GeV$. A second region, denoted by \textit{\VH\ topo}, is defined by $p_\mathrm{T}^j < 200\,\GeV$ and the presence of two jets with an invariant mass $m_{jj}$ in the range $60 \leq m_{jj} < 120 \,\GeV$, selecting events originating from \VH\ production in particular. The remaining events are grouped into a third bin, denoted by \textit{\VBF\ topo + Rest}, which includes mainly the \VBF-topology region (\textit{\VBF\ topo}) defined by the presence of two jets with $m_{jj} \geq 400 \,\GeV$ and a pseudorapidity difference $|\Delta\eta_{jj}| \geq 2.8$, as well as events that fall in none of the above selections (\textit{Rest}). The measurement sensitivity for the corresponding cross section is provided mainly by the \VBF-topology region, within which the cross section is measured precisely by the analyses targeting \VBF\ production.
 
\item \textbf{\VHlep} is split into the two processes \qqtoWH\ and \pptoZH, the latter including both quark-initiated and gluon-initiated production. These regions are further split according to $\pTV$, the transverse momentum of the $W$ or $Z$ boson. For the \qqtoWH\ process two bins are defined for $\pT^V < 250\,\GeV$ and $\pT^V \geq 250\,\GeV$, while for \pptoZH\ three bins are defined for $\pT^V < 150\,\GeV$, $150\,\GeV \leq \pT^V < 250\,\GeV$ and $\pT^V \geq 250\,\GeV$. This definition deviates from the one given in Ref.~\cite{YR4}, where the \qqtoZH\ and \ggtoZH\ processes are measured separately and no splitting is performed at $\pT^V = 250\,\GeV$ for \ggtoZH, given the limited sensitivity of the current measurements to separating the \qqtoZH\ and \ggtoZH\ processes.
\end{itemize}
 
The above merging scheme of Stage 1 bins is summarized in Figure~\ref{fig:STXS_merge1}.
 
Sensitivity to the 0-jet and 1-jet, $\ptH <60\,\GeV$ regions of the \ggtoH\ process is provided mainly by the \hfourl, \hgg\ and \hwwenmun\ analyses, with the leading contribution in each region coming from \hwwenmun\ and \hgg\ respectively. For the 1-jet, $60 \leq \ptH <120\,\GeV$ region, the main contributions to the sensitivity are from \hfourl\ and \hgg, dominated by the latter. The \hgg\ analysis also provides the largest sensitivity in the rest of the \ggtoH\ regions as well as in the \qqtoHqq\ sector, apart from the $p_\mathrm{T}^j > 200\,\GeV$ region for which \htt\ dominates the sensitivity. The \VH, \Hbb\ analysis provides the most sensitive measurements in the \VHlep\ regions. Finally, the \hgg\ and \ttH\ multilepton analyses provide the leading contributions to the measurement of the \ttH+\tH\ region.
 
\begin{figure}[tp!]
\centering
\includegraphics[width=.975\textwidth]{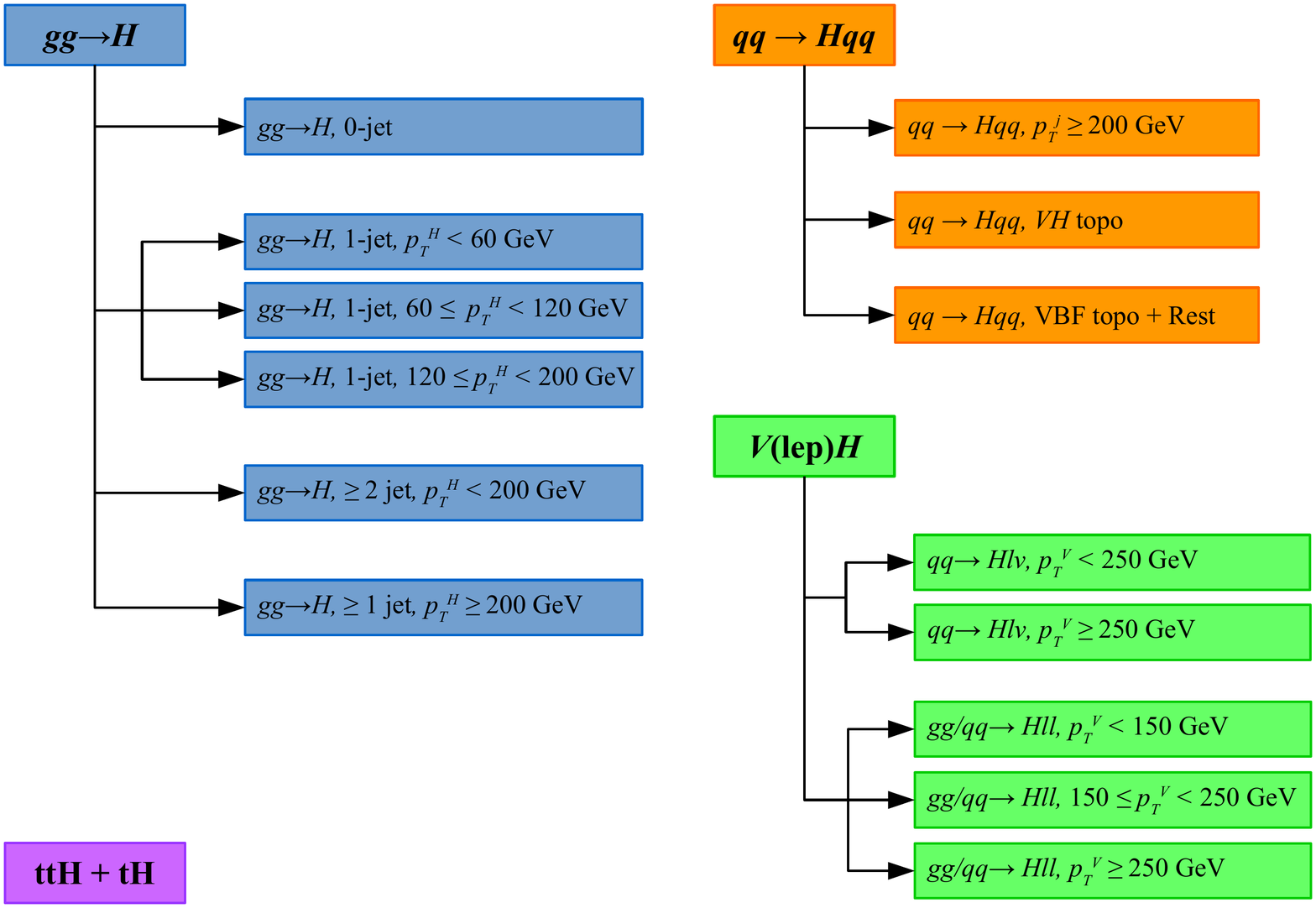}
\caption{Definition of the STXS measurement regions used in this paper. For each Higgs boson production process, the regions are defined starting from the top of the corresponding schematic, with regions nearer the top taking precedence if the selections overlap. The \bbH\ production mode is considered as part of \ggtoH.}
\label{fig:STXS_merge1}
\end{figure}
 
The measured event yields are described by Eq.~(\ref{eq:yields}), with parameters of interest of the form $(\sigma \times \BR)_{if}$ denoting the cross section times branching fraction in STXS region $i$ and decay channel~$f$. The acceptance factors $(\epsilon \times A)_{if}^k$ for each analysis category~$k$ are determined from SM Higgs boson production processes, modeled using the samples described in Section~\ref{sec:sample}, and act as templates in the fits of the STXS cross sections to the data. The dependence on the theory assumptions is less than in the measurement of the total cross sections in each production mode, since the $(\epsilon \times A)_{if}^k$ are computed over smaller regions. Assumptions about the kinematics within a given STXS region lead to some model-dependence, which can be reduced further by using a finer splitting of the phase space, as allowed by experimental precision. Results using a splitting finer than the one described in this section are presented in Appendix~\ref{sec:stxs_weak}.
 
Theory uncertainties for the \ggtoH\ and \qqtoHqq\ processes are defined as in Ref.~\cite{HIGG-2016-21}, while those of the \VHlep\ process follow the scheme described in Ref.~\cite{ATL-PHYS-PUB-2018-035}. For the measurement bins defined by merging several bins of the STXS Stage-1 framework, the $(\epsilon \times A)$ factors are determined assuming that the relative fractions of each Stage-1 bin are as in the SM, and SM uncertainties in these fractions are taken into account.
 
\subsection{Results}
\label{sec:stxs_results}
The fit parameters chosen for the combined STXS measurements are the cross sections for Higgs boson production in STXS region $i$ times the branching fraction for the \hzz\ decay, $(\sigma \times \BR)_{i, \zz}$, and the ratios of branching fractions $\BR_f/\BR_{\zz}$ for the other final states $f$. Similarly to the ratio model in Section~\ref{sec:ratios}, the cross sections times branching fractions for final states other than \zz\ are parameterized as

\begin{equation*}
(\sigma \times \BR)_{if} = (\sigma \times \BR)_{i,\zz} \cdot \left(\frac{\BR_f}{\BR_{\zz}}\right).
\end{equation*}
 
The results are shown in Figures~\ref{fig:stxs_results_norm} and~\ref{fig:stxs_results_log} and in Table~\ref{tab:stxs_results}. The observed upper limits at 95\% CL on the cross sections in the $qq\rightarrow Hqq$, \VH\ topo and $~qq\rightarrow Hqq,~\pT^{j} \geq 200~\GeV$ bins are found to be 1.45 pb and 0.59 pb, respectively, taking into account the physical bound on the parameter values as discussed in Section~\ref{sec:model}. The corresponding expected upper limits are 1.53 pb and 0.80 pb, respectively.

The correlations between the measured parameters are shown in Figure~\ref{fig:stxs_correlation}.  The largest anti-correlations are between $\BR_{\bbbar}/\BR_{\zz}$ and the cross-section measurements in the \VHlep\ region, since the \VH, \Hbb\ analysis is sensitive to products of these quantities; between the cross-section measurement in the \ggtoH\ 0-jet region and both $\BR_{\yy}/\BR_{\zz}$ and  $\BR_{\ww}/\BR_{\zz}$, since the \hgg, \hfourl\ and \hwwenmun\ decay channels provide the most precise measurements in this region; between $\BR_{\yy}/\BR_{\zz}$ and the cross-section measurement in the $qq\rightarrow Hqq$, \VBF\ topo + Rest region, since there is a tension between the \hgg\ and \hfourl\ measurements in this region; between $\BR_{\tau\tau}/\BR_{\zz}$ and the cross-section measurement in the $\pT^H  > 200$\,\GeV\ region, since the high-\ptH\ channels of the \htt\ analysis are sensitive to their product; and between the cross-section measurements in the $qq\rightarrow Hqq,~\pT^{j} \geq 200~\GeV$ and $gg\rightarrow H,~\geq\textrm{1-jet}, \ptH \geq 200~\GeV$ regions on the one hand, and the $qq\rightarrow Hqq,~\pT^{j} \geq 200~\GeV$ and $gg\rightarrow H,~\textrm{1-jet}, 120 \leq \ptH < 200~\GeV$ regions on the other hand, since in both cases there is cross-contamination between these processes in the experimental selections.
 
The largest positive correlations are between the $(W \rightarrow \ell\nu)H$ and $(Z \rightarrow \ell\ell)H$ measurement regions, related to their strong anti-correlation with $\BR_{\bbbar}/\BR_{\zz}$; and between $\BR_{\yy}/\BR_{\zz}$ and  $\BR_{\ww}/\BR_{\zz}$, due to their strong anti-correlation with the cross-section measurement in the 0-jet region.
 
The results show good overall agreement with the SM predictions in a range of kinematic regions of Higgs boson production processes.  The probability of compatibility between the measurement and the
SM prediction corresponds to a $p$-value of $p_{\textrm{SM}}=89$\%, computed using the procedure outlined in Section~\ref{sec:model} with 19 degrees of freedom.

\begin{figure}[tp!]
\centering
\includegraphics[width=.9\textwidth]{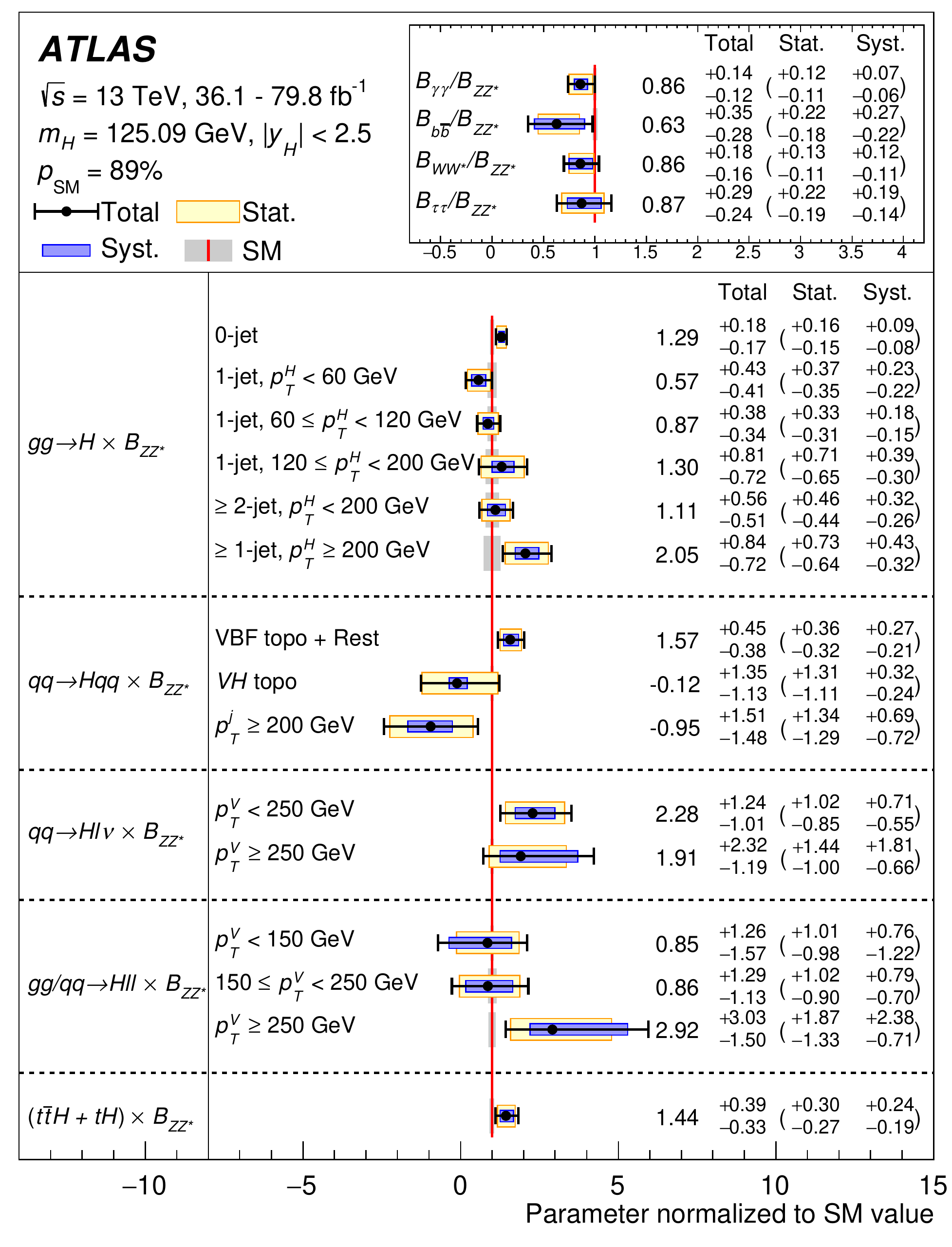}
\caption{Best-fit values and uncertainties for the cross sections in each measurement region
and of the ratios of branching fractions ${\BR}_f/{\BR}_{\zz}$, normalized to the SM predictions for the various parameters. The parameters directly extracted from the fit are the products $(\sigma_i \times \BR_{\zz})$ and the ratios ${\BR}_f/{\BR}_{\zz}$. The black error bar shows the total uncertainty in each measurement.}
\label{fig:stxs_results_norm}
\end{figure}
 
\begin{figure}[tp!]
\centering
\includegraphics[width=.95\textwidth]{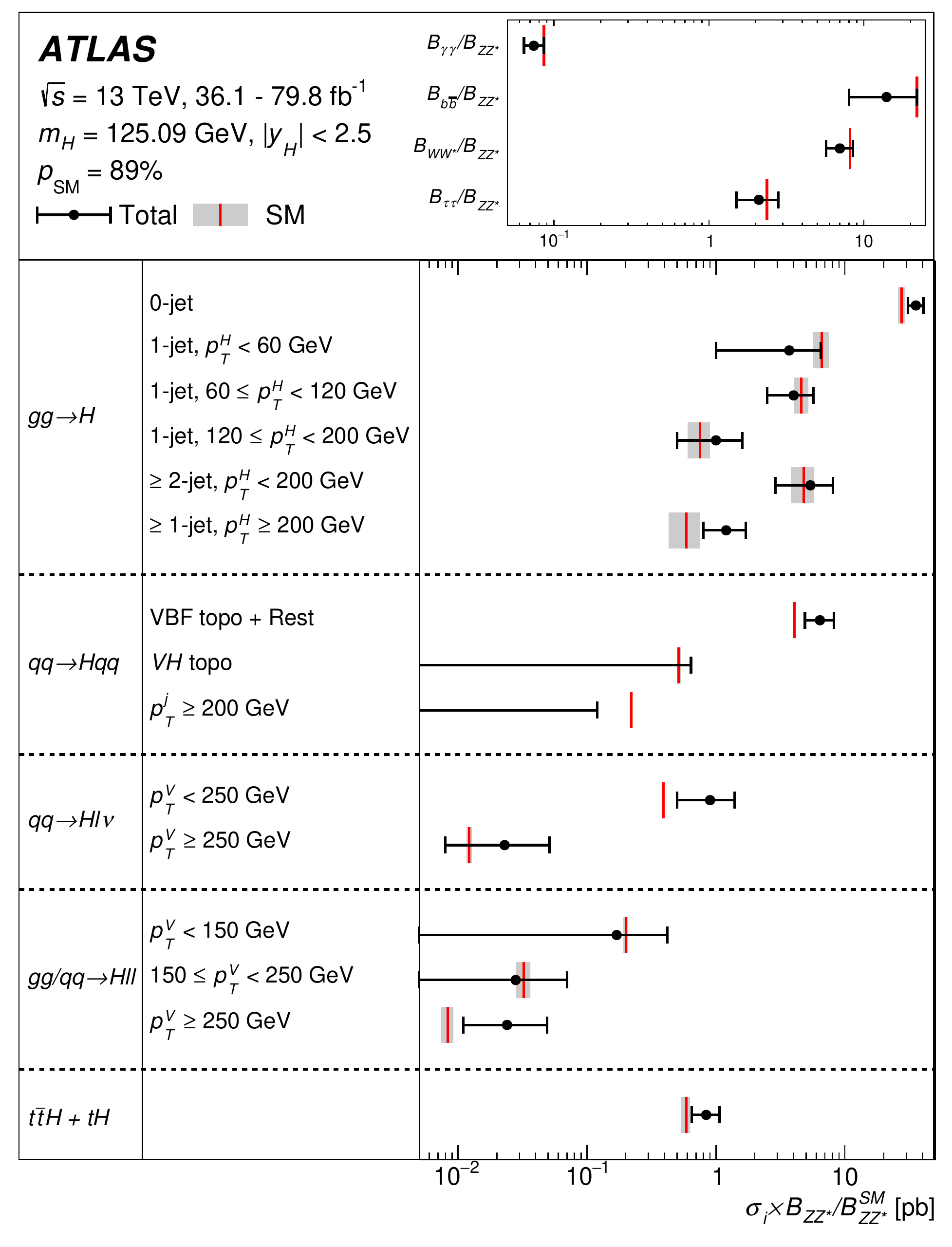}
\caption{Best-fit values and uncertainties for the cross sections in each measurement region
and of the ratios of branching fractions ${\BR}_f/{\BR}_{\zz}$. The parameters directly extracted from the fit are the products $(\sigma_i \times \BR_{\zz})$ and the ratios ${\BR}_f/{\BR}_{\zz}$; the former are shown divided by the SM value of ${\BR}_{\zz}$.
The black error bar shows the total uncertainty in each measurement.}
\label{fig:stxs_results_log}
\end{figure}
 
\begin{table}[!ht]
\caption{
Best-fit values and uncertainties for the cross sections in each measurement region, and of the ratios of branching fractions ${\BR}_f/{\BR}_{\zz}$. The total uncertainties are decomposed into components for data statistics (Stat.) and systematic uncertainties (Syst.). The SM predictions~\cite{YR4} are also shown for each quantity with their total uncertainties. The parameters directly extracted from the fit are the products $(\sigma_i \times \BR_{\zz})$ and the ratios ${\BR}_f/{\BR}_{\zz}$; the former are shown divided by the SM value of ${\BR}_{\zz}$.
}
\begin{center}
\renewcommand{\arraystretch}{1.35}
\begin{tabular}{l|rlll@{ }|l@{ }}
\hline \hline
\multirow{2}{*}{Measurement region $\displaystyle \Big( (\sigma_i \times \BR_{\zz}) / \BR^\text{SM}_{\zz} \Big)$} & \multicolumn{1}{c}{Value} & \multicolumn{3}{c|}{ Uncertainty [pb]} & \multicolumn{1}{c}{SM prediction} \\
&  \multicolumn{1}{c}{[pb]}                & Total   & Stat.                & Syst.   & \multicolumn{1}{c}{[pb]}   \\
\hline
$gg\rightarrow H,~\textrm{0-jet}$                           & $35.5$   & $\,\,^{+\;\,5.0}_{-\;\,4.7}$  & $\,\,^{+\;\,4.4}_{-\;\,4.1}$     & $\,\,^{+\;\,2.5}_{-\;\,2.2}$     & $\0\0 27.5 \pm 1.8$          \\
$gg\rightarrow H,~\textrm{1-jet}, \ptH < 60~\GeV$           & $3.7$ & $\,\,^{+\;\,2.8}_{-\;\,2.7}$     & $\,\,^{+\;\,2.4}_{-\;\,2.3}$     & $\,\,^{+\;\,1.5}_{-\;\,1.4}$     & $\0\0\0 6.6 \pm 0.9$           \\
$gg\rightarrow H,~\textrm{1-jet}, 60 \leq \ptH < 120~\GeV$  & $4.0$ & $\,\,^{+\;\,1.7}_{-\;\,1.5}$     & $\,\,^{+\;\,1.5}_{-\;\,1.4}$     & $\,\,^{+\;\,0.8}_{-\;\,0.7}$     & $\0\0\0 4.6 \pm 0.6$           \\
$gg\rightarrow H,~\textrm{1-jet}, 120 \leq \ptH < 200~\GeV$ & $1.0$ & $\,\,^{+\;\,0.6}_{-\;\,0.5}$     & ${}\pm{} 0.5$            & $\,\,^{+\;\,0.3}_{-\;\,0.2}$     & $\0\0 0.75 \pm 0.15$         \\
$gg\rightarrow H,~\geq\textrm{1-jet}, \ptH \geq 200~\GeV$   & $1.2$ & $\,\,^{+\;\,0.5}_{-\;\,0.4}$     & ${}\pm{} 0.4$            & $\,\,^{+\;\,0.3}_{-\;\,0.2}$     & $\0\0 0.59 \pm 0.16$         \\
$gg\rightarrow H,~\geq \textrm{2-jet}, \ptH < 200~\GeV$     & $5.4$ & $\,\,^{+\;\,2.7}_{-\;\,2.5}$     & $\,\,^{+\;\,2.2}  _{-\;\,2.1}$     & $\,\,^{+\;\,1.5}_{-\;\,1.3}$     & $\0\0\0 4.8 \pm 1.0$           \\
$qq\rightarrow Hqq,~\VBF\text{ topo + Rest}$                & $6.4$ & $\,\,^{+\;\,1.8}_{-\;\,1.5}$     & $\,\,^{+\;\,1.5}  _{-\;\,1.3}$     & $\,\,^{+\;\,1.1}_{-\;\,0.9}$     & $\0\0 4.07 \pm 0.09$         \\
$qq\rightarrow Hqq,~\VH\text{ topo}$                        & $-0.06$     & $\,\,^{+\;\,0.70}_{-\;\,0.58}$   & $\,\,^{+\;\,0.68} _{-\;\,0.57}$   & $\,\,^{+\;\,0.16}_{-\;\,0.12}$   & $\0 0.515 \pm 0.019$       \\
$qq\rightarrow Hqq,~\pT^{j} \geq 200~\GeV$                  & $-0.21$     & ${}\pm{} 0.33$                   & $\,\,^{+\;\,0.29} _{-\;\,0.28}$   & $\,\,^{+\;\,0.15}_{-\;\,0.16}$   & $\0 0.220 \pm 0.005$       \\
$qq\rightarrow H\ell\nu,~\pT^{V} < 250~\GeV$                & $0.90$ & $\,\,^{+\;\,0.49} _{-\;\,0.40}$  & $\,\,^{+\;\,0.40} _{-\;\,0.33}$   & $\,\,^{+\;\,0.28}_{-\;\,0.22}$   & $\0 0.393 \pm 0.009$       \\
$qq\rightarrow H\ell\nu,~\pT^{V} \geq 250~\GeV$             & $0.023$       & $\,\,^{+\;\,0.028}_{-\;\,0.015}$ & $\,\,^{+\;\,0.018}_{-\;\,0.012}$ & $\,\,^{+\;\,0.022}_{-\;\,0.008}$ & $0.0122 \pm 0.0006$     \\
$gg/qq\rightarrow H\ell\ell,~\pT^{V} < 150~\GeV$            & $ 0.17$    & $\,\,^{+\;\,0.25} _{-\;\,0.31}$  & ${}\pm{} 0.20$           & $\,\,^{+\;\,0.15}_{-\;\,0.24}$   & $\0 0.200 \pm 0.008$       \\
$gg/qq\rightarrow H\ell\ell,~150 \leq \pT^{V} < 250~\GeV$   & $0.028$       & $\,\,^{+\;\,0.042}_{-\;\,0.037}$ & $\,\,^{+\;\,0.033}_{-\;\,0.029}$ & $\,\,^{+\;\,0.026}_{-\;\,0.023}$ & $0.0324 \pm 0.0041$     \\
$gg/qq\rightarrow H\ell\ell,~\pT^{V} \geq 250~\GeV$         & $0.024$       & $\,\,^{+\;\,0.025}_{-\;\,0.013}$ & $\,\,^{+\;\,0.016}_{-\;\,0.011}$ & $\,\,^{+\;\,0.020}_{-\;\,0.006}$  & $0.0083 \pm 0.0009$     \\
$\ttH$+$\tH$                                                & $0.84$    & $\,\,^{+\;\,0.23} _{-\;\,0.19}$  & $\,\,^{+\;\,0.18} _{-\;\,0.16}$   & $\,\,^{+\;\,0.14}_{-\;\,0.11}$   & $\0\0 0.59\,\, ^{+\;\,0.04}_{-\;\,0.05}$ \\
\hline
\multirow{2}{*}{Branching fraction ratio}                   & \multirow{2}{*}{Value} & \multicolumn{3}{c|}{ Uncertainty } & \multicolumn{1}{c}{\multirow{2}{*}{SM prediction}} \\
&                         & Total   & Stat.                & Syst.               & \\
\hline
${\BR}_{\gamma\gamma}/{\BR}_{\zz}$ & $0.074$ & $\,\,^{+\;\,0.012}_{-\;\,0.010}$ & $\,\,^{+\;\,0.010}_{-\;\,0.009}$ & $\,\,^{+\;\,0.006}_{-\;\,0.005}$  & $0.0860 \pm 0.0010$     \\
${\BR}_{\bb}/{\BR}_{\zz}$          & $14$       & $\,\,^{+\;\,8}_{-\;\,6}$         & $\,\,^{+\;\,5}_{-\;\,4}$         & $\,\,^{+\;\,6}_{-\;\,5}$          & $\0\0 22.0 \pm 0.5$          \\
${\BR}_{\ww}/{\BR}_{\zz}$          & $7.0$   & $\,\,^{+\;\,1.5}_{-\;\,1.3}$     & $\,\,^{+\;\,1.1}_{-\;\,0.9}$     & $\,\,^{+\;\,1.0}_{-\;\,0.9}$      & $\0\0 8.15 \pm{} <0.01$        \\
${\BR}_{\tau\tau}/{\BR}_{\zz}$     & $2.1$   & $\,\,^{+\;\,0.7}_{-\;\,0.6}$     & ${}\pm{} 0.5$            & $\,\,^{+\;\,0.5}_{-\;\,0.3}$      & $\0\0 2.37 \pm 0.02$         \\
\hline \hline
\end{tabular}
\end{center}
\label{tab:stxs_results}
\end{table}
 
\begin{figure}[tp!]
\centering
\includegraphics[width=1.0\textwidth]{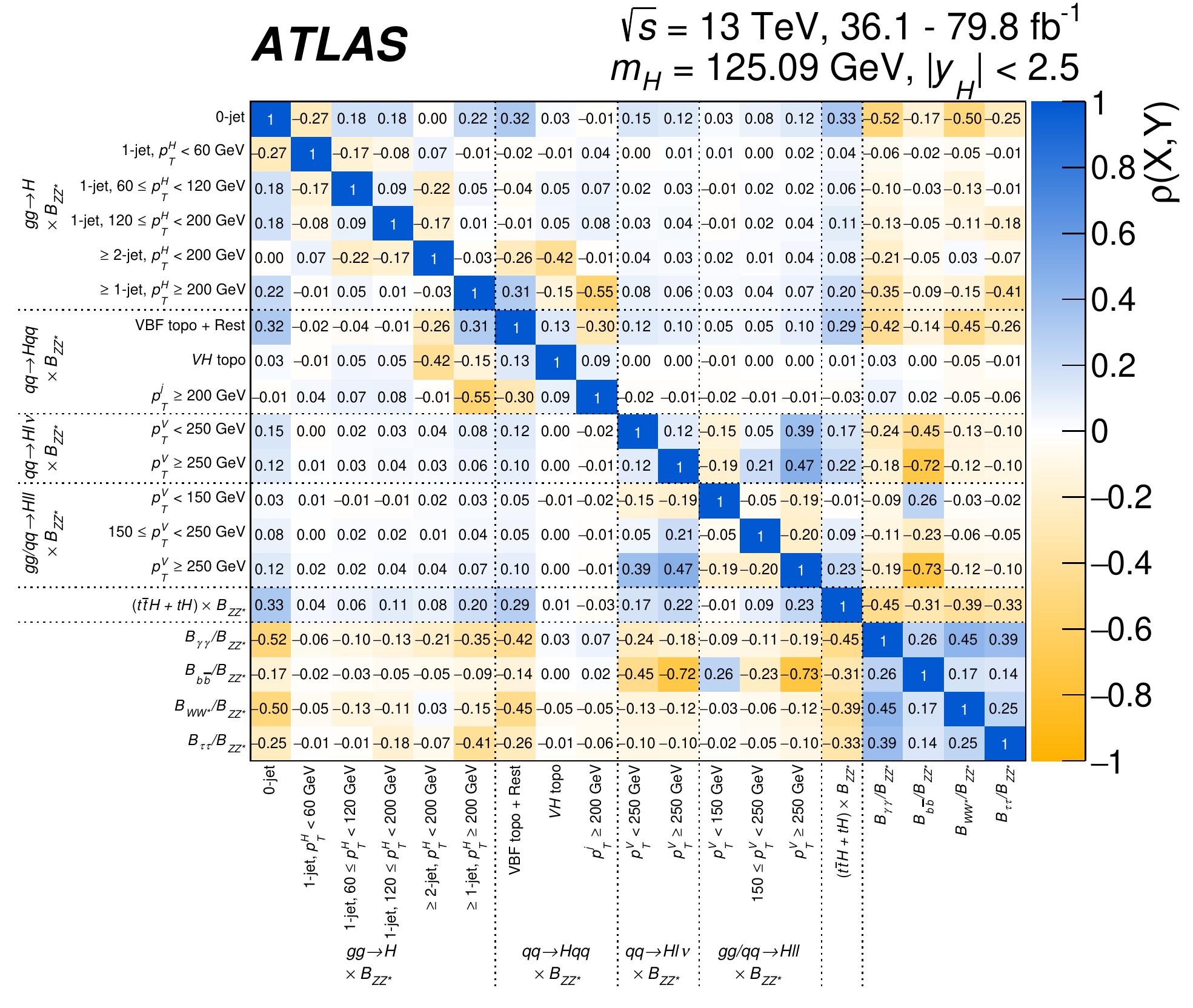}
\caption{Correlation matrix for the measured values of the simplified template cross sections and ratios of branching fractions. The fit parameters are the products $(\sigma_i \times \BR_{\zz})$ and the ratios ${\BR}_f/{\BR}_{\zz}$. 
}
\label{fig:stxs_correlation}
\end{figure}

\FloatBarrier

\section{Interpretation of results in the $\kappa$ framework}
\label{sec:kappa}
When testing the Higgs boson coupling strengths, the production cross sections $\sigma_i$, decay branching fractions $\BR_f$ and the signal-strength parameters $\mu_{if}$ defined in Eq.(~\ref{eq:mu}) cannot be treated independently, as each observed process involves at least two Higgs boson coupling strengths. Scenarios with a consistent treatment of coupling strengths in Higgs boson production and decay modes are presented in this section.
 
\subsection{Framework for coupling-strength measurements}
\label{sec:kappa:framework}
Coupling-strength modifiers $\vec{\kappa}$ are introduced to study modifications of the Higgs boson couplings related to BSM physics, within a framework~\cite{Heinemeyer:2013tqa} ($\kappa$-framework) based on the leading-order contributions to each production and decay process. Within the assumptions made in this framework, the Higgs boson production and decay can be factorized, such that the cross section times branching fraction of an individual channel $\sigma(i \to H \to f)$ contributing to a measured signal yield is parameterized as
 
\begin{equation}
\sigma_i \times \BR_f = \frac{\sigma_i(\vec{\kappa}) \times \Gamma_f(\vec{\kappa})}{\Gamma_H} ,
\label{eq:kappa:sigma_BR}
\end{equation}
 
where $\Gamma_H$ is the total width of the Higgs boson and $\Gamma_f$ is the partial width for Higgs boson decay into the final state $f$. For a given production process or decay mode $j$, the corresponding coupling-strength modifier $\kappa_j$ is defined by
 
\begin{equation*}
\kappa_j^2 = \frac{\sigma_j}{\sigma_j^\text{SM}} \quad \text{or} \quad \kappa_j^2 = \frac{\Gamma_j}{\Gamma_j^\text{SM}} .
\end{equation*}
 
The SM expectation, denoted by the label \textquote{SM}, by definition corresponds to $\kappa_j = 1$.
 
The total width of the Higgs boson is affected both by modifications of the $\kappa_{j}$, and contributions from two additional classes of Higgs boson decays: invisible decays, which are identified through an \ETmiss\ signature in the analyses described in Section~\ref{sec:channels:hinv}; and undetected decays, to which none of the analyses included in this combination are sensitive (the latter includes for instance Higgs boson decays into light quarks, or to BSM particles to which none of the input analyses provide appreciable sensitivity).
In the SM, the branching fraction for decays into invisible final states is $\sim 0.1\%$, from the \Hvvvv\ process. BSM contributions to this branching fraction and to the branching fraction to undetected final states are denoted by \Binv\ and \Bund\ respectively, with the SM corresponding to $\Binv=\Bund=0$. The Higgs boson total width is then expressed as $\Gamma_{H}(\vec{\kappa}, \Binv, \Bund) = \kappa_H^2(\vec{\kappa}, \Binv, \Bund) \, \Gamma_{H}^{\text{SM}}$ with
 
\begin{equation}
\kappa_H^2(\vec{\kappa}, \Binv, \Bund) = \frac{\sum_j \BR_f^\text{SM} \kappa_j^2}{(1 - \Binv - \Bund)} .
\label{eq:CH:1}
\end{equation}
 
Constraints on $\Binv$ are provided by the analyses described in Section~\ref{sec:channels:hinv}, but no direct constraints are included for $\Bund$. Since its value scales all observed cross sections of on-shell Higgs boson production $\sigma(\mathit{i}\to H\to\mathit{f})$ through Eqs.~\ref{eq:kappa:sigma_BR} and~\ref{eq:CH:1}, further assumptions about undetected decays must be included in order to interpret these measurements in terms of absolute coupling-strength scale factors $\kappa_{j}$. The simplest assumption is that there are no undetected Higgs boson decays and the invisible branching fraction is as predicted by the SM. An alternative, weaker assumption, is to require $\kappa_{W} \le 1$ and $\kappa_{Z} \le 1$~\cite{Heinemeyer:2013tqa}. A second alternative uses the assumption that the signal strength of off-shell Higgs boson production only depends on the coupling-strength scale factors and not on the total width~\cite{Kauer:2012hd,Caola:2013yja}, $\sigma^\text{off}(i \to H^\ast \to f) \sim \kappa_{i,\text{off}}^2 \times \kappa_{f,\text{off}}^2$. If the coupling strengths in off-shell Higgs boson production are furthermore assumed to be identical to those for on-shell Higgs boson production, $\kappa_{j,\text{off}} = \kappa_{j,\text{on}}$, and under the assumptions given in Section~\ref{sec:channels:offshell}, the Higgs boson total width can be determined from the ratio of off-shell to on-shell signal strengths~\cite{HIGG-2014-10,HIGG-2017-06}. These assumptions can also be extended to apply to $\Binv$ as well as $\Bund$, as an alternative to the measurements of Section~\ref{sec:channels:hinv}.
 
An alternative approach is to rely on measurements of ratios of coupling-strength scale factors, which can be measured without assumptions about the Higgs boson total width, since the dependence on $\Gamma_{H}$ of each coupling strength cancels in their ratios.
 
The current LHC data are insensitive to the coupling-strength modifiers $\kappa_c$ and $\kappa_s$. Thus, in the following it is assumed that $\kappa_c$ varies as $\kappa_t$ and $\kappa_s$ varies as $\kappa_b$. Other coupling modifiers ($\kappa_u$, $\kappa_d$, and $\kappa_e$) are irrelevant for the combination provided they are of order unity. The $gg \to H$, $H \to gg$, \ggZH, $H \to \gamma\gamma$ and $H \to Z\gamma$ processes are loop-induced in the SM. The $ggH$ vertex and the $H \to \gamma\gamma$ process are treated either using effective scale factors $\kappa_g$ and $\kappa_\gamma$, respectively, or expressed in terms of the more fundamental coupling-strength scale factors corresponding to the particles that contribute to the loop, including all interference effects. The \ggZH\ process is never described using an effective scale factor and always resolved in terms of modifications of the SM Higgs boson couplings to the top quark and the $Z$~boson. This assumption impacts the description of BSM effects in \ggZH, since these lead to modified production kinematics~\cite{Englert:2013vua}. However, the effect of introducing an explicit dependence on the transverse momentum of the $Z$~boson in the parameterization was found to have a negligible impact on the results at the current level of experimental precision. Similarly, the $H \to Z\gamma$ decay is always expressed in terms of the Higgs boson couplings to the $W$~boson and the $t$-quark as no analysis targeting this decay mode is included in the combination.
These relations are summarized in Table~\ref{tab:kappas}. All uncertainties in the best-fit values shown in the following take into account both the experimental and theoretical systematic uncertainties, following the procedures outlined in Section~\ref{sec:model}.
 
\begin{table}[!htbp]
\caption{Parametrizations of Higgs boson production cross sections $\sigma_i$, partial decay widths $\Gamma^f$, and the total width $\Gamma_H$, normalized to their SM values, as functions of the coupling-strength modifiers~$\kappa$. The effect of invisible and undetected decays is not considered in the expression for $\Gamma_H$.
For effective $\kappa$~parameters associated with loop processes, the resolved scaling in terms of the modifications of the Higgs boson couplings to the fundamental SM particles is given. The coefficients are derived following the methodology in Ref.~\cite{Heinemeyer:2013tqa}.}
\begin{center}
\renewcommand{\arraystretch}{1.3}
\begin{tabular}{lcccl}
\hline \hline
\multirow{2}{*}{Production}                                                          & \multirow{2}{*}{Loops}        & \multirow{2}{*}{Interference} & Effective                     & \multirow{2}{*}{Resolved modifier}                                 \\
&                               &                               & modifier                      &                                                                    \\
\hline
$\sigma({\ggF})$                                                                     & $\checkmark$                  & $t$--$b$                      & $\kappa_g^2$                  & $1.04\,\kappa_t^2 + 0.002\,\kappa_b^2 - 0.04 \, \kappa_t \kappa_b$ \\
$\sigma({\VBF})$                                                                     & -                             & -                             & -                             & $0.73\,\kappa_W^2 + 0.27\,\kappa_Z^2$                              \\
$\sigma({\qqZH})$                                                                    & -                             & -                             & -                             & $\kappa_Z^2$                                                       \\
$\sigma({\ggZH})$                                                                    & $\checkmark$                  & $t$--$Z$                      & $\kappa_{(ggZH)}$             & $2.46\,\kappa_Z^2 + 0.46\,\kappa_t^2 - 1.90\,\kappa_Z \kappa_t$    \\
$\sigma({\WH})$                                                                      & -                             & -                             & -                             & $\kappa_W^2$                                                       \\
$\sigma({\ttH})$                                                                     & -                             & -                             & -                             & $\kappa_t^2$                                                       \\
$\sigma({\WtH})$                                                                     & -                             & $t$--$W$                      & -                             & $2.91\,\kappa_t^2 + 2.31\,\kappa_W^2 - 4.22\,\kappa_t \kappa_W$    \\
$\sigma({\tHq})$                                                                     & -                             & $t$--$W$                      & -                             & $2.63\,\kappa_t^2 + 3.58\,\kappa_W^2 - 5.21\,\kappa_t \kappa_W$    \\
$\sigma({\bbH})$                                                                     & -                             & -                             & -                             & $\kappa_b^2$                                                       \\
\hline\hline
\multicolumn{5}{l}{Partial decay width}                                             \\
\hline
$\Gamma^{bb}$                                                                        & -                             & -                             & -                             & $\kappa_b^2$                                                       \\
$\Gamma^{\ww}$                                                                       & -                             & -                             & -                             & $\kappa_W^2$                                                       \\
$\Gamma^{gg}$                                                                        & $\checkmark$                  & $t$--$b$                      & $\kappa_g^2$                  & $1.11\,\kappa_t^2 + 0.01\,\kappa_b^2 - 0.12 \, \kappa_t \kappa_b$  \\
$\Gamma^{\tau\tau}$                                                                  & -                             & -                             & -                             & $\kappa_\tau^2$                                                    \\
$\Gamma^{\zz}$                                                                       & -                             & -                             & -                             & $\kappa_Z^2$                                                       \\
$\Gamma^{cc}$                                                                        & -                             & -                             & -                             & $\kappa_c^2\ (= \kappa_t^2)$                                      \\
$\Gamma^{\yy}$                                                                       & $\checkmark$                  & $t$--$W$                      & $\kappa_\gamma^2$             & $1.59\,\kappa_W^2 + 0.07\,\kappa_t^2 - 0.67\,\kappa_W \kappa_t$    \\
$\Gamma^{Z\gamma}$                                                                   & $\checkmark$                  & $t$--$W$                      & $\kappa_{(Z\gamma)}^2$        & $1.12\,\kappa_W^2 - 0.12\,\kappa_W \kappa_t$                       \\
$\Gamma^{ss}$                                                                        & -                             & -                             & -                             & $\kappa_s^2\ (= \kappa_b^2)$                                      \\
$\Gamma^{\mu\mu}$                                                                    & -                             & -                             & -                             & $\kappa_\mu^2$                                                     \\
\hline\hline
\multicolumn{5}{l}{Total width ($\Binv = \Bund = 0$)} \\
\hline
\multirow{5}{*}{$\Gamma_H$}                                                          & \multirow{5}{*}{$\checkmark$} & \multirow{5}{*}{-}            & \multirow{5}{*}{$\kappa_H^2$} & $0.58\,\kappa_b^2 + 0.22\,\kappa_W^2 $ \\
&                               &                               &                               & $\; + 0.08\,\kappa_g^2 + 0.06\,\kappa_\tau^2 $ \\
&                               &                               &                               & $\; + 0.03\,\kappa_Z^2 + 0.03\,\kappa_c^2 $                           \\
&                               &                               &                               & $\; + 0.0023\,\kappa_\gamma^2 + 0.0015\,\kappa_{(Z\gamma)}^2 $ \\
&                               &                               &                               & $\; + 0.0004\,\kappa_s^2 + 0.00022\,\kappa_\mu^2$                       \\
\hline \hline
\end{tabular}
\end{center}
\label{tab:kappas}
\end{table}
 
\FloatBarrier
 
\subsection{Fermion and gauge boson couplings}
The model studied in this section probes the universal coupling-strength scale factors $\kappa_V = \kappa_W = \kappa_Z$ for all vector bosons and $\kappa_F = \kappa_t = \kappa_b = \kappa_\tau = \kappa_\mu$ for all fermions. The effective couplings corresponding to the $ggH$ and $H \to \gamma\gamma$ vertex loops are resolved in terms of the fundamental SM couplings. It is assumed that there are no invisible or undetected Higgs boson decays, i.e.~$\Binv = \Bund = 0$. Only the relative sign between $\kappa_V$ and $\kappa_F$ is physical. As a negative relative sign has been excluded~\cite{HIGG-2015-07}, $\kappa_V \ge 0$ and $\kappa_F \ge 0$ are assumed. These definitions can be applied either globally, yielding two parameters, or separately for each of the five major decay channels, yielding ten parameters, $\kappa_V^f$ and $\kappa_F^f$ with the superscript $f$ indicating the decay mode. The best-fit values and uncertainties from a combined fit are
 
\begin{align*}
\kappa_V &= 1.05 \pm 0.04 \\
\kappa_F &= 1.05 \pm 0.09.
\end{align*}
 
Figure~\ref{fig:kf_kv} shows the results of the combined fit in the ($\kappa_V$,~$\kappa_F$)~plane as well as those of the individual decay modes in this benchmark model. Both $\kappa_V$ and $\kappa_F$ are measured to be compatible with the SM expectation. The probability of compatibility between the SM hypothesis with the best-fit point corresponds to a $p$-value of $p_\text{SM}=41\%$, computed using the procedure outlined in Section~\ref{sec:model} with two degrees of freedom. In the combined measurement a linear correlation of $44\%$ between $\kappa_V$ and $\kappa_F$ is observed.
 
\begin{figure}[h!]
\centering
\includegraphics[width=.8\textwidth]{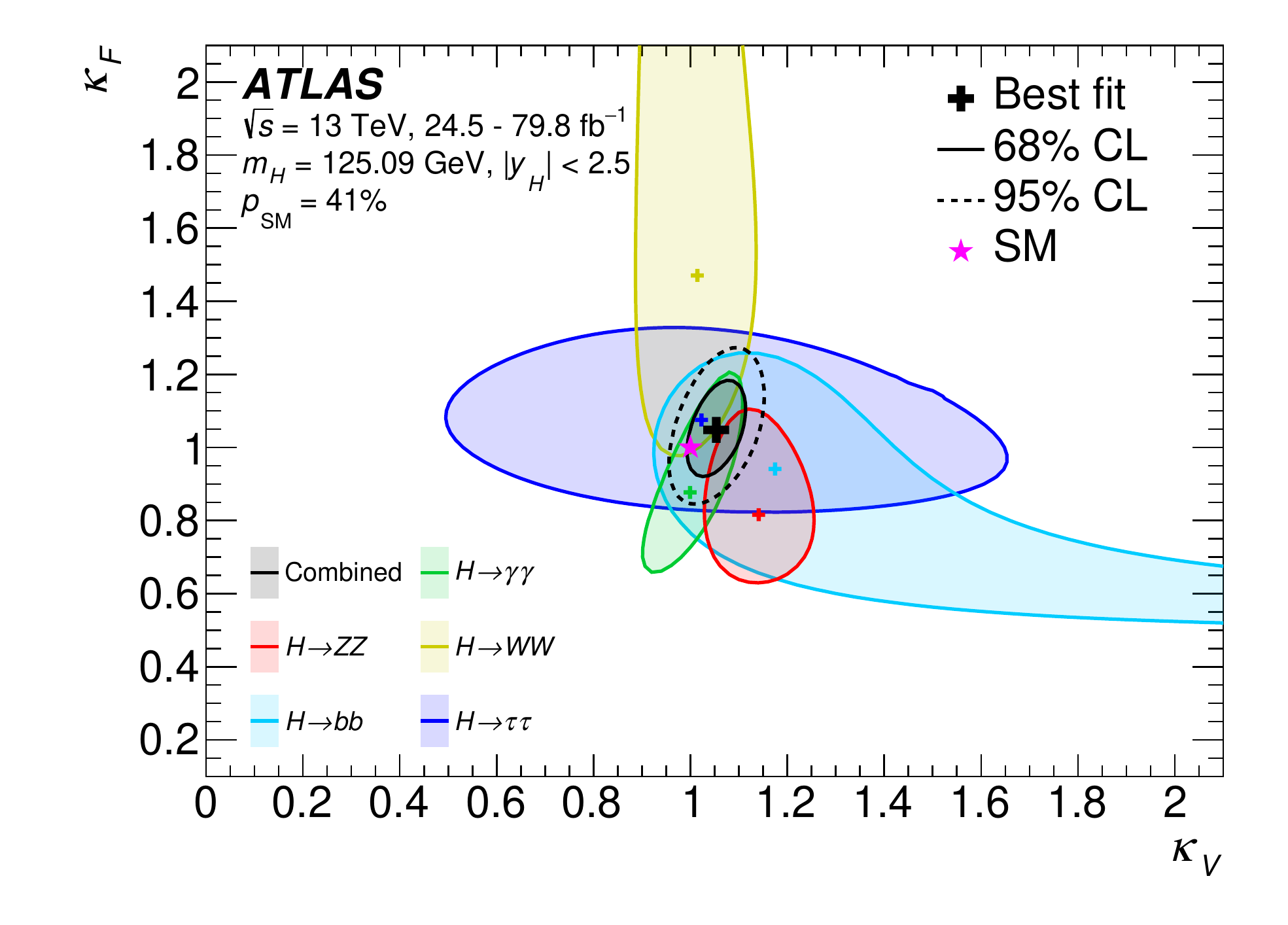}
\caption{Negative log-likelihood contours at 68\% and 95\% CL in the ($\kappa_V^f$,~$\kappa_F^f$)~plane for the individual decay modes and their combination ($\kappa_F$ versus $\kappa_V$ shown in black) assuming the coupling strengths to fermions and vector bosons to be positive. No contributions from invisible or undetected Higgs boson decays are assumed. The best-fit value for each measurement is indicated by a cross while the SM hypothesis is indicated by a star.
\label{fig:kf_kv}}
\end{figure}
 
\FloatBarrier
 
\subsection{Probing BSM contributions in loops and decays}
\label{sec:kappa:kg_ky}
To probe contributions of new particles either though loops or new final states, the effective coupling strengths to photons and gluons $\kappa_\gamma$ and $\kappa_g$ are measured. These parameters are defined to be positive as there is by construction no sensitivity to the sign of these coupling strengths. The modifiers corresponding to other loop-induced processes are resolved. The potential new particles contributing to these vertex loops may or may not contribute to the total width of the Higgs boson through direct invisible or undetected decays. In the former case, the total width is parameterized in terms of the branching fractions $\Binv$ and $\Bund$ defined in Section~\ref{sec:kappa:framework}. Furthermore, the benchmark models studied in this section assume that all coupling-strength modifiers of known SM particles are unity, i.e.~they follow the SM predictions, and that the kinematics of the Higgs boson decay products are not altered significantly.
 
Assuming $\Binv = \Bund = 0$, the best-fit values and uncertainties from a combined fit are
 
\begin{align*}
\kappa_\gamma &= 1.00 \pm 0.06 \\
\kappa_g      &= 1.03^{+0.07}_{-0.06}.
\end{align*}
 
Figure~\ref{fig:kgkgam} shows negative log-likelihood contours obtained from the combined fit in the ($\kappa_\gamma$,~$\kappa_g$)~plane. Both $\kappa_\gamma$ and $\kappa_g$ are measured to be compatible with the SM expectation. The probability of compatibility between the SM hypothesis with the best-fit point corresponds to a $p$-value of $p_\text{SM}=88\%$, computed using the procedure outlined in Section~\ref{sec:model} with two degrees of freedom. A linear correlation of $-44\%$ between $\kappa_\gamma$ and $\kappa_g$ is observed, in part due to the constraint on their product from the rate of \Hyy\ decays in the \ggF\ channel.
 
To also consider additional contributions to the total width of the Higgs boson, the assumption of no invisible or undetected decays is dropped and \Binv\ and \Bund\ are included as independent parameters in the model. The measurements sensitive to Higgs boson decays into invisible final states described in Section~\ref{sec:channels:hinv} are included in the combination and used to constrain $\Binv$. The \Bund\ parameter is constrained by decay modes that do not involve a loop process. The results from this model are
 
\begin{align*}
\kappa_\gamma &= 0.97 \pm 0.06 \\
\kappa_g      &= 0.95 \pm 0.08 \\
\Binv\        & <0.43 \text{ at 95\% CL} \\
\Bund\        & <0.12 \text{ at 95\% CL}.
\end{align*}
 
Limits on \Binv\ and \Bund\ are set using the $\tilde{t}_{\mu}$ prescription presented in Section~\ref{sec:model}. The expected upper limits at 95\% CL on \Binv\ and \Bund\ are $0.20$ and $0.31$ respectively.
The probability of compatibility between the SM hypothesis with the best-fit point corresponds to a $p$-value of $p_\text{SM}=19\%$, computed using the procedure outlined in Section~\ref{sec:model} with four degrees of freedom.
 
The results for both models are summarized in Figure~\ref{fig:kappa:summary_kgky}.
 
\clearpage
 
\enlargethispage{2cm}
 
\begin{figure}[H]
\centering
\includegraphics[width=.8\textwidth]{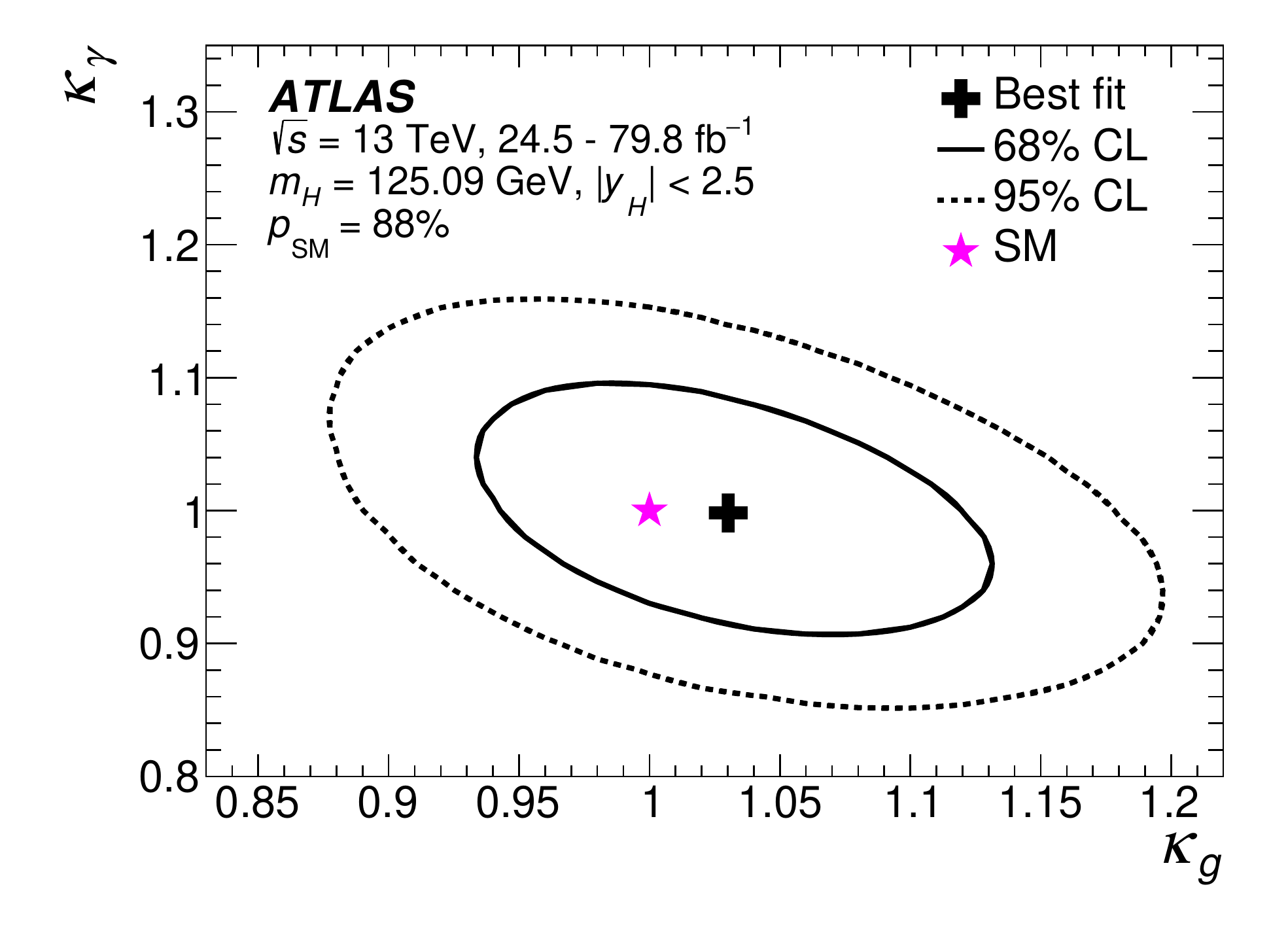}
\caption{Negative log-likelihood contours at 68\% and 95\% CL in the ($\kappa_\gamma$,~$\kappa_g$)~plane obtained from a combined fit, constraining all other coupling-strength modifiers to their SM values and assuming no contributions from invisible or undetected Higgs boson decays. The best-fit value for each measurement is indicated by a cross while the SM hypothesis is indicated by a star.
\label{fig:kgkgam}}
\end{figure}
 
\begin{figure}[H]
\centering
\includegraphics[width=.8\textwidth]{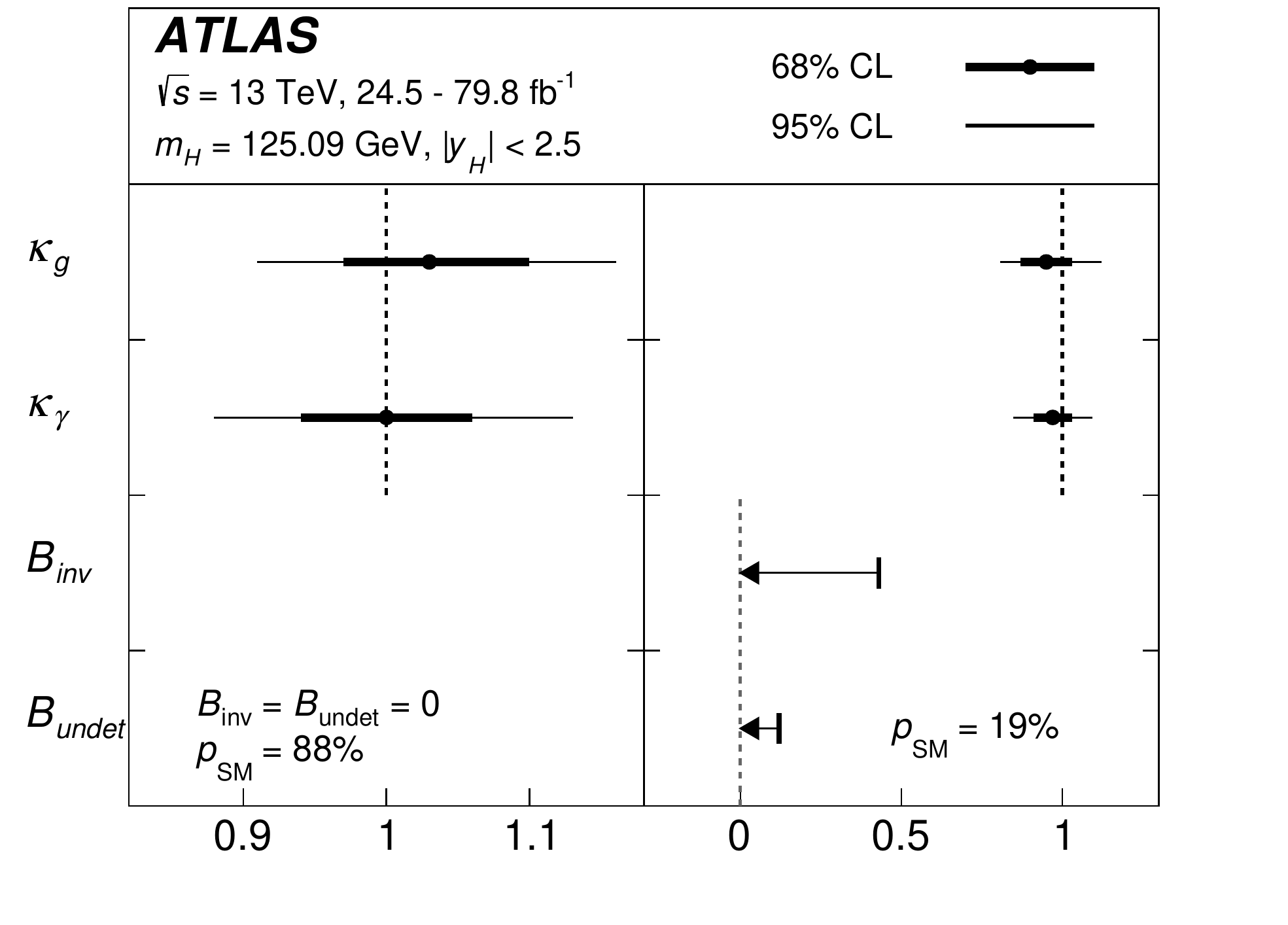}
\caption{Best-fit values and uncertainties for effective modifiers to the photon and gluon couplings of the Higgs boson, with either $\Binv=\Bund=0$ (left), or \Binv\ and \Bund\ included as free parameters (right). In the latter case, the measurements of the Higgs boson decay rate into invisible final states are included in the combination. The SM corresponds to $\kappa_\gamma = \kappa_g = 1$ and $\Binv=\Bund=0$. All coupling-strength modifiers of known SM particles are assumed to be unity, i.e.~they follow the SM predictions.}
\label{fig:kappa:summary_kgky}
\end{figure}
 
\FloatBarrier
 
\subsection{Generic parameterization assuming no new particles in loops and decays}
\label{sec:kappa:moneyplot}
In this model the scale factors for the coupling strengths to $W$, $Z$, $t$, $b$, $\tau$ and $\mu$ are treated independently. The Higgs boson couplings to second-generation quarks are assumed to scale as the couplings to the third-generation quarks. SM values are assumed for the couplings to first-generation fermions. Furthermore, it is assumed that only SM particles contribute to Higgs boson vertices involving loops, and modifications of the coupling-strength scale factors for fermions and vector bosons are propagated through the loop calculations. Invisible or undetected Higgs boson decays are assumed not to exist. All coupling-strength scale factors are assumed to be positive. The results of the \hmm\ analysis are included for this specific benchmark model. The results are shown in Table~\ref{tab:kappa:all}. The expected 95\% CL upper limit on $\kappa_\mu$ is 1.79. All measured coupling-strength scale factors in this generic model are found to be compatible with their SM expectation. The probability of compatibility between the SM hypothesis with the best-fit point corresponds to a $p$-value of $p_\text{SM}=78\%$, computed using the procedure outlined in Section~\ref{sec:model} with six degrees of freedom. Figure~\ref{fig:kappamass} shows the results of this benchmark model in terms of reduced coupling-strength scale factors, defined as
 
\begin{equation*}
y_V = \sqrt{ \kappa_V \frac{g_V}{2v} } = \sqrt{\kappa_V}\frac{m_V}{v}
\end{equation*}
 
for weak bosons with a mass $m_V$, where $g_V$ is the absolute Higgs boson coupling strength and $v=246\,\GeV$ is the vacuum expectation value of the Higgs field, and
 
\begin{equation*}
y_F = \kappa_F\frac{g_F}{\sqrt{2}} = \kappa_F\frac{m_F}{v}
\end{equation*}
 
for fermions with a mass $m_F$. For the $b$ quark and the top quark, the $\overline{MS}$ running mass evaluated at a scale of $125.09\,\GeV$ is used.
 
\begin{table}[!htbp]
\caption{
Fit results for $\kappa_{Z}$, $\kappa_{W}$, $\kappa_{b}$, $\kappa_{t}$, $\kappa_{\tau}$ and $\kappa_\mu$, all assumed to be positive. In this benchmark model BSM contributions to Higgs boson decays are assumed not to exist and Higgs boson vertices involving loops are resolved in terms of their SM content. The upper limit on $\kappa_{\mu}$ is set using the CL$_\text{s}$ prescription.}
\begin{center}
\renewcommand{\arraystretch}{1.3}
\begin{tabular}{ll}
\hline \hline
Parameter       & Result                      \\
\hline
$\kappa_{Z}$    & $1.10 \pm 0.08$      \\
$\kappa_{W}$    & $1.05 \pm 0.08$             \\
$\kappa_{b}$    & $1.06\,\,^{+\;\,0.19}_{-\;\,0.18}$      \\
$\kappa_{t}$    & $1.02\,\,^{+\;\,0.11}_{-\;\,0.10}$      \\
$\kappa_{\tau}$ & $1.07 \pm 0.15$      \\
$\kappa_\mu$    & $<1.53 \text{ at 95\% CL}$ \\
\hline \hline
\end{tabular}
\end{center}
\label{tab:kappa:all}
\end{table}
 
\begin{figure}[h!]
\centering
\includegraphics[width=.8\textwidth]{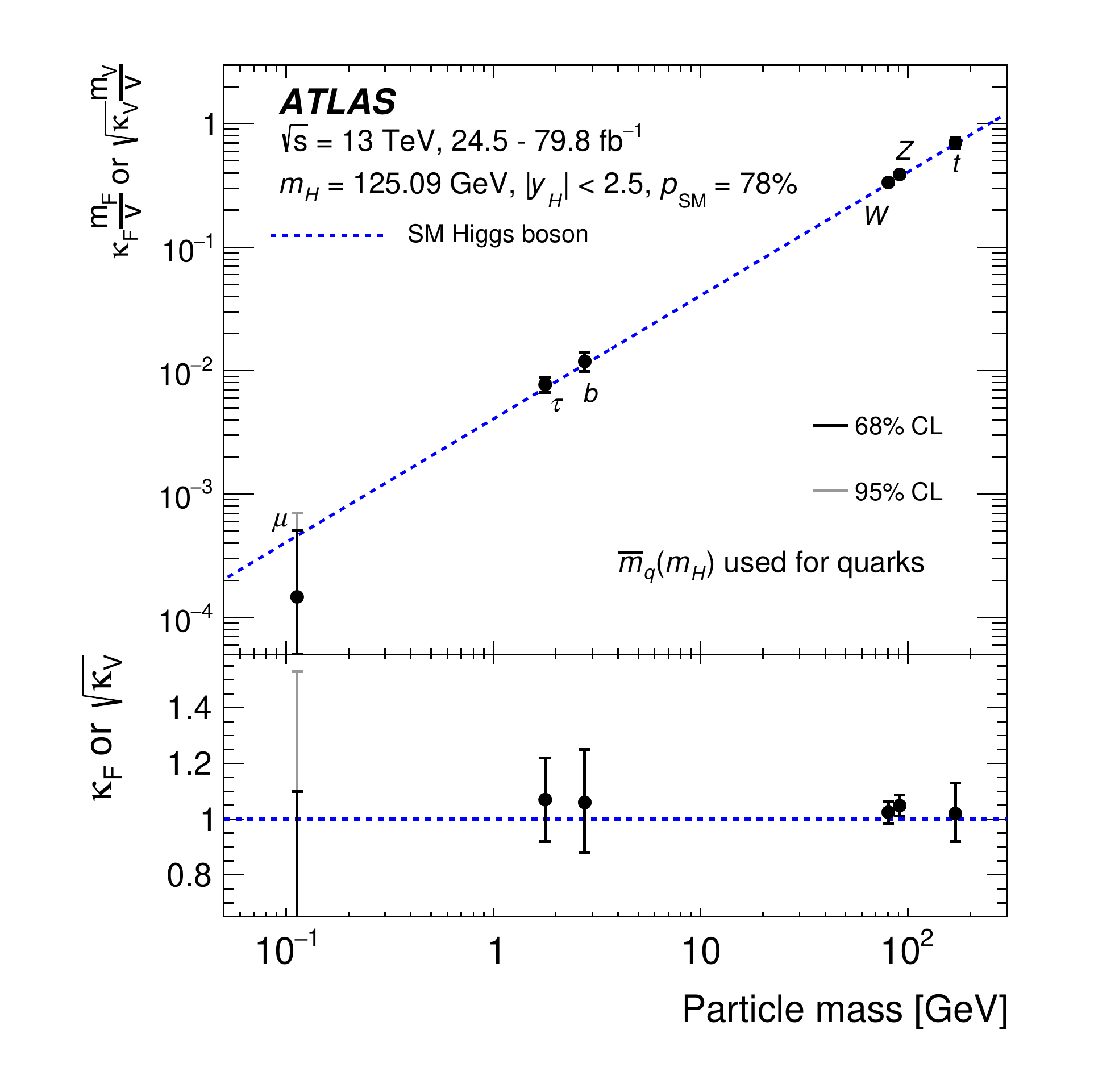}
\caption{Reduced coupling-strength modifiers $\kappa_F \frac{m_F}{v}$ for fermions ($F=t,b,\tau,\mu$) and $\sqrt{\kappa_V}\frac{m_V}{v}$ for weak gauge bosons ($V=W,Z$) as a function of their masses $m_F$ and $m_V$, respectively, and the vacuum expectation value of the Higgs field $v=246\,\GeV$. The SM prediction for both cases is also shown (dotted line). The black error bars represent $68\%$ CL intervals for the measured parameters. For $\kappa_\mu$ the light error bars indicate the $95\%$ CL interval. The coupling modifiers $\kappa_F$ and $\kappa_V$ are measured assuming no BSM contributions to the Higgs boson decays, and the SM structure of loop processes such as \ggF, \Hyy\ and \Hgg. The lower inset shows the ratios of the values to their SM predictions.}
\label{fig:kappamass}
\end{figure}
 
\FloatBarrier
 
\subsection{Generic parameterization including effective photon and gluon couplings with and without BSM contributions in decays}
\label{sec:kappa:all}
The models considered in this section are based on the same parameterization as the one in Section~\ref{sec:kappa:moneyplot} but the \ggF, \Hgg\ and \Hyy\ loop processes are parameterized using the effective coupling-strength modifiers $\kappa_g$ and $\kappa_\gamma$, similar to the benchmark model probed in Section~\ref{sec:kappa:kg_ky}.
 
The measured parameters include $\kappa_{Z}$, $\kappa_{W}$, $\kappa_b$, $\kappa_t$, $\kappa_\tau$, $\kappa_\gamma$ and $\kappa_g$. The sign of $\kappa_t$ can be either positive or negative, while $\kappa_Z$ is assumed to be positive without loss of generality. All other model parameters are also assumed to be positive. Furthermore it is assumed that the probed for BSM effects do not affect the kinematics of the Higgs boson decay products significantly. Three alternative scenarios are considered for the total width of the Higgs boson:
\begin{enumerate}[label=(\alph*)]
\item No BSM contributions to the total width ($\Binv = \Bund = 0$).
\item Both $\Binv$ and $\Bund$ are added as free parameters to the model. The measurements of Higgs boson decays into invisible final states described in Section~\ref{sec:channels:hinv} are included in the combination, for these results only, and used to provide a constraint on $\Binv$. The conditions $\kappa_W \le 1$ and $\kappa_Z \le 1$ are used to provide a constraint on $\Bund$ as discussed in Section~\ref{sec:kappa:framework}.
\item A single free parameter $\BBSM = \Binv + \Bund$ is added to the model. The measurements of off-shell production described in Section~\ref{sec:channels:offshell} are included in the combination, for these results only, and used to provide a constraint on \BBSM\ under the assumptions listed in Section~\ref{sec:kappa:framework}.
\end{enumerate}
 
The numerical results for the various scenarios are summarized in Table~\ref{tab:kappa:particlegeneric} and illustrated in Figure~\ref{fig:kappa_particle_loop_summary}.
Limits on \Binv, \Bund\ and \BBSM\ are set using the $\tilde{t}_{\mu}$ prescription presented in Section~\ref{sec:model}.
All probed fundamental coupling-strength scale factors, as well as the probed loop-induced coupling scale factors are measured to be compatible with their SM expectation for all explored assumptions. Upper limits are set on the fraction of Higgs boson decays into invisible or undetected decays. In scenario (b) the observed (expected) $95\%$ CL upper limits on the branching fractions are $\Binv < 0.30$ ($0.16$) and $\Bund < 0.21$ ($0.36$), and the lower limits on the couplings to vector bosons are $\kappa_Z > 0.88$ ($0.76$) and $\kappa_W > 0.85$ ($0.77$). In scenario (c), the observed (expected) upper limit on \BBSM\ is 0.49 (0.51).
The probability of compatibility between the SM hypothesis with the best-fit point in scenario (a) corresponds to a p-value of $p_\text{SM}=88\%$, computed using the procedure outlined in Section~\ref{sec:model} with seven degrees of freedom.
 
\begin{table}[!htbp]
\caption{Fit results for Higgs boson coupling modifiers per particle type with effective photon and gluon couplings and either (a) $\Binv = \Bund = 0$, (b) $\Binv$ and $\Bund$ included as free parameters, the conditions $\kappa_{W,Z} \leq 1$ applied and the measurement of the Higgs boson decay rate into invisible final states included in the combination, or (c) $\BBSM = \Binv + \Bund$ included as a free parameter, the measurement of off-shell Higgs boson production included in the combination, and the assumptions described in the text applied to the off-shell coupling-strength scale factors. The SM corresponds to $\Binv=\Bund=\BBSM=0$ and all $\kappa$ parameters set to unity. All parameters except $\kappa_t$ are assumed to be positive.}
\begin{center}
\renewcommand{\arraystretch}{1.3}
\resizebox{\columnwidth}{!}{
\begin{tabular}{llll}
\hline \hline
Parameter & (a) $\Binv = \Bund = 0$ & (b) \Binv\ free, $\Bund \ge 0$, $\kappa_{W,Z} \le 1$  & (c) $\BBSM \ge 0$, $\kappa_{\text{off}} = \kappa_{\text{on}}$ \\
\hline
$\kappa_{Z}$    & $1.11 \pm 0.08$                    & $>0.88 \text{ at 95\% CL}$                            & $1.20\,\,^{+\;\,0.18}_{-\;\,0.17}$      \\
$\kappa_{W}$    & $1.05 \pm 0.09$                    & $>0.85 \text{ at 95\% CL}$                            & $1.15 \pm 0.18$        \\
$\kappa_{b}$    & $1.03\,\,^{+\;\,0.19}_{-\;\,0.17}$ & $0.85\,\,^{+\;\,0.15}_{-\;\,0.13}$                    & $1.14\,\,^{+\;\,0.21}_{-\;\,0.25}$      \\
$\kappa_{t}$    & $1.09\,\,^{+\;\,0.15}_{-\;\,0.14}$ & $[-1.08, -0.77] \cup [0.96, 1.23] \text{ at 68\% CL}$ & $1.18 \pm 0.23$     \\
$\kappa_{\tau}$ & $1.05\,\,^{+\;\,0.16}_{-\;\,0.15}$ & $0.99 \pm 0.14$                                       & $1.16\,\,^{+\;\,0.22}_{-\;\,0.24}$      \\
$\kappa_\gamma$ & $1.05 \pm 0.09$                    & $0.96\,\,^{+\;\,0.08}_{-\;\,0.06}$                    & $1.16\,\,^{+\;\,0.17}_{-\;\,0.18}$      \\
$\kappa_g$      & $0.99\,\,^{+\;\,0.11}_{-\;\,0.10}$ & $1.05\,\,^{+\;\,0.12}_{-\;\,0.14}$                    & $1.08\,\,^{+\;\,0.17}_{-\;\,0.18}$      \\
\Binv\          & -                                  & $ <0.30 \text{ at 95\% CL}$                           & -                           \\
\Bund\          & -                                  & $ <0.21 \text{ at 95\% CL}$                           & -                           \\
\BBSM\          & -                                  & -                                                     & $ <0.49 \text{ at 95\% CL}$ \\
\hline \hline
\end{tabular}
}
\end{center}
\label{tab:kappa:particlegeneric}
\end{table}
 
\begin{figure}[!htbp]
\centering
\includegraphics[width=.75\textwidth]{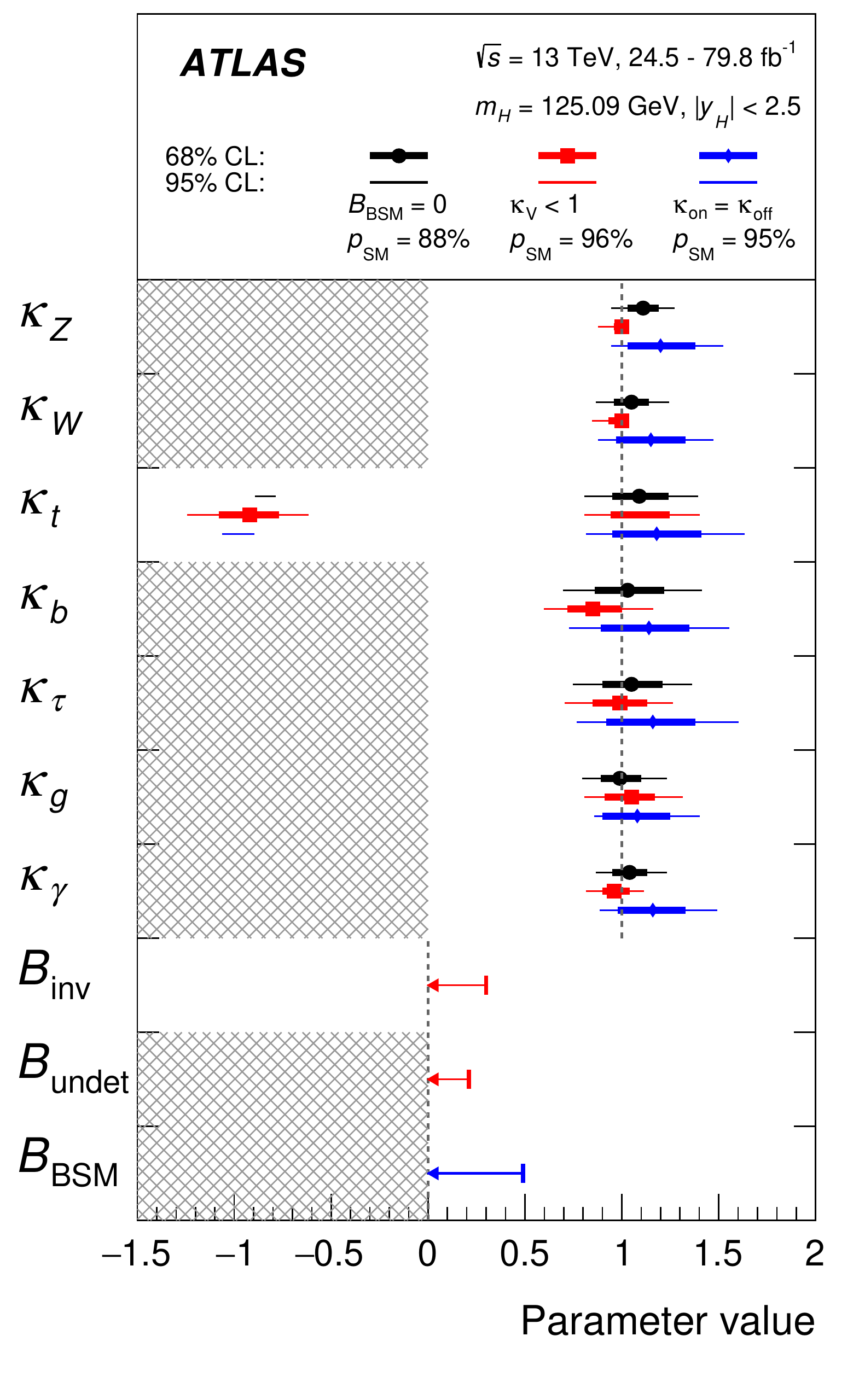}
\caption{Best-fit values and uncertainties for Higgs boson coupling modifiers per particle type with effective photon and gluon couplings and either $\Binv = \Bund = 0$ (black); $\Binv$ and $\Bund$ included as free parameters, the conditions $\kappa_{W,Z} \leq 1$ applied and the measurement of the Higgs boson decay rate into invisible final states included in the combination (red); or $\BBSM = \Binv + \Bund$ included as a free parameter, the measurement of off-shell Higgs boson production included in the combination, and the assumptions described in the text applied to the off-shell coupling-strength scale factors (blue). The SM corresponds to $\Binv=\Bund=0$ and all $\kappa$ parameters set to unity. All parameters except $\kappa_t$ are assumed to be positive.}
\label{fig:kappa_particle_loop_summary}
\end{figure}
 
\FloatBarrier
 
\subsection{Generic parameterization using ratios of coupling modifiers}
\label{sec:kappa:ratios}
The five absolute coupling-strength scale factors and two effective loop-coupling scale factors measured in the previous benchmark model are expressed as ratios of scale factors that can be measured independent of any assumptions about the Higgs boson total width. The model parameters are defined in Table~\ref{tab:kappa:generic}. All parameters are assumed to be positive. This parameterization represents the most model-independent determination of coupling-strength scale factors that is currently possible in the $\kappa$-framework. The numerical results from the fit to this benchmark model are summarized in Table~\ref{tab:kappa:generic} and visualized in Figure~\ref{fig:summary_kappa_generic}. All model parameters are measured to be compatible with their SM expectation. The probability of compatibility between the SM hypothesis with the best-fit point corresponds to a $p$-value of $p_\text{SM}=85\%$, computed using the procedure outlined in Section~\ref{sec:model} with seven degrees of freedom.
 
The parameter $\lambda_{WZ}$ in this model is of particular interest: identical coupling-strength scale factors for the $W$ and $Z$ bosons are required within tight bounds by the $\mathrm{SU(2)}$ custodial symmetry and the $\rho$ parameter measurements at LEP and at the Tevatron~\cite{ALEPH:2010aa}.  The ratio $\lambda_{\gamma Z}$ is sensitive to new charged particles contributing to the {\hgg} loop unlike in {\hzz} decays. Similarly, the ratio $\lambda_{tg}$ is sensitive to new colored particles contributing through the {\ggF} loop unlike in {\ttH} events. The observed values are in agreement with the SM expectation.
 
\begin{table}[!htbp]
\caption{
Best-fit values and uncertainties for ratios of coupling modifiers. The second column provides the expression of the measured parameters in terms of the coupling modifiers defined in previous sections. All parameters are defined to be unity in the SM.}
\begin{center}
\renewcommand{\arraystretch}{1.3}
\begin{tabular}{lll}
\hline \hline
Parameter &  \begin{tabular}{@{}c@{}} Definition in terms \\ of $\kappa$ modifiers \end{tabular} & Result \\
\hline
$\kappa_{gZ}$        & $  \kappa_g\kappa_Z/\kappa_H$ & $1.06 \pm 0.07$ \\
$\lambda_{tg}$       & $  \kappa_t/\kappa_g$         & $1.10\,\,^{+\;\,0.15}_{-\;\,0.14}$ \\
$\lambda_{Zg}$       & $  \kappa_Z/\kappa_g$         & $1.12\,\,^{+\;\,0.15}_{-\;\,0.13}$ \\
$\lambda_{WZ}$       & $  \kappa_W/\kappa_Z$         & $0.95 \pm 0.08$ \\
$\lambda_{\gamma Z}$ & $  \kappa_\gamma/\kappa_Z$    & $0.94 \pm 0.07$ \\
$\lambda_{\tau Z}$   & $  \kappa_\tau/\kappa_Z$      & $0.95 \pm 0.13$ \\
$\lambda_{bZ}$       & $  \kappa_b/\kappa_Z$         & $0.93\,\,^{+\;\,0.15}_{-\;\,0.13}$ \\
\hline \hline
\end{tabular}
\end{center}
\label{tab:kappa:generic}
\end{table}
 
\begin{figure}[!htbp]
\centering
\includegraphics[width=.8\textwidth]{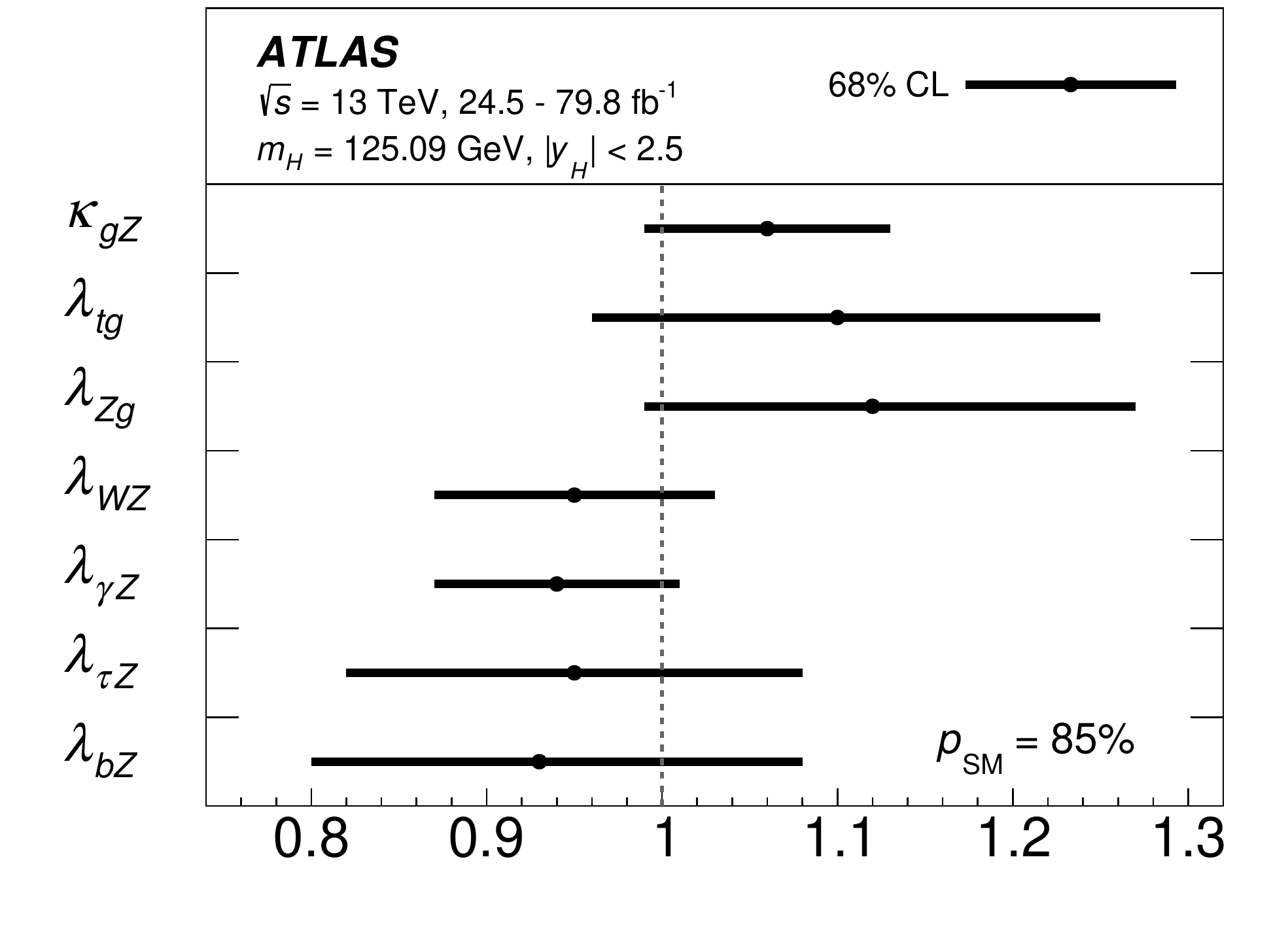}
\caption{Measured ratios of coupling modifiers. The dashed line indicates the SM value of unity for each parameter.}
\label{fig:summary_kappa_generic}
\end{figure}
 
\FloatBarrier

\section{Constraints on new phenomena}
\label{sec:bsm}
Two-Higgs-doublet models (2HDMs)~\cite{Lee:1973iz,Gunion:2002zf,Branco:2011iw,Heinemeyer:2013tqa} and supersymmetry~\cite{Golfand:1971iw,Volkov:1973ix,Wess:1974tw,Wess:1974jb,Ferrara:1974pu,Salam:1974ig} are promising extensions of the SM. The measurements are interpreted in these benchmark models, providing indirect limits on their parameters that are complementary to those obtained by direct searches for new particles. The interpretations presented in this section follow the procedure discussed in Ref.~\cite{HIGG-2015-03}.
 
\subsection{Two-Higgs-doublet model}
In 2HDMs, the SM Higgs sector is extended by introducing an additional complex isodoublet scalar field with weak hypercharge one. Four types of 2HDMs satisfy the Paschos--Glashow--Weinberg condition~\cite{Glashow:1976nt,Paschos:1976ay}, which prevents the appearance of tree-level flavor-changing neutral currents:
 
\begin{itemize}
\item Type~I: One Higgs doublet couples to vector bosons, while the other one couples to fermions. The first doublet is \textit{fermiophobic} in the limit where the two Higgs doublets do not mix.
\item Type~II: One Higgs doublet couples to up-type quarks and the other one to down-type quarks and charged leptons.
\item Lepton-specific: The Higgs bosons have the same couplings to quarks as in the Type~I model and to charged leptons as in Type~II.
\item Flipped: The Higgs bosons have the same couplings to quarks as in the Type~II model and to charged leptons as in Type~I.
\end{itemize}
 
The observed Higgs boson is identified with the light CP-even neutral scalar $h$ predicted by 2HDMs, and its accessible production and decay modes are assumed to be the same as those of the SM Higgs boson. Its couplings to vector bosons, up-type quarks, down-type quarks and leptons relative to the corresponding SM predictions are expressed as functions of the mixing angle~$\alpha$ between~$h$ and the heavy CP-even neutral scalar, and the ratio of the vacuum expectation values of the Higgs doublets,~$\tan\beta$~\cite{HIGG-2015-03}.
 
Figure~\ref{fig:bsm:2hdm} shows the regions of the $( \cos(\beta-\alpha), \tan\beta )$~plane that are excluded at a confidence level of $95\, \%$ or higher, for each of the four types of 2HDMs. The expected exclusion limits in the SM hypothesis are also overlaid. The data are consistent with the alignment limit~\cite{Branco:2011iw} at~$\cos(\beta - \alpha) = 0$, in which the couplings of~$h$ match those of the SM Higgs boson, within one standard deviation or better in each of the tested models. The allowed regions also include  narrow, curved \textit{petal} regions at positive $\cos(\beta - \alpha)$ and moderate $\tan \beta$ in the Type~II, Lepton-specific, and Flipped~models. These correspond to regions with~$\cos(\beta + \alpha) \approx 0$, for which some fermion couplings have the same magnitude as in the SM, but the opposite sign.
 
\begin{figure}[tbp]
\centering
\subfloat[]{\includegraphics[width=0.49\textwidth]{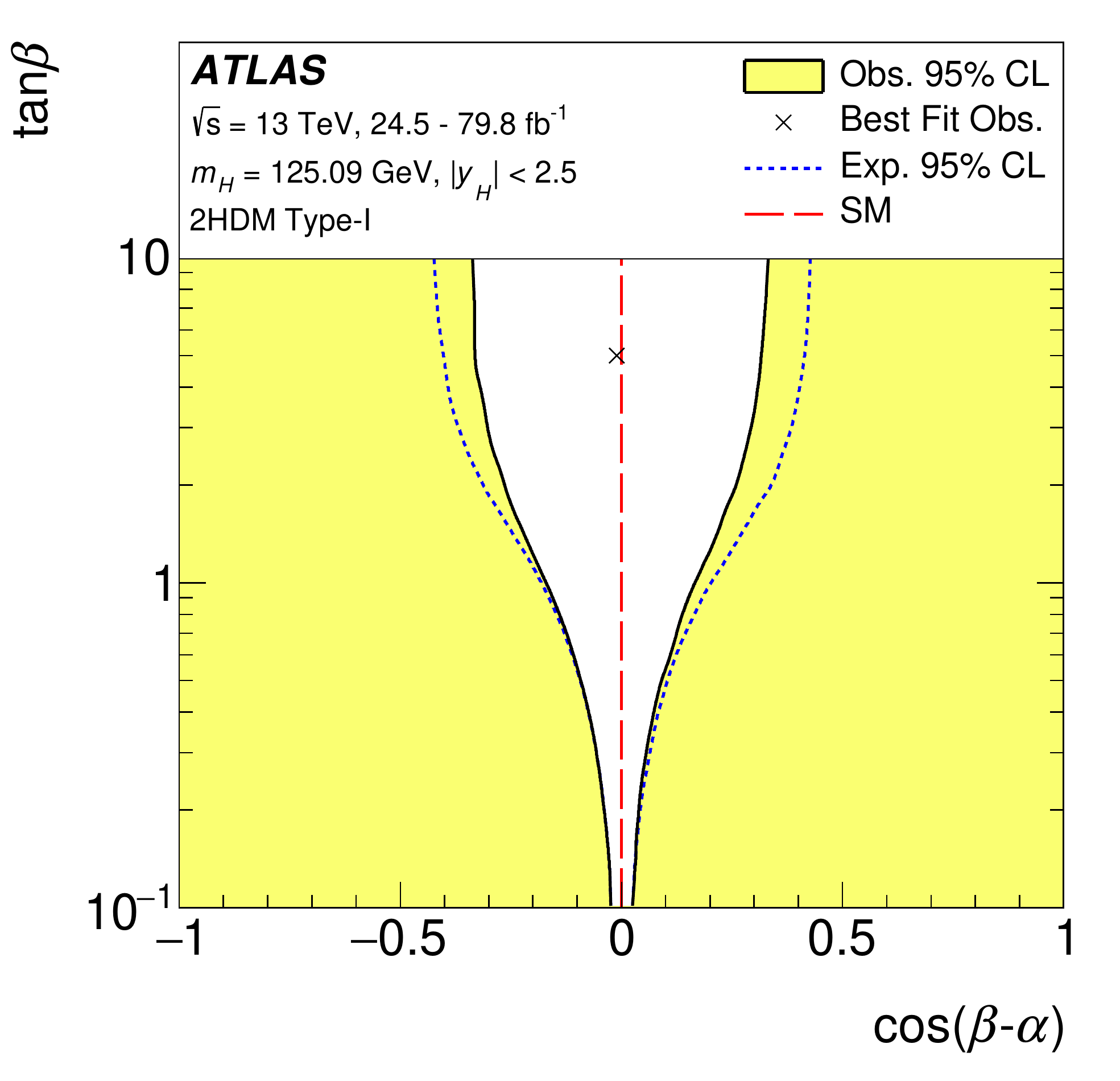}}
\subfloat[]{\includegraphics[width=0.49\textwidth]{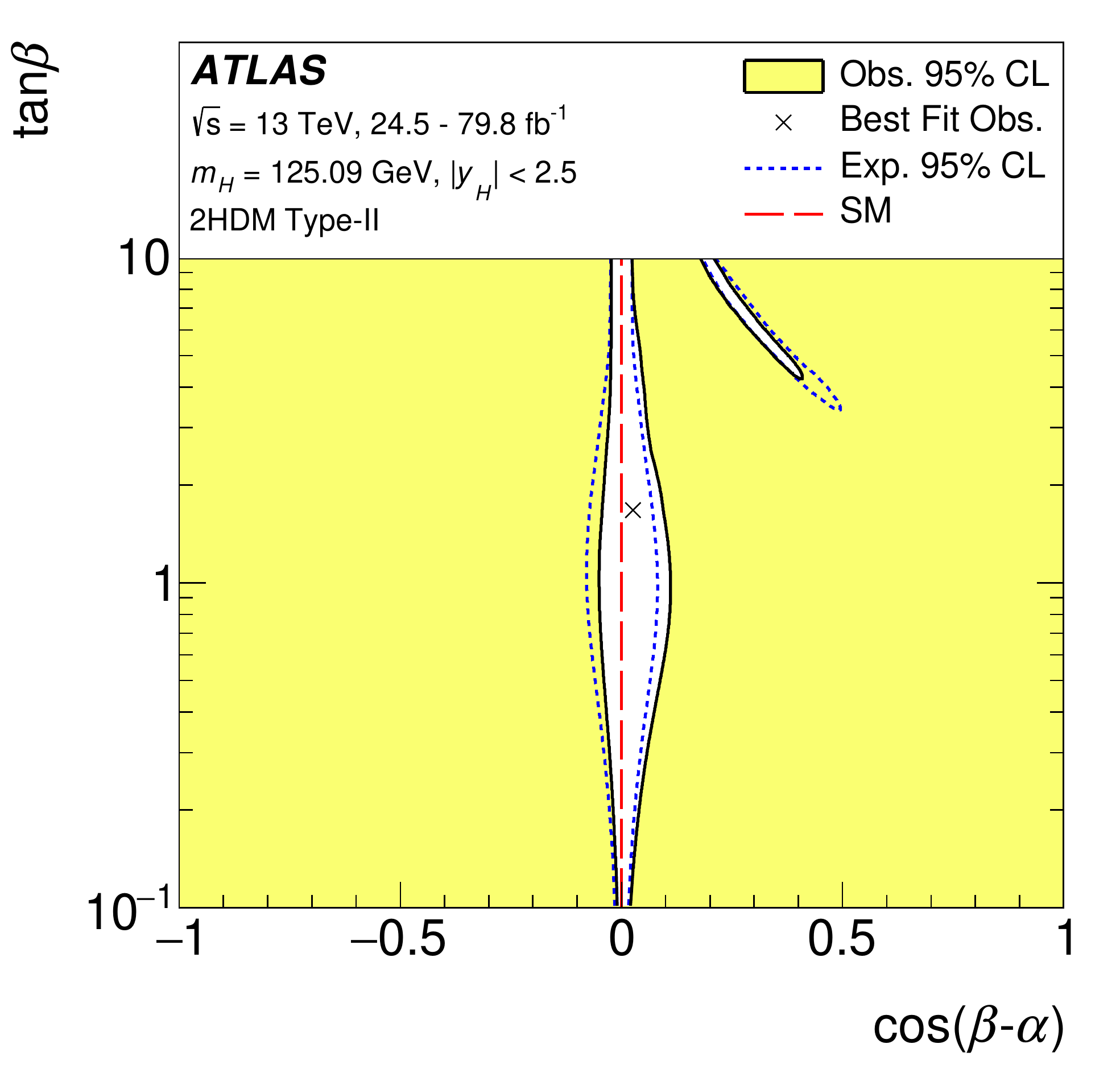}} \\
\subfloat[]{\includegraphics[width=0.49\textwidth]{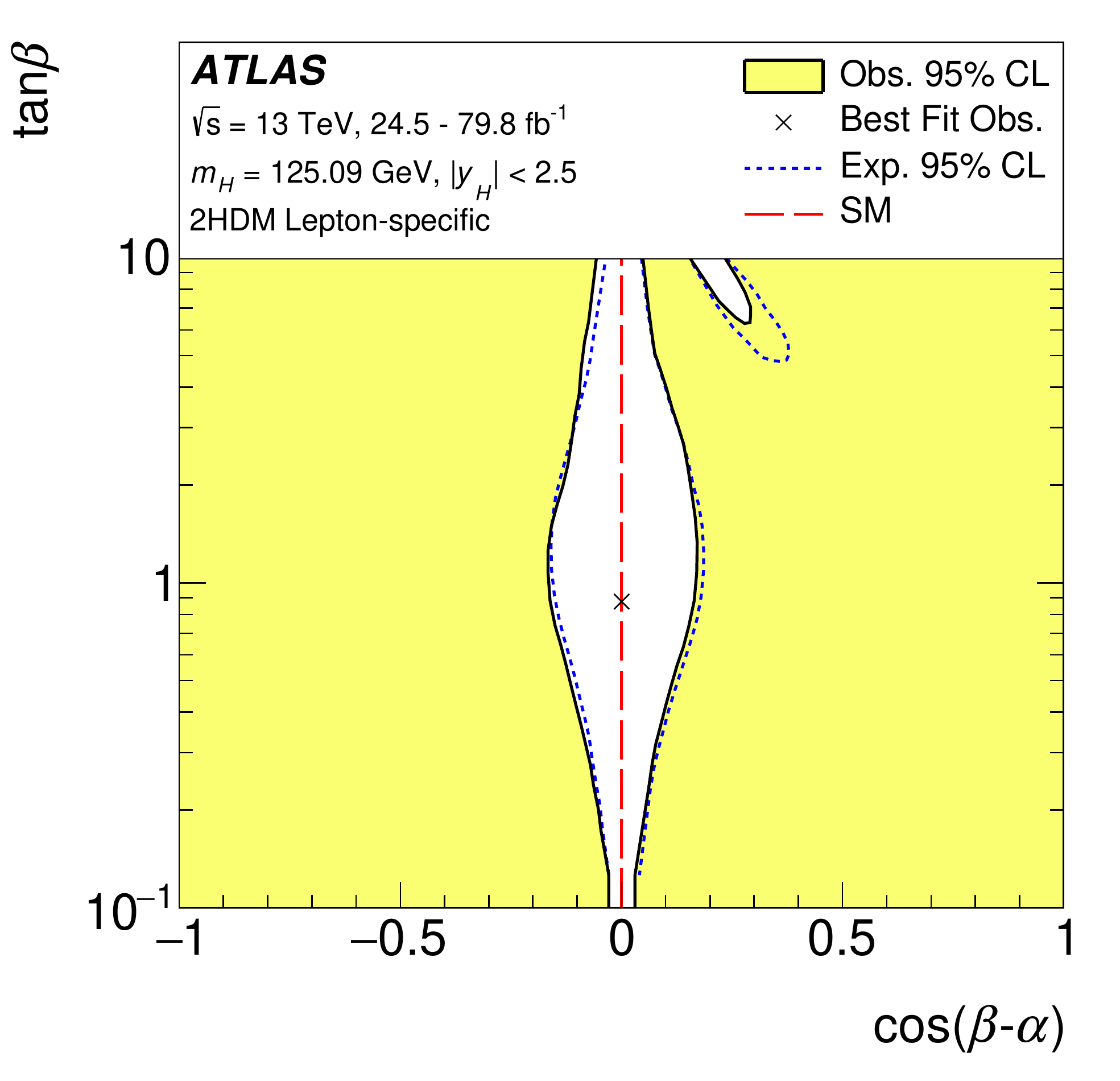}}
\subfloat[]{\includegraphics[width=0.49\textwidth]{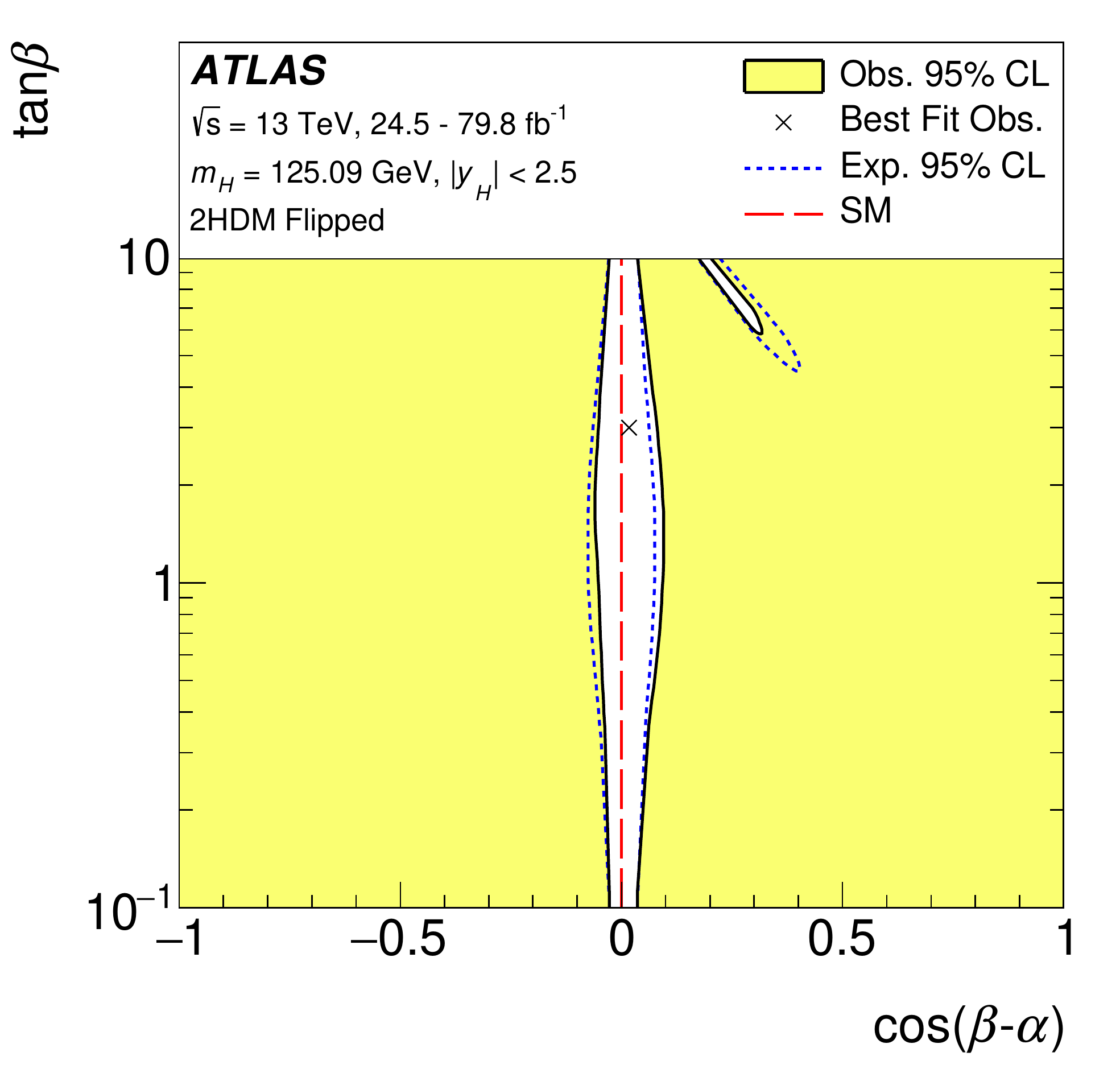}}
\caption{Regions of the $(\cos(\beta - \alpha), \tan \beta)$ plane of four types of 2HDMs excluded by fits to the measured rates of Higgs boson production and decays. Contours at 95\% CL, defined in the asymptotic approximation by $-2 \ln \Lambda = 5.99$, are drawn for both the data and the expectation for the SM Higgs sector. The cross in each plot marks the observed best-fit value.
The angles $\alpha$ and $\beta$ are taken to satisfy $0 \leq \beta \leq \pi/2$ and $0 \leq \beta - \alpha \leq \pi$ without loss of generality. The alignment limit at $\cos(\beta - \alpha) = 0$, in which all Higgs boson couplings take their SM values, is indicated by the dashed red line.}
\label{fig:bsm:2hdm}
\end{figure}
 
\subsection{Simplified Minimal Supersymmetric Standard Model}
The scalar sector of the Minimal Supersymmetric Standard Model (MSSM)~\cite{Fayet:1974pd,Fayet:1976et,Fayet:1977yc} is a realization of a Type~II 2HDM. As a benchmark, a simplified MSSM model in which the Higgs boson is identified with the light CP-even scalar~$h$, termed hMSSM~\cite{Maiani:2013hud,Djouadi:2013uqa,Djouadi:2015jea}, is studied. The assumptions made in this model are discussed in Ref.~\cite{HIGG-2015-03}. Notably, the hMSSM is a good approximation of the MSSM only for moderate values of $\tan\beta$. For $\tan\beta \gtrsim 10$ the scenario is approximate due to missing supersymmetry corrections in the Higgs boson coupling to $b$-quarks, and for $\tan\beta$ of $O(1)$ the precision of the approximation depends on $m_A$, the mass of the CP-odd scalar~\cite{YR4}. The production and decay modes accessible to $h$ are assumed to be the same as those of the SM Higgs boson.
 
The Higgs boson couplings to vector bosons, up-type fermions and down-type fermions relative to the corresponding SM predictions are expressed as functions of the ratio of the vacuum expectation values of the Higgs doublets, $\tan\beta$, $m_A$, and the masses of the $Z$~boson and of~$h$.
 
Figure~\ref{fig:bsm:mssm} shows the regions of the hMSSM parameter space that are indirectly excluded by the measurement of the Higgs boson production and decay rates. The data are consistent with the SM decoupling limit at large $m_A$, where the $h$~couplings tend to those of the SM Higgs boson. The observed (expected) lower limit at $95\, \%$ CL on the CP-odd Higgs boson mass is at least $m_A > 480\,\GeV$ ($m_A > 400\, \GeV$) for $1 \leq \tan\beta \leq 25$, increasing to $m_A > 530\,\GeV$ ($m_A > 450\, \GeV$) at $\tan\beta = 1$. The observed limit is stronger than the expected limit because the hMSSM model exhibits a physical boundary $\kappa_V \le 1$, but the Higgs boson coupling to vector bosons is measured to be larger than the SM value, as presented in Section~\ref{sec:kappa}.
 
\begin{figure}[tbp]
\centering
\includegraphics[width=0.7\textwidth]{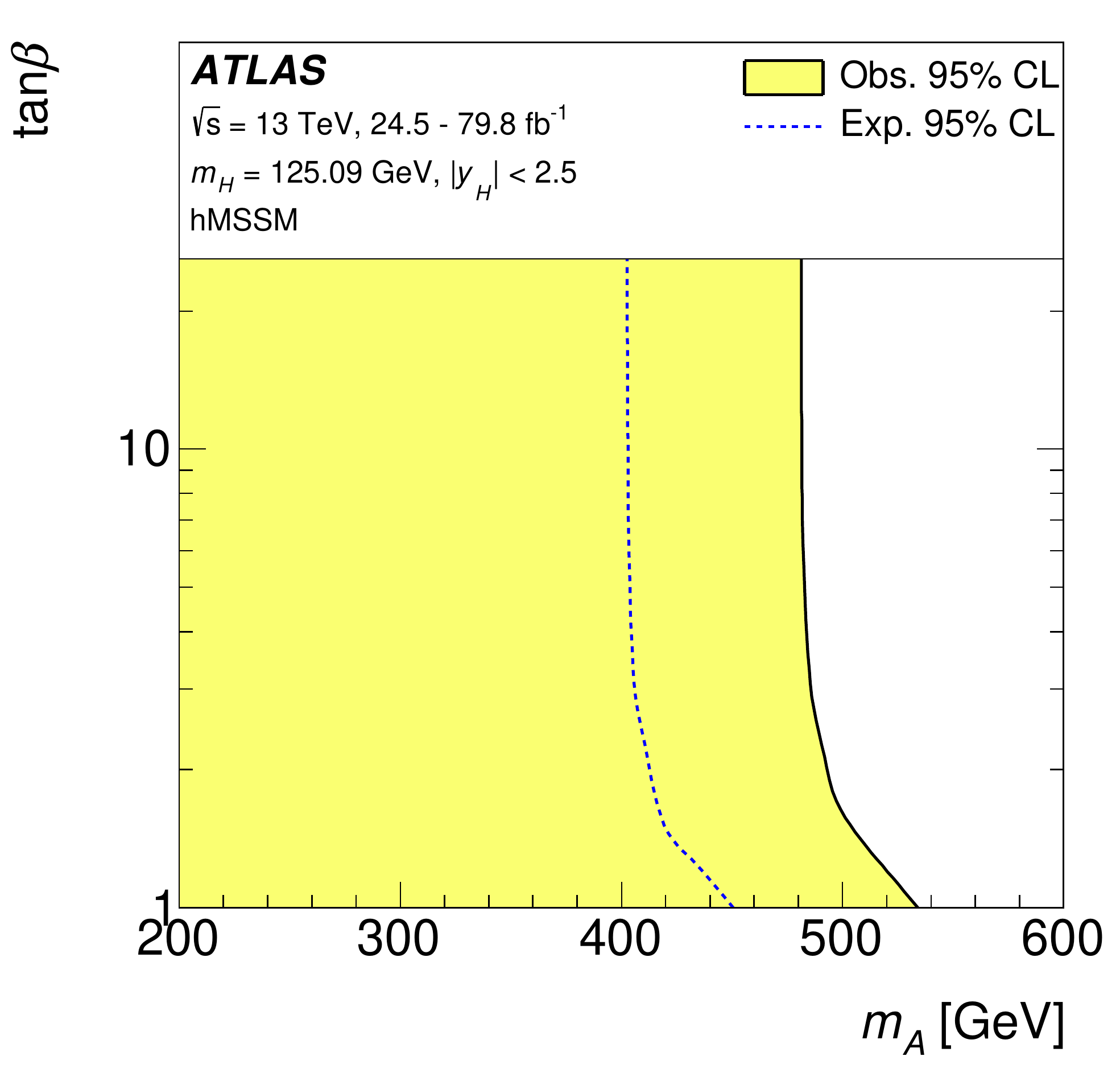}
\caption{Regions of the $(m_A, \tan \beta)$ plane in the hMSSM excluded by fits to the measured rates of Higgs boson production and decays. Likelihood contours at 95\% CL, defined in the asymptotic approximation by $-2 \ln \Lambda = 5.99$, are drawn for both the data and the expectation of the SM Higgs sector. The regions to the left of the solid contour are excluded. The decoupling limit, in which all Higgs boson couplings tend to their SM value, corresponds to $m_A \to \infty$.}
\label{fig:bsm:mssm}
\end{figure}
\FloatBarrier

\section{Conclusions}
\label{sec:conclusion}
Measurements of Higgs boson production cross sections and branching fractions have been performed using up to $79.8\, \ifb$ of $pp$ collision data produced by the LHC at $\sqrt{s}=$~13\,\TeV~and recorded by the ATLAS detector. The results presented in this paper are based on the combination of analyses of the \Hyy, \hzz, \hww, \htt, \hbb\ and \Hmm\ decay modes, searches for decays into invisible final states, as well as on measurements of off-shell Higgs boson production.
 
The global signal strength is determined to be $\mu = 1.11^{+0.09}_{-0.08}$.

The Higgs boson production cross sections within the region $|y_H|<2.5$ are measured in a combined fit for the gluon--gluon fusion process, vector-boson fusion, the associated production with a $W$ or $Z$ boson and the associated production with top quarks, assuming the SM Higgs boson branching fractions. The combined measurement leads to an observed (expected) significance for the vector-boson fusion production process of $6.5\sigma$ ($5.3\sigma$). For the $\VH$ production mode the observed (expected) significance is $5.3\sigma$ ($4.7\sigma$). The $\ttH + \tH$ processes are measured with an observed (expected) significance of $5.8\sigma$ ($5.3\sigma$).
 
Removing the assumption of SM branching fractions, a combined fit is performed for the production cross section times branching fraction for each pair of production and decay processes to which the combined analyses are sensitive. Results are also presented for a model in which these quantities are expressed using the cross section of the $gg \to H \to ZZ^*$ process, ratios of production cross sections relative to that of \ggF\ production, and ratios of branching fractions relative to that of $\hzz$.
 
Cross sections are measured in 15~regions of Higgs boson production kinematics defined within the simplified template cross-section framework, which primarily characterize the transverse momentum of the Higgs boson, the topology of associated jets and the transverse momentum of associated vector bosons. The measurements in all regions are found to be compatible with SM predictions.
 
The observed Higgs boson yields are used to obtain confidence intervals for $\kappa$ modifiers to the couplings of the SM Higgs boson to fermions, weak vector bosons, gluons, and photons and to the branching fraction of the Higgs boson into invisible and undetected decay modes. A variety of physics-motivated constraints on the Higgs boson total width are explored: Using searches for $\Hinv$ and constraints on couplings to vector bosons, the branching fraction of invisible Higgs boson decays into BSM particles is constrained to be less than $30\%$ at $95\%$~CL, while the branching fraction of decays into undetected particles is less than $22\%$ at $95\%$~CL. The overall branching fraction of the Higgs boson into BSM decays is determined to be less than $47\%$ at $95\%$~CL using measurements of off-shell Higgs boson production in combination with measurements of SM Higgs boson production and rates. No significant deviation from the SM predictions is observed in any of the benchmark models studied.
 
Finally, the results are interpreted in the context of two-Higgs-doublet models and the hMSSM. Constraints are set in the $(m_A, \tan \beta)$ plane of the hMSSM and the $(\cos(\beta - \alpha), \tan \beta)$ plane in 2HDM Type-I, Type-II, Lepton-specific and Flipped models.


\section*{Acknowledgments}
 

We thank CERN for the very successful operation of the LHC, as well as the
support staff from our institutions without whom ATLAS could not be
operated efficiently.
 
We acknowledge the support of ANPCyT, Argentina; YerPhI, Armenia; ARC, Australia; BMWFW and FWF, Austria; ANAS, Azerbaijan; SSTC, Belarus; CNPq and FAPESP, Brazil; NSERC, NRC and CFI, Canada; CERN; CONICYT, Chile; CAS, MOST and NSFC, China; COLCIENCIAS, Colombia; MSMT CR, MPO CR and VSC CR, Czech Republic; DNRF and DNSRC, Denmark; IN2P3-CNRS and CEA-DRF/IRFU, France; SRNSFG, Georgia; BMBF, HGF and MPG, Germany; GSRT, Greece; RGC and Hong Kong SAR, China; ISF and Benoziyo Center, Israel; INFN, Italy; MEXT and JSPS, Japan; CNRST, Morocco; NWO, Netherlands; RCN, Norway; MNiSW and NCN, Poland; FCT, Portugal; MNE/IFA, Romania; MES of Russia and NRC KI, Russia Federation; JINR; MESTD, Serbia; MSSR, Slovakia; ARRS and MIZ\v{S}, Slovenia; DST/NRF, South Africa; MINECO, Spain; SRC and Wallenberg Foundation, Sweden; SERI, SNSF and Cantons of Bern and Geneva, Switzerland; MOST, Taiwan; TAEK, Turkey; STFC, United Kingdom; DOE and NSF, United States of America. In addition, individual groups and members have received support from BCKDF, CANARIE, Compute Canada and CRC, Canada; ERC, ERDF, Horizon 2020, Marie Sk{\l}odowska-Curie Actions and COST, European Union; Investissements d'Avenir Labex, Investissements d'Avenir Idex and ANR, France; DFG and AvH Foundation, Germany; Herakleitos, Thales and Aristeia programmes co-financed by EU-ESF and the Greek NSRF, Greece; BSF-NSF and GIF, Israel; CERCA Programme Generalitat de Catalunya and PROMETEO Programme Generalitat Valenciana, Spain; G\"{o}ran Gustafssons Stiftelse, Sweden; The Royal Society and Leverhulme Trust, United Kingdom.
 
The crucial computing support from all WLCG partners is acknowledged gratefully, in particular from CERN, the ATLAS Tier-1 facilities at TRIUMF (Canada), NDGF (Denmark, Norway, Sweden), CC-IN2P3 (France), KIT/GridKA (Germany), INFN-CNAF (Italy), NL-T1 (Netherlands), PIC (Spain), ASGC (Taiwan), RAL (UK) and BNL (USA), the Tier-2 facilities worldwide and large non-WLCG resource providers. Major contributors of computing resources are listed in Ref.~\cite{ATL-GEN-PUB-2016-002}.
 

\clearpage
\appendix
\part*{Appendix}
\addcontentsline{toc}{part}{Appendix}
\section{Simplified template cross-section measurement results with finer granularity}\label{sec:stxs_weak}
This section presents measurements of STXS parameters in a model that has finer granularity than the model of Section~\ref{sec:stxs_results}, and is thus closer to the original proposal of Stage 1 STXS in Refs.~\cite{YR4,LesHouches}. The changes relative to the model of Section~\ref{sec:stxs_results} are as follows: in the \ggtoH\ process, the region defined by $\ptH \ge 200\,\gev$ and $\ge 1$ jets is split into separate bins for 1 jet and $\ge 2$ jets; a VBF-topology (\textit{VBF topo}) region is defined for events with $\geq 2$ jets using the same selection as in the \qqtoHqq\ process; the remaining $\ge 2$ jet events are separated into three bins of \ptH\ in the same way as the 1-jet events; in the \qqtoHqq\ process, the \textit{VBF topo + Rest} region is split into separate bins for \textit{VBF topo} and \textit{Rest}; and in the \qqtoWH\ process, the $\pT^V < 250\,\GeV$ region is split into two bins for $\pT^V < 150\,\GeV$ and $150 \leq \pT^V < 250\,\GeV$, matching the binning used in \pptoZH.
The results are shown in Figures~\ref{fig:stxs_results_weak} and \ref{fig:stxs_correlation_weak}.
 
\begin{figure}[H]
\centering
\includegraphics[width=.55\textwidth]{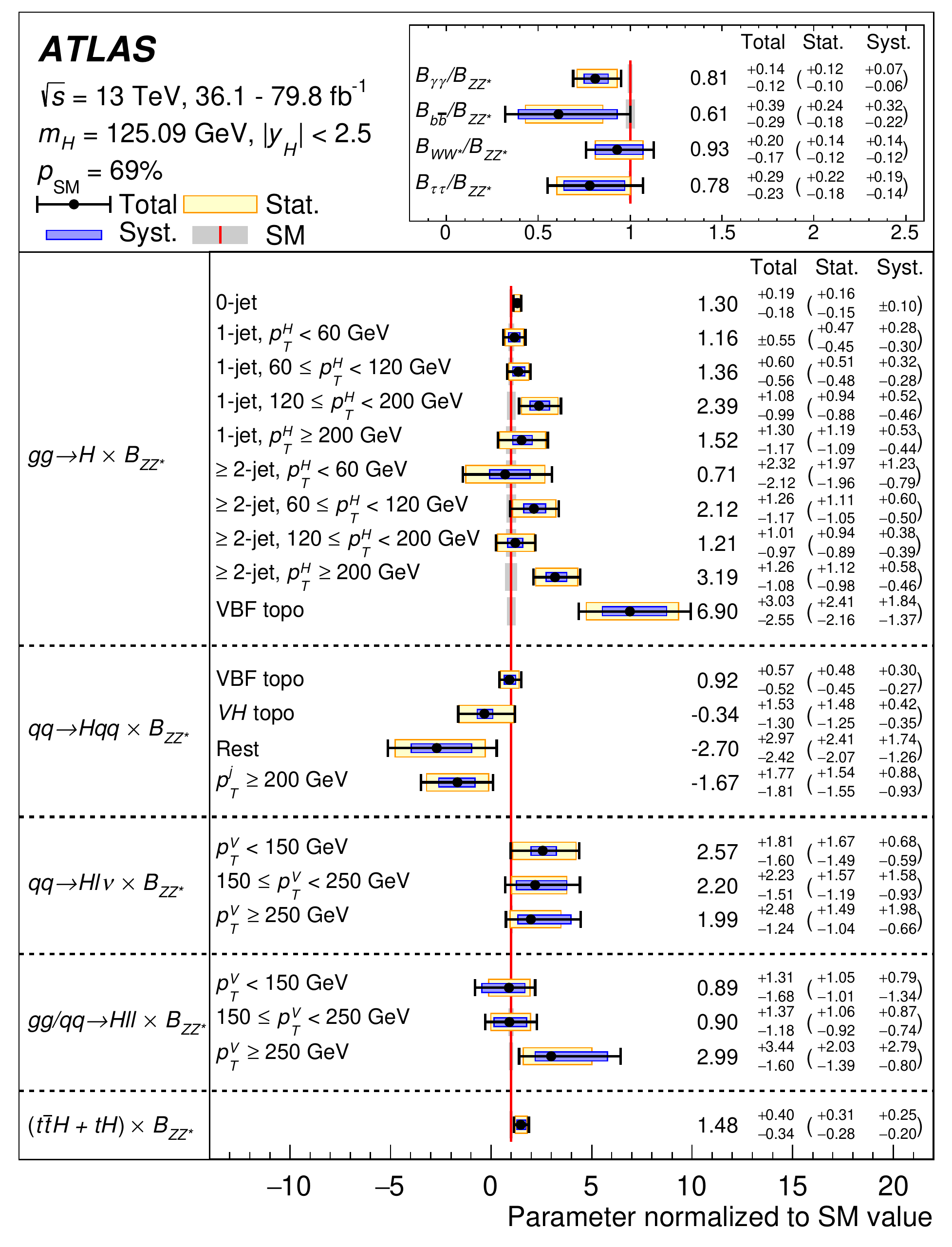}
\caption{Best-fit values and uncertainties for the cross sections in each measurement region times the \hzz\ branching fraction in a model with finer granularity.
The results are shown normalized to the SM predictions for the various parameters. The black error bar shows the total uncertainty in each measurement.}
\label{fig:stxs_results_weak}
\end{figure}

\begin{figure}[H]
\centering
\includegraphics[width=.9\textwidth]{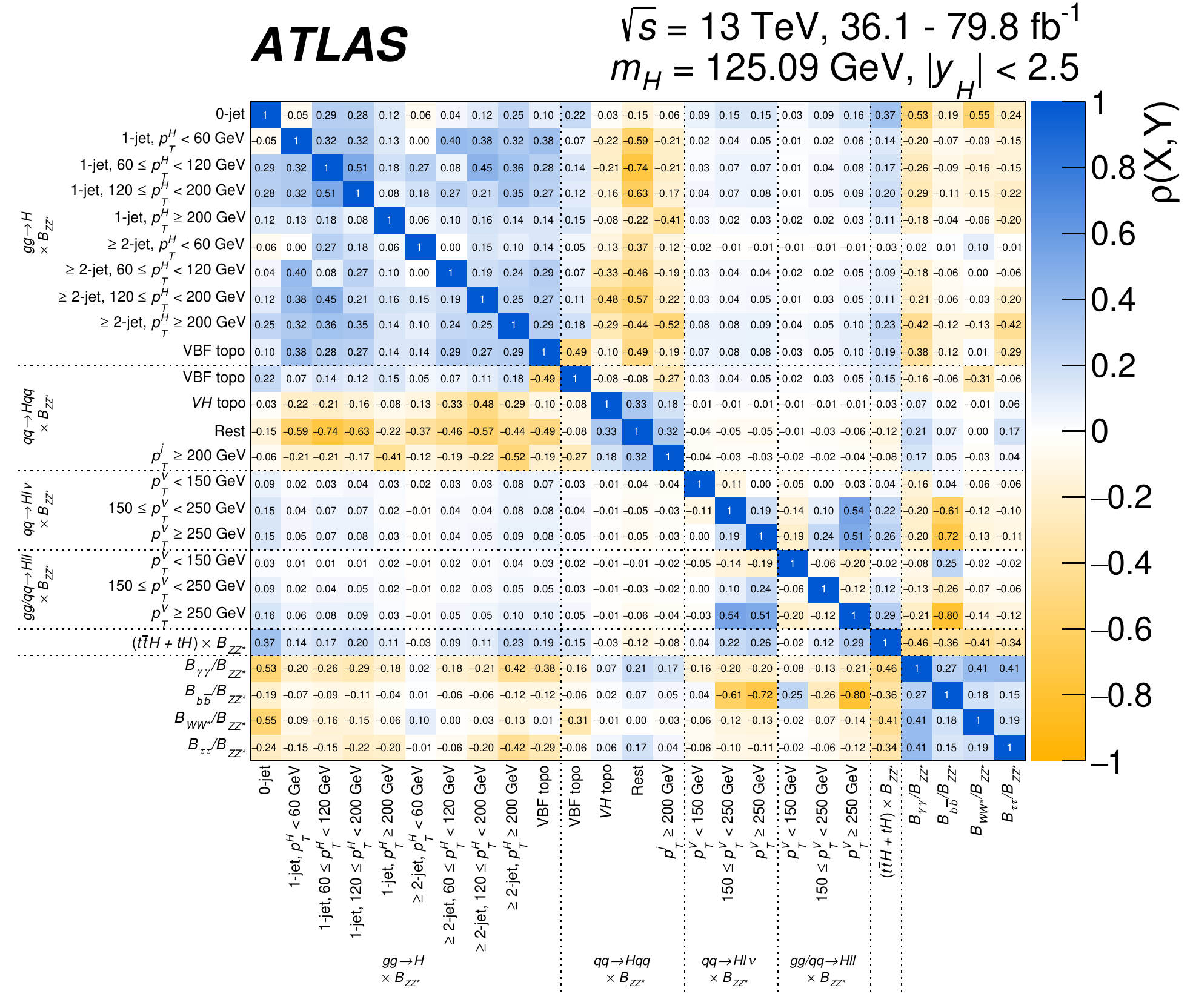}
\caption{Correlation matrix for the measured values of the simplified template cross sections in each measurement region times the \hzz\ branching fraction in a model with finer granularity.
}
\label{fig:stxs_correlation_weak}
\end{figure}
\FloatBarrier
\clearpage
 
\printbibliography
 
\clearpage
 
\begin{flushleft}
{\Large The ATLAS Collaboration}

\bigskip

G.~Aad$^\textrm{\scriptsize 102}$,    
B.~Abbott$^\textrm{\scriptsize 129}$,    
D.C.~Abbott$^\textrm{\scriptsize 103}$,    
O.~Abdinov$^\textrm{\scriptsize 13,*}$,    
A.~Abed~Abud$^\textrm{\scriptsize 71a,71b}$,    
K.~Abeling$^\textrm{\scriptsize 53}$,    
D.K.~Abhayasinghe$^\textrm{\scriptsize 94}$,    
S.H.~Abidi$^\textrm{\scriptsize 167}$,    
O.S.~AbouZeid$^\textrm{\scriptsize 40}$,    
N.L.~Abraham$^\textrm{\scriptsize 156}$
% ...
\bigskip
\\

% ...
$^{13}$Institute of Physics, Azerbaijan Academy of Sciences, Baku; Azerbaijan.\\
% ...
$^{40}$Niels Bohr Institute, University of Copenhagen, Copenhagen; Denmark.\\
% ...
$^{53}$II. Physikalisches Institut, Georg-August-Universit\"{a}t G\"ottingen, G\"ottingen; Germany.\\
% ...
$^{71}$$^{(a)}$INFN Sezione di Pavia;$^{(b)}$Dipartimento di Fisica, Universit\`a di Pavia, Pavia; Italy.\\
% ...
$^{94}$Department of Physics, Royal Holloway University of London, Egham; United Kingdom.\\
% ...
$^{102}$CPPM, Aix-Marseille Universit\'e, CNRS/IN2P3, Marseille; France.\\
$^{103}$Department of Physics, University of Massachusetts, Amherst MA; United States of America.\\
% ...
$^{129}$Homer L. Dodge Department of Physics and Astronomy, University of Oklahoma, Norman OK; United States of America.\\
% ...
$^{156}$Department of Physics and Astronomy, University of Sussex, Brighton; United Kingdom.\\
% ...
$^{167}$Department of Physics, University of Toronto, Toronto ON; Canada.\\
% ...
$^{a}$ Also at Borough of Manhattan Community College, City University of New York, New York NY; United States of America.\\
$^{b}$ Also at Centre for High Performance Computing, CSIR Campus, Rosebank, Cape Town; South Africa.\\
% ...
$^{*}$ Deceased

\end{flushleft}

 
\setcounter{section}{0}

\end{document}